\def\FullVersion{} % 
\def\ShowAuthor{} %
\def\AckSection{} %
\ifdefined\LLNCS
\documentclass{style/llncs}
\renewcommand{\paragraph}{\subsubsection}
\else
\documentclass[12pt,letterpaper]{article}
\usepackage{amsthm}
\usepackage[margin=1in]{geometry}
\fi
\usepackage{graphicx,color,url,booktabs,comment}
\usepackage{authblk}
\usepackage{amsmath, amssymb,amsbsy}
\usepackage{multicol}%%
\usepackage{framed}%%
\usepackage{enumitem}%%
\usepackage{bbm}
\usepackage{mathrsfs}
\usepackage{braket}
\usepackage{algorithm}
\usepackage{multicol}
\usepackage{algpseudocode}
\usepackage[dvipsnames]{xcolor}
\usepackage{float}
\usepackage{pagecolor}
\usepackage[utf8]{inputenc}
\usepackage{tabularx}
\usepackage[hyperfootnotes=false,,pdfstartview=FitH,colorlinks,linkcolor=blue,filecolor=blue,citecolor=blue,urlcolor=blue]{hyperref}
\usepackage{footnotebackref}
\MakeRobust{\Call}
\makeatletter
\LetLtxMacro{\BHFN@Old@footnotemark}{\@footnotemark}
\renewcommand*{\@footnotemark}{%
    \refstepcounter{BackrefHyperFootnoteCounter}%
    \xdef\BackrefFootnoteTag{bhfn:\theBackrefHyperFootnoteCounter}%
    \label{\BackrefFootnoteTag}%
    \BHFN@Old@footnotemark
}
\makeatother
\usepackage{xspace}
\usepackage{cleveref} %% for referencing use \cref
\usepackage{mdframed}
\usepackage{amsfonts}
\usepackage{tikz}
\usepackage{tikz-cd}
\usepackage{tikz-qtree}
\usepackage{adjustbox}
\usepackage{blindtext}
\usepackage{makecell}
\usepackage{hhline}
\usepackage[noadjust]{cite}
\usepackage{graphicx}
\usepackage[normalem]{ulem}

\usetikzlibrary{arrows,chains,matrix,positioning,scopes,quantikz2}
\ifdefined\LLNCS
\else % Cannot be used with LLNCS
\fi

\setitemize{itemsep=0pt}
\setenumerate{itemsep=0pt}

\newcommand{\customnote}[4]{{#4\color{#1}[#2: #3]}}
\newcommand{\setnote}[4]{\ifdefined\Draft\newcommand{#1}[1]{\customnote{#3}{#2}{##1}{#4}}\else\newcommand{#1}[1]{}\fi}

\newcommand{\linefill}{\rule{\linewidth}{0.8pt}}

\newcounter{protocol}

\newcounter{securityGame}

\newcounter{balgorithm}

\newcounter{numConstruction}
\newenvironment{Construction}[1]
{
\refstepcounter{numConstruction}
\par\setlength{\parindent}{0pt}
\rule{\textwidth}{0.3mm}
\vspace{-2.6mm}\textbf{Construction~\thenumConstruction} #1 

\hrulefill %\break
}
{
\hrulefill \break
\par
}

\newcounter{simulator}

\ifdefined\LLNCS
\newenvironment{thm}{\begin{theorem}}{\end{theorem}}

\newenvironment{rmk}{\begin{remark}}{\end{remark}}
\newenvironment{lem}{\begin{lemma}}{\end{lemma}}
\newenvironment{cor}{\begin{corollary}}{\end{corollary}}
\newtheorem{dfn}[definition]{Definition}
\else
\newtheorem{thm}{Theorem}[section]

\newtheorem{rmk}{Remark}
\newtheorem{Mycircuit}{Circuit}

\newtheorem{cor}[thm]{Corollary}

\newtheorem{lemma}[thm]{Lemma}
\newtheorem{claim}[thm]{Claim}

\theoremstyle{definition}
\newtheorem{dfn}[thm]{Definition}
\fi

\ifdefined\LLNCS
\else
\newtheorem*{thm*}{Theorem}
\newenvironment{theorem}{\begin{thm}}{\end{thm}}
\makeatletter
\newtheorem*{rep@theorem}{\rep@title}
\newcommand{\newreptheorem}[2]{%
\newenvironment{rep#1}[1]{%
 \def\rep@title{#2 \ref{##1}}%
 \begin{rep@theorem}}%
 {\end{rep@theorem}}}
\makeatother

\newreptheorem{thm}{Theorem}
\newreptheorem{lem}{Lemma}
\fi

\ifdefined\LLNCS
\newaliascnt{claiml}{theorem}
\newtheorem{claiml}[claiml]{Claim}
\aliascntresetthe{claiml}

\renewenvironment{claim}{\begin{claiml}}{\end{claiml}}

\crefname{claiml}{Claim}{Claims}
\else\fi

\crefname{lemma}{Lemma}{Lemmas}
\crefname{claim}{Claim}{Claims}
\crefname{figure}{Figure}{Figures}
\crefname{corollary}{Corollary}{Corollaries}
\crefname{proposition}{Proposition}{Propositions}
\crefname{conjecture}{Conjecture}{Conjectures}
\crefname{definition}{Definition}{Definitions}
\crefname{remark}{Remark}{Remarks}
\crefname{example}{Example}{Examples}
\crefname{algorithm}{Algorithm}{Algorithms}
\crefname{bAlgorithm}{Algorithm}{Algorithms}
\crefname{protocol}{Protocol}{Protocols}
\Crefformat{algorithm}{\textup{Algorithm~#2#1#3}}
\renewcommand{\cref}{\Cref} %make all reference start with uppercase

\ifdefined\LLNCS
\newenvironment{proofof}[1]{\begin{proof}[of~#1]}{\end{proof}}
\newenvironment{proofsketch}{\begin{trivlist} \item {\it Proof sketch.}} {\qed\end{trivlist}}
\newenvironment{proofsketchof}[1]{\begin{proofsketch}[of~#1]}{\end{proofsketch}}
\else

\newenvironment{proofsketch}{\begin{trivlist} \item {\it Proof sketch.}} {\qed\end{trivlist}}

\fi

\let\mathbb\relax % remove the definition by unicode-math
\DeclareMathAlphabet{\mathbb}{U}{msb}{m}{n}

\newcommand{\ie}{{i.e.,\ }}
\newcommand{\eg}{{e.g.,\ }}

\def\eps{\epsilon}

\def\cB{\mathcal{B}}

\def\cD{\mathcal{D}}

\def\cF{\mathcal{F}}

\def\cK{\mathcal{K}}
\def\cG{\mathcal{G}}

\def\cL{{\mathcal{L}}}

\def\cO{{\cal O}}

\def\cT{{\cal T}}
\def\cU{\mathcal{U}}

\def\cX{{\cal X}}
\def\cY{{\cal Y}}

\def\sF{\mathsf{F}}
\def\sG{\mathsf{G}}

\def\bbC{\mathbb{C}}
\def\bbF{\mathbb{F}}

\def\bbN{\mathbb{N}}

\newcommand{\nstep}{t}
\renewcommand{\mark}[1]{\textcolor{Red}{#1}} %! Set to `red` in the future

\newcommand{\zo}{\{0,1\}}
\newcommand{\abs}[1]{\left\lvert#1\right\rvert}
\renewcommand{\set}[1]{\left\{#1\right\}}
\newcommand{\norm}[1]{\left\lVert#1\right\rVert}

\newcommand{\inner}[1]{\langle{#1}\rangle}
\DeclareMathOperator*{\E}{\mathbb{E}}

\newcommand{\given}{\ensuremath{\;\middle|\;}}
\newcommand{\from}{\leftarrow}

\newcommand{\intext}{{\operatorname{in}}}
\newcommand{\out}{{\operatorname{out}}}
\newcommand{\outtext}{\out}

\newcommand{\inp}{\mathbbm{i}}

\newcommand{\secparam}{\lambda}
\newcommand{\negl}{\operatorname{negl}}
\DeclareMathOperator{\poly}{poly}

\newcommand{\adv}{{\mathcal{A}}}

\newcommand{\vk}{\mathsf{vk}}

\newcommand{\Gen}{\mathsf{Gen}}
\newcommand{\KeyGen}{\mathsf{KeyGen}}
\newcommand{\Auth}{\mathsf{Auth}}
\newcommand{\Enc}{\mathsf{Enc}}

\newcommand{\Dec}{\mathsf{Dec}}

\newcommand{\Eval}{\mathsf{Eval}}
\newcommand{\Verify}{\mathsf{Ver}}
\newcommand{\Ver}{\Verify}
\newcommand{\Sim}{\mathsf{Sim}}

\newcommand{\Sign}{\mathsf{Sign}}
\newcommand{\Token}{\mathsf{Token}}
\newcommand{\token}{\mathsf{token}}

\newcommand{\auth}{\mathsf{auth}}
\newcommand{\fDec}{\mathsf{fDec}}
\newcommand{\fVer}{\mathsf{fVer}}

\newcommand{\TP}{\mathsf{TP}}
\newcommand{\Send}{\mathsf{Send}}
\newcommand{\Recv}{\mathsf{Recv}}
\newcommand{\TPSend}{\TP.\Send}
\newcommand{\TPRecv}{\TP.\Recv}

\newcommand{\PRF}{\text{PRF}}
\newcommand{\QObf}{\mathsf{QObf}}
\newcommand{\QEval}{\mathsf{QEval}}

\newcommand{\Den}[1]{\cD^{#1}}
\newcommand{\gray}[1]{\textcolor{gray}{#1}}
\newcommand{\reg}{\gray}
\newcommand{\onreg}[2]{\ensuremath{{#1}^{\gray{#2}}}}
\newcommand{\ketbra}[1]{\ket{#1}\!\bra{#1}}
\newcommand{\identity}{\mathbbm{1}}
\newcommand{\Ig}{I}
\newcommand{\Tg}{\ensuremath{\mathsf{T}}}
\newcommand{\Sg}{\ensuremath{\mathsf{S}}}
\newcommand{\Hg}{\ensuremath{\mathsf{H}}}
\newcommand{\Xg}{\ensuremath{\mathsf{X}}}
\newcommand{\Yg}{\ensuremath{\mathsf{Y}}}
\newcommand{\Zg}{\ensuremath{\mathsf{Z}}}
\newcommand{\CNOT}{\ensuremath{\mathsf{CNOT}}}
\newcommand{\SWAP}{\ensuremath{\mathsf{SWAP}}}
\newcommand{\CTRL}{\mathsf{ctrl}}
\newcommand{\rooti}{\ensuremath{e^{i \pi /4}}}
\def\ot{\otimes}

\newcommand{\PauliGroup}{\mathscr{P}}
\newcommand{\CliffordGroup}{\mathscr{C}}
\DeclareMathOperator{\Tr}{Tr}

\newcommand{\meas}[1]{\mathcal{M}\left[#1\right]}
\newcommand{\measureto}[2]{\mathcal{M}_{#1}\left[#2\right]}

\newcommand{\hybrid}{\mathsf{Hyb}}%{\mathcal{H}}

\newcommand{\wh}{\widehat}
\newcommand{\wt}{\widetilde}

\newcommand{\epr}{\mathsf{epr}}

\newcommand{\mixednospace}[1]{\mathsf{Mixed}\left[#1\right]}
\newcommand{\mixed}[1]{\mixednospace{#1}}
\newcommand{\aux}{{\mathsf{aux}}}

\newcommand{\Has}{\mathsf{Has}}
\newcommand{\Only}{\mathsf{Only}}

\newcommand{\Write}{\mathsf{Write}}
\newcommand{\WriteH}{\mathsf{WriteH}}
\newcommand{\update}{\mathsf{update}}
\newcommand{\EqualH}{\mathsf{EqualH}}

\newcommand{\PLM}{\mathsf{PLM}}
\newcommand{\pub}{\mathsf{pub}}
\newcommand{\priv}{\mathsf{priv}}
\title{Obfuscation of Unitary Quantum Programs}
\setnote{\TODO}{TODO}{red}{}
\setnote{\Enote}{Ercheng}{teal}{\footnotesize}
\setnote{\Mnote}{Miryam}{cyan}{\footnotesize}
\ifdefined\ShowAuthor
\author[1]{Mi-Ying (Miryam) Huang\thanks{Email: miying.huang@usc.edu \qquad\qquad This is the full version of a paper that appears in FOCS 2025.}}\affil[1]{University of Southern California}
\author[2]{Er-Cheng Tang\thanks{Email: erchtang@uw.edu}}\affil[2]{University of Washington}
\date{}
\else
\author{}
\date{}
\fi

\begin{document}
\maketitle
\begin{abstract}
Program obfuscation aims to hide the inner workings of a program while preserving its functionality. In the quantum setting, recent works have obtained obfuscation schemes for specialized classes of quantum circuits. For instance, Bartusek, Brakerski, and Vaikuntanathan (STOC 2024) constructed a quantum state obfuscation scheme, which supports the obfuscation of quantum programs represented as quantum states for pseudo-deterministic quantum programs with classical inputs and outputs in the classical oracle model.

In this work, we improve upon existing results by constructing the first quantum state obfuscation scheme for \emph{unitary} (or \emph{approximately unitary}) quantum programs supporting \emph{quantum} inputs and outputs in the classical oracle model. 
At the core of our obfuscation scheme are two novel ingredients: a \emph{functional} quantum authentication scheme that allows key holders to learn specific functions of the authenticated quantum state with \emph{simulation-based} security, and a compiler that represents an arbitrary quantum circuit as a \emph{projective} linear-plus-measurement quantum program described by a sequence of non-adaptive Clifford gates interleaved with adaptive and \emph{compatible} measurements.

\end{abstract}

\ifdefined\FullVersion
%\blfootnote{}
\newpage
\tableofcontents
\clearpage
\fi

\section{Introduction}
\label{Sec:introduction}
Program obfuscation is an important problem in cryptography and complexity, where the goal is to obscure the internal workings of a program while preserving its functionality.
The study of obfuscation has a long tradition in classical cryptography. It is well-known that the notion of virtual black-box (VBB) obfuscation, which guarantees that the obfuscated program discloses no additional information beyond what can be learned from its input-output behavior, is impossible to achieve for general circuits~\cite{barak2001possibility}. On the one hand, the impossibility has led to the study of weaker notions, such as indistinguishability obfuscation (iO) \cite{barak2001possibility, garg2013candidate,ananth2019indistinguishability,jain2021indistinguishability,jain2022indistinguishability}, which only requires the obfuscations of functionally equivalent circuits to be computationally indistinguishable. On the other hand, it has led to the consideration of stronger models, such as the pseudorandom oracle model \cite{jain2023pseudorandom}, where VBB obfuscation, and even the stronger notion of ideal obfuscation, is feasible.

Motivated by the importance of obfuscation in classical cryptography, a more recent line of work has emerged that studies the feasibility of obfuscating quantum programs. Similarly to the classical setting, it is known that VBB obfuscation of general quantum circuits is impossible in the plain model under standard cryptographic assumptions~\cite{alagic2021impossibility}, so efforts have naturally turned to weaker notions of obfuscation or stronger models. For example, \cite{broadbent2021constructions} achieves iO for quantum circuits with logarithmically many non-Clifford gates supporting quantum inputs and outputs. \cite{bartusek2023obfuscation} achieves obfuscation for pseudo-deterministic quantum circuits with classical inputs and outputs in the classical oracle model. More recently, \cite{coladangelo2024use} introduces the concept of ``quantum state obfuscation,'' which additionally supports obfuscation of programs equipped with a quantum state. The same work demonstrates its applicability for constructing other quantum cryptographic primitives, e.g., \ copy-protection. Informally, a quantum state obfuscation scheme takes a quantum program $(\psi,Q)$, comprising a quantum circuit $Q$ and an auxiliary quantum state $\psi$, and outputs an obfuscated quantum program with the same functionality. Later on, \cite{bartusek2024quantum} constructs ideal quantum state obfuscation in the classical oracle model, again for pseudo-deterministic quantum programs with classical inputs and outputs.

While these works represent significant progress, they fall short of supporting obfuscation of general quantum input-output functionalities: they are either restricted to pseudo-deterministic quantum programs with classical inputs and outputs, or to quantum circuits with a logarithmic number of non-Clifford gates. These limitations lead to the following compelling question:

\begin{quote}
    \centering
    \it Is it possible to obfuscate a broad class of quantum programs\\ that support quantum inputs and outputs?
\end{quote}

\subsection{Our Results}
We affirmatively address the aforementioned question by presenting the first quantum state obfuscation scheme for \emph{unitary} (or negligibly close to unitary) quantum programs supporting \emph{quantum} inputs and outputs in the classical oracle model. Our scheme resolves an open question raised by Bartusek, Brakerski, and Vaikuntanathan~\cite{bartusek2024quantum}, who presented a scheme that only obfuscates pseudo-deterministic\footnote{Every pseudo-deterministic quantum program can be directly converted into an approximately unitary quantum program by coherent execution of the program. Therefore, our scheme also obfuscates pseudo-deterministic quantum programs.} quantum programs supporting classical inputs and outputs. To capture the security of our scheme, we propose a definition of ideal obfuscation for quantum programs that implement unitary transformations. Moreover, when restricted to programs with classical functionality, our definition reduces to the standard notion of ideal obfuscation. The obfuscated program of our scheme can be used to evaluate multiple quantum inputs.

\begin{thm}[Informal]
There exists a quantum state obfuscation scheme for the class of all approximately\footnote{As an improvement on the allowed approximation factor, our result can obfuscate programs with up to $\negl(\secparam)$ approximation error, whereas \cite{bartusek2024quantum} only obfuscates programs that have $2^{-2n}\negl(\lambda)$ error.} unitary quantum programs supporting quantum inputs and outputs, achieving the notion of ideal obfuscation (Definition \ref{dfn:obf}) in the classical oracle model.
\end{thm}

Our quantum state obfuscation is proven secure in the classical oracle model, where the obfuscator can output a classical oracle that is accessible to all parties. This can be viewed as an idealized model in which ideal obfuscation is achievable for all classical functions. Our result achieves unconditional security when the constructed oracle internally uses a truly random function, and additionally relies on the existence of a post-quantum one-way function when the classical oracle is required to be efficiently computable, as in \cite{bartusek2024quantum}.

Historically, such oracle-assisted results have often foreshadowed oracle-free constructions, as demonstrated by public-key quantum money \cite{zhandry2021quantum} and copy-protection \cite{coladangelo2021hidden}. The classical oracle model can be heuristically instantiated with post-quantum iO in the plain model. In light of a recent work \cite{jain2023pseudorandom}, it also seems plausible to provably instantiate the classical oracle model with post-quantum security from the pseudorandom oracle model, a setting that only requires modeling idealized hash functions. Below, we summarize the comparison between our result and prior quantum obfuscation constructions in Table~\ref{table:sum}.

\begin{table}[!ht]
\scriptsize
\centering
\resizebox{\textwidth}{!}{
\begin{tabular}{|c|c|c|c|c|c|c|c|}
    \hline
    {} & \makecell{Obfuscator \\ input} & \makecell{Obfuscator \\ output} & \makecell{Program \\ input} & \makecell{Program \\ output} & Program class & \makecell{Assumption/ \\ Model} & Result \\
    \hhline{|=|=|=|=|=|=|=|=|}
    \cite{broadbent2021constructions} & Classical & Quantum & Quantum & Quantum & \makecell{Unitaries with log-many\\ non-Clifford gates} & \makecell{iO for classical \\ circuits} & iO*\\
    \hline
    \cite{bartusek2021indistinguishability} & Classical & Classical & Quantum* & Classical & Null circuits & \makecell{Classical oracle \\ model + LWE} & iO \\
    \hline 
    \cite{bartusek2023obfuscation} & Classical & Quantum & Classical & Classical & \makecell{(Pseudo-)deterministic \\ circuits} & \makecell{Classical oracle \\ model + LWE} & VBB \\
    \hline
    \cite{coladangelo2024use} & Quantum & Quantum & Classical & Classical & Deterministic circuits & \makecell{Quantum oracle \\ model} & iO \\
    \hline
    \cite{bartusek2024quantum} & Quantum & Quantum & Classical & Classical & \makecell{(Pseudo-)deterministic \\ circuits} &  \makecell{Classical oracle \\ model} & Ideal\\
    \hline
    This paper& Quantum & Quantum & Quantum & Quantum & \makecell{(Approximately-)unitary circuits} &  \makecell{Classical oracle \\ model} & Ideal\\
    \hline
\end{tabular}
}
\caption{Construction of quantum obfuscation schemes. Note the caveats: (1) The obfuscator in \cite{broadbent2021constructions} produces a quantum state that enables evaluation of the program on a single input, after which the state may become destroyed. While suitable for certain applications, this behavior deviates from the standard notion of program obfuscation, which typically requires support for multiple evaluations on different inputs. (2) The construction in \cite{bartusek2021indistinguishability} supports quantum inputs but requires multiple identical copies of the quantum input.}
\label{table:sum}
\end{table}

\paragraph{Related Work.}
Very recently, \cite{zhandry2025model} showed that every \emph{classically described} quantum circuit implementing a unitary operator $U$, potentially using ancilla qubits (i.e., $\ket{0}$), can be used to implement its conjugate $U^*$, transpose $U^\top$, and controlled operator $\CTRL$-$U$. Based on this observation, \cite{zhandry2025model} proposed that having access to a unitary $U$ should be modeled as having oracle access to the full collection $\CTRL$-$U$, $\CTRL$-$U^\dag$, $\CTRL$-$U^*$, $\CTRL$-$U^\top$. In this work, we show how to obfuscate every program that implements a unitary transformation $\mathcal{U}:\rho \mapsto U \rho U^\dag$ into a \emph{quantumly described} quantum program with an auxiliary quantum state\footnote{Looking ahead, our obfuscated program is designed to carry several authenticated $\Tg$ magic states to assist the implementation of each individual $\Tg$ gate. These states can only be used to compute $\Tg$ but not its conjugate $\Tg^*$. Consequently, the obfuscated program cannot be used to implement $\mathcal{U}^*$. Furthermore, since each magic state is protected under a distinct Pauli key and the classical oracle that holds the keys rejects out-of-order executions, the obfuscated program cannot be used to implement $\mathcal{U}^\top$. It is worth noting that $\mathcal{U}^\dagger$ is achievable (unlike $\mathcal{U}^*$ and $\mathcal{U}^\top$) by performing a rewinding execution after a normal forward execution.}. Moreover, the latter allows \emph{only} the computation of $\cU$ and $\cU^\dag$, or equivalently, the computation of $\CTRL$-$(U^\dag \SWAP^{\ot n} U)$ where $U$ acts on $n$ qubits (see \cref{rmk:2}). This highlights a key distinction between two flavors of program access to a unitary transformation: The computational power of program access to a quantumly described quantum program can be encapsulated using a single oracle, whereas the power of program access to a classically described quantum circuit needs to be captured using multiple oracles.

\paragraph{Applications.} 
\cite{coladangelo2024use} demonstrated that quantum state obfuscation implies a best-possible copy-protection scheme. Accordingly, our construction realizes such a scheme and possibly extends to copy-protecting unitary functionalities. \cite{bartusek2023obfuscation} employed quantum obfuscation to construct a functional encryption scheme. By instantiating their construction with our obfuscator, we enable a functional encryption scheme supporting unitary functionalities.

\paragraph{Open Problems.}
Here, we discuss some open problems motivated by this work. 
\begin{enumerate}
    \item One interesting problem is to obfuscate isometries, which map a quantum input $\rho_\intext$ to a quantum output of the form $V(\rho_\intext \ot \ketbra{0^\kappa})V^\dag$ for some efficient unitary operator $V$, while ensuring reusability of the obfuscated program. Moreover, a related question would be the possibility of obfuscating classical randomized functions.
    \begin{figure}[H]
    \begin{center}
    \begin{quantikz}[row sep=1.2em, column sep=1.2em]
      \lstick{$\rho_{\intext}$} & \qw & \gate[2, style={minimum width=3.5em, minimum height=3em}]{{\Large Q}} & \qw & \rstick{$\rho_\out = U \rho_{\intext} U^\dag$} \\
      \lstick{$\psi_\aux$} & \qw & \qw & \qw & \rstick{$\psi'_\aux$} 
    \end{quantikz}
    \hfill
    \begin{quantikz}[row sep=1.2em, column sep=1.2em]
      \lstick{$\rho_{\intext}$} & \qw & \gate[3, style={minimum width=3.5em, minimum height=6em}]{{\Large Q}} & \qw & \rstick[2]{$\rho_\out = V(\rho_{\intext}\ot\ketbra{0^\kappa})V^\dag$}  \\
      \lstick{} & \qw & \qw & \qw & \rstick{ } \\
      \lstick{$\psi_\aux$} & \qw & \qw & \qw & \rstick{$\psi'_\aux$} 
    \end{quantikz}
    \end{center}
    \caption{The quantum program $(Q,\psi_\aux)$ on the left implements a unitary on its input, whereas the quantum program $(Q,\psi_\aux)$ on the right implements an isometry on its input. Our scheme obfuscates unitaries, and it remains open to obfuscate isometries.}
    \end{figure}
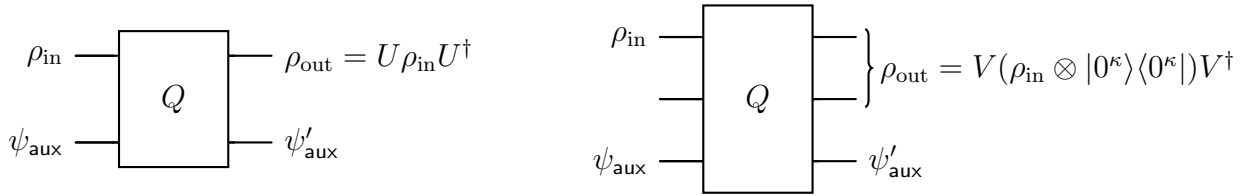

    \item Can one achieve quantum state indistinguishability obfuscation in the plain model assuming post-quantum indistinguishability obfuscation?
    \item The use of classical obfuscation in conjunction with pseudorandom functions has found great applications in classical cryptography. Can the combination of our obfuscation scheme for unitary programs and recent studies on pseudorandom unitaries \cite{ma2025construct,schuster2025random} open new avenues in quantum cryptography?
    \item In quantum complexity, several oracle separation results have been established relative to quantum oracles but remain unresolved relative to classical oracles.
    Can obfuscation of unitaries in the classical oracle model help bridge the gap by instantiating the quantum oracles using classical oracles and auxiliary quantum states?
\end{enumerate}

\section{Technical Overview}\label{Sec:overview}
In this section, we start by revisiting the work of \cite{bartusek2024quantum}, which served as a key inspiration for our work. We then explore the challenges of obfuscating general unitary programs with quantum inputs and outputs. Afterward, we present our new techniques, which overcome these challenges and substantially strengthen the results of \cite{bartusek2024quantum}.

We begin by discussing an approach for obfuscating quantum circuits. A common method to perform quantum computation without revealing the computation's internal details is to apply homomorphic operations on authenticated quantum states.\footnote{It is well-known that quantum authentication implies quantum encryption. Such a method also appeared in works that aimed for quantum one-time programs \cite{broadbent2013quantum}.} 
However, no known quantum authentication scheme can fully support universal quantum computation through homomorphic operations. A common workaround is to implement the unsupported homomorphic gates through ``magic state injection,'' which appends auxiliary magic states to the input state, performs measurements on them in the middle of the computation, and applies adaptive correction gates based on the measurement outcomes. Nevertheless, this method faces a key challenge in the context of obfuscation: revealing the adaptive correction gates may leak the intermediate measurement outcomes. For instance, one approach is to compile a quantum circuit as an adaptive sequence of Clifford gates, applied to the input along with some magic states. However, this approach runs precisely into the issue that subsequent choices of Clifford gates reveal the previous measurement outcomes.

To overcome this issue, \cite{bartusek2024quantum} develops a new representation of quantum programs (including those with auxiliary quantum states) with classical inputs and outputs as what they call a ``linear-plus-measurement'' (LM) quantum program. The latter consists of a quantum state $\psi_+$, a predetermined sequence of Clifford gates $U_1,\dots,U_\nstep$, and some classical functions $f_1,\dots,f_\nstep,g$ for performing \emph{adaptive measurements}. The LM quantum program takes a classical input $\inp$, and produces a classical output as follows.
\begin{itemize}
    \item Initialize a quantum register $\reg{V}$ with the quantum state $\psi_+$. 
    \item For $j=1,\dots,\nstep$, do:
    \begin{itemize}
        \item Apply $U_j$ on register $\reg{V}$.
        \item Take the classical function $f_j^{\inp,r_1,\dots,r_{j-1}}(\cdot)$ that depends on the classical input $\inp$ and all previous measurement outcomes $r_1,\dots,r_{j-1}$. Coherently compute $f_j^{\inp,r_1,\dots,r_{j-1}}(\reg{V})$ and measure the function output in the computational basis. Denote the measurement result by $r_j$.
        \item Apply $U_j^\dag$ on register $\reg{V}$.
    \end{itemize}
    \item Output $g(\inp,r_1,\dots,r_\nstep)$.
\end{itemize}
Representing a quantum computation as an LM quantum program brings a key benefit: the adaptive component of the quantum computation is now reduced to the coherent evaluation of classical functions, which can be protected using classical obfuscation.
\cite{bartusek2024quantum} shows that any quantum program can be compiled into an LM quantum program, where the state $\psi_+$ consists of the original program along with ancillary zero states and magic states.

With the compiler of LM quantum program in hand, \cite{bartusek2024quantum} realizes the aforementioned obfuscation approach as follows. The obfuscator starts by running the compiler to produce an LM quantum program,  where all information about the original program is encoded in the quantum state $\psi_+$. Then the obfuscator authenticates $\psi_+$, prepares classical oracles $\sF_j, \sG$ that internally computes $f_j, g$,
%to enable \emph{homomorphic} adaptive measurements, 
and outputs the authenticated state and the classical oracles as the obfuscated program. To evaluate the obfuscated program on classical input $\inp$, one can homomorphically apply each Clifford gate $U_j$ on the authenticated states on their own. Meanwhile, each homomorphic adaptive measurement $f_j$ can be performed by querying the classical oracle $\sF_j$ in superposition on the tuple\footnote{The tuple also contains a tokenized signature for ensuring input consistency across oracle calls, but we omit it here for presentation simplicity.} consisting of the input $\inp$, the authenticated state, and the outputs of $\sF_1,\dots,\sF_{j-1}$. Crucially, each oracle $\sF_j$ is designed to output only the encoding of previous intermediate measurement results $r_1,\dots,r_{j}$, which allows future measurements to adaptively depend on them without having the measurement results revealed in plaintext. At the end, one queries the oracle $\sG$ on the encodings produced by $\sF_1,\dots,\sF_\nstep$ to obtain the final output $g(r_1,\dots,r_\nstep)$ in the clear.
\cite{bartusek2024quantum} shows that these steps could be done assuming ideal classical obfuscation and obtains a quantum state obfuscation scheme for a restricted class of quantum circuits, whose inputs and outputs are required to be classical and pseudo-deterministic. 

\subsection{Hurdles in Quantum Functionalities}
We now examine in more detail why \cite{bartusek2024quantum} confines its scope to classical pseudo-deterministic functionalities. Then, we highlight the challenges that arise when extending obfuscation to quantum programs with quantum inputs and outputs, as well as to more general quantum circuits beyond pseudo-deterministic ones. Here, we explain the difficulties from the perspectives of construction, functionality, and security analysis.

In the construction of \cite{bartusek2024quantum}, the classical oracle plays a central role in interpreting the execution of the obfuscated quantum program. In particular, the classical oracle learns the classical input $\inp$, checks whether the same classical input is being used throughout the execution, and returns the computed classical output. However, when the inputs and outputs are quantum states, it is unclear whether a classical oracle can interpret the quantum input, ensure the consistency of quantum input, and return the correct quantum output to the evaluator. Therefore, restricting to programs with classical inputs and outputs seems inherent in their approach.

From the functionality standpoint, the obfuscator has to produce a program that can be evaluated on as many inputs as needed. Such a reusablity requirement can be difficult to achieve for general quantum programs that are equipped with quantum states due to the no-cloning theorem. To ensure that the obfuscated program remains reusable, prior works \cite{bartusek2023obfuscation,bartusek2024quantum} focus on quantum programs where the output is classical and pseudo-deterministic. In this case, measuring the output of the obfuscated program in the standard basis would barely collapse the quantum state, so the obfuscated program can be restored after each evaluation through rewinding. 
Nevertheless, such an argument does not work beyond classical pseudo-deterministic quantum programs, and existing techniques fail to obfuscate more general quantum programs (with quantum states) in a way that ensures reusability of the obfuscated program.

Regarding security analysis, \cite{bartusek2024quantum} heavily relies on the functionality being classical and pseudo-deterministic, which allows classifying all potential intermediate measurement results into possible and impossible ones, according to whether they may occur in honest executions. Following this idea, \cite{bartusek2024quantum} reduces the security proof to the simpler task of showing the infeasibility of producing impossible measurement results. However, such a classification is no longer effective when going beyond classical and pseudo-deterministic functionalities. Instead of tackling only the support of the distribution, one has to analyze the entire distribution of homomorphic measurement results that the adversary produces dynamically, a technique that is not present in \cite{bartusek2024quantum}. Even if the distribution can be analyzed dynamically, one must still demonstrate that the quantum state always corresponds to some honestly computed state. However, correctly interpreting a quantum state as an intermediate state in the LM quantum program of \cite{bartusek2024quantum} requires identifying the exact stage of execution to which the state belongs. This is particularly challenging when different stages of the LM quantum program are only loosely related or not directly comparable. Moreover, it would be difficult to characterize quantum states that correspond to superpositions of intermediate states across multiple stages of the LM quantum program execution.

\subsection{Our Solutions}

\paragraph{Improved representation of quantum computation.} 
We offer a new perspective on quantum computation by introducing the notion of \emph{projective} LM (PLM) quantum programs, a variant of LM quantum programs where the adaptive measurements involved are all compatible\footnote{Recall that two measurements are considered compatible if their corresponding projectors commute. In our case, each adaptive measurement can be specified by a pair $(U_j, f_j^{\inp,r_1,\dots,r_{j-1}})$. We require compatibility only between any two pairs $(U_i, f_i^{\inp,r_1,\dots,r_{i-1}})$ and $(U_j, f_j^{\inp,r_1,\dots,r_{j-1}})$ that share the same prefix $(\inp,r_1,\dots)$.
} with each other and effectively projects the quantum state onto a known orthonormal basis.
Formally, an LM quantum program is called \emph{projective}, if for every classical input $\inp$, there exists an orthonormal basis $\{\ket{\Phi_{\inp,(r_1,\cdots,r_\nstep)}}\}$ such that the cumulative effect of the adaptive measurements is to apply the projective measurement $\{\proj{\Phi_{\inp,(r_1,\cdots,r_\nstep)}}\}_{r_1,\dots,r_\nstep}$, where $r_j$ stands for the $j$-th adaptive measurement outcome. Following this definition, on classical input $\inp$, executing the PLM quantum program whose quantum state is changed from $\psi_+$ to $\ket{\Phi_{\inp,(r_1,\cdots,r_\nstep)}}$ always yields deterministic adaptive measurement outcomes $(r_1,\dots,r_\nstep)$.

The projective property provides a unified basis for analyzing the quantum state across different stages of program execution, thereby eliminating the need to identify the stage to which the state corresponds. In addition, PLM quantum programs offer a natural way to interpret adaptive measurements as a process of revealing how $\psi_+$ decomposes into orthonormal basis vectors $\{\ket{\Phi_{\inp,(r_1,\cdots,r_\nstep)}}\}_{r_1,\dots,r_\nstep}$ without altering the amplitudes. Moreover, this basis is a convenient basis for analysis. For example, performing the measurement $\set{\sum_{r: g(\inp,r)=y} \ketbra{\Phi_{\inp,r}}}_y$ on $\psi_+$ is equivalent to measuring the classical output $y$ of the original quantum program on input $\inp$. To explain, LM quantum program on input $\inp$ performs the adaptive measurements to obtain $r_1,\dots,r_\nstep$ and outputs $y= g(r_1,\dots,r_\nstep)$. In PLM quantum program, $r_1,\dots,r_\nstep$ can be obtained alternatively by performing the projective measurement $\set{ \ketbra{\Phi_{\inp,r}}}_r$ on $\psi_+$. If one only wants to obtain the final output and not the intermediate measurement results, then they can directly apply the projective measurement $\set{\sum_{r: g(\inp,r)=y} \ketbra{\Phi_{\inp,r}}}_y$ on $\psi_+$ to obtain the output $y$. 
Furthermore, such an argument continues to hold even if $\psi_+$ is entangled with the adversary's states. Thus, decomposing in this basis enables a clean analysis of the adversary's behavior on each component while preserving the entanglement structure.

\paragraph{Handling quantum inputs and outputs.} 
To obfuscate programs with quantum inputs and outputs, we integrate quantum teleportation into the obfuscation construction. In more detail, the obfuscator additionally prepares EPR pairs $(\epr_{\intext,\pub}, \epr_{\intext,\priv}), (\epr_{\outtext,\priv}, \epr_{\outtext,\pub})$, provides $(\epr_{\intext,\pub}, \epr_{\outtext,\pub})$ in the clear, and provides the authentication of $(\epr_{\intext,\priv}, \epr_{\outtext,\priv})$ as part of the obfuscated quantum program.\footnote{A similar idea was considered in \cite{broadbent2013quantum} in the context of quantum one-time programs.} To evaluate the obfuscated program, the evaluator teleports a quantum input using $\epr_{\intext,\pub}$ and obtains a classical teleportation result $\inp$. 
Next, the evaluator performs the following computation homomorphically over authenticated states, where only $\inp$ and $y$ are represented as plaintext:
\begin{itemize}
    \item Interpret $\inp$ as a Pauli correction $P_\inp$ and apply it on $\epr_{\intext,\priv}$.
    \item Use the resulting state as the quantum input.
    \item Perform the quantum computation of interest.
    \item Teleport the quantum output using $\epr_{\outtext,\priv}$ and output the teleportation result $y$.
\end{itemize}
The evaluator can apply the Pauli correction $P_y$ on $\epr_{\outtext,\pub}$ to recover the quantum output.

\paragraph{Enabling program reusability.}
The idea above runs into a natural challenge. Equipped only with a limited number of EPR pairs, how can the obfuscated program be reused for multiple evaluations? Fortunately, we show that reusability can be guaranteed whenever the quantum program implements a unitary or approximately-unitary functionality, \ie the quantum output is close to the result of applying a unitary to the quantum input. Intuitively, such programs can be reused because the quantum output would be unentangled with the auxiliary state, so one can retrieve the output without damaging the auxiliary state. We refer the reader to \cref{sec:approximate-unitary} for further details.

\paragraph{Providing functional security.}
Although the quantum authentication scheme of \cite{bartusek2024quantum} supports homomorphic measurements, its security guarantees only capture static distributional properties of each measurement result, but do not account for the joint distribution with quantum states or any dynamic behavior.
We remedy this shortcoming by upgrading the authentication into a functional authentication scheme with simulation-based security. Syntactically, the functional authentication scheme consists of the following components:
\begin{itemize}
    \item $k \from \KeyGen(1^\lambda, 1^n)$ generates a classical key $k$ with security parameter $\lambda$.
    \item $\wt{\rho} \from \Enc_k(\rho)$ authenticates an $n$-qubit state $\rho$ into a ciphertext state $\wt{\rho}$.
    \item $\wt{\rho}_{\theta,G} \from \Eval_{\theta,G}(\wt{\rho})$ homomorphically applies the unitary $\Hg^{\theta} G$ over the ciphertext $\wt{\rho}$, where $\Hg$ is a Hadamard gate, $\theta \in \zo^n$, and $G$ is a unitary consisting of $\CNOT$ gates.
    \item $\top/\bot \gets \Verify_{k, \theta, G}(\wt{m})$ classically verifies whether $\wt{m}$ is a valid measurement result of the ciphertext state in the column basis of $\Hg^\theta G$. It outputs either $\top$ (valid) or $\bot$ (invalid).
    \item $m/\bot \gets \Dec_{k, \theta, G}(\wt{m})$ classically decodes $\wt{m}$ into the measurement result $m$ of the plaintext state in the column basis of $\Hg^\theta G$. It outputs $\bot$ if $\wt{m}$ is an invalid string.
\end{itemize}
We can view the classical oracle $\Dec_{k, \theta, G}$ as a functional key for learning the plaintext measurement result of the ciphertext state in the column basis of $\Hg^\theta G$. Our notion of security ensures that an adversary with the ciphertext state and multiple functional keys learns nothing beyond what is permitted by the combined functionality of the keys. More precisely, our simulation-based definition guarantees that access to the ciphertext state and multiple functional keys is indistinguishable from access to a dummy ciphertext state and multiple simulated oracles, each conditionally revealing the corresponding measurement information of the plaintext state. That is, the security is guaranteed even if the adversary has partial decryption ability.

\subsection{Obfuscation of Approximately Unitary Quantum Programs}
\label{sec:overview:ideal:obfuscation}
How should ideal obfuscation be defined for quantum programs? We begin by exploring the implications of white-box access to a quantum program that implements a unitary transformation $\cU: \rho \mapsto U \rho U^\dag$. While a circuit that implements a unitary operator $U$ naturally yields a circuit that implements $\CTRL$-$U$, a quantum program with auxiliary state that implements $\mathcal{U}$ does not necessarily admit implementing $\CTRL$-$U$ for any choice of $U$'s global phase.\footnote{$U$ is determined by $\mathcal{U}$, but only up to a global phase.} The latter situation could happen if the auxiliary state of the program becomes a different state after program execution. In such a case, attempting to use the program to implement $\CTRL$-$U$ would entangle the control qubit with the auxiliary register, thereby failing to implement $\CTRL$-$U$. Fortunately, if one considers the combined process of executing the program, applying an efficient unitary $A$, and then rewinding the program, the resulting auxiliary state will always be restored to its original state. Therefore, we can show that white-box access to a quantum program that implements $\cU$ inherently allows one to perform the transformations associated\footnote{Associated with the operator $U$ is the transformation $\cU: \rho \mapsto U \rho U^\dag$.} with $U$, $U^\dag$, and $\CTRL$-$(U^\dag A U)$ for any number of times, provided that $A$ itself can be efficiently implemented.\footnote{$\CTRL$-$(U^\dag A U)$ is invariant under the choice of $U$'s global phase.} This motivates a tentative definition of ideal obfuscation for approximately unitary quantum programs: the obfuscated program should only grant black-box access to the unitaries $U$, $U^\dag$, and $\CTRL$-$(U^\dag A U)$.
In this work, we show that such a notion of ideal obfuscation can, in fact, be realized in the classical oracle model. Hence, we define ideal obfuscation as follows.
\begin{itemize}
    \item An ideal obfuscation scheme consists of an obfuscator $\QObf(1^\secparam,(\psi,Q)) \to (\wt{\psi},\sF)$ transforming a quantum program $(\psi,Q)$ into an obfuscated program specified by state $\wt{\psi}$ and classical oracle $\sF$, and an algorithm $\QEval^{\sF}(\cdot,\wt{\psi})$ for executing the obfuscated program.
    \item $\QEval^{\sF}(\cdot,\wt{\psi})$ approximately implements the same functionality as $(\psi,Q)$.
    \item $(\wt{\psi},\sF)$ can be efficiently simulated with oracle access to some unitary $\CTRL$-$(U^\dag A U)$.\footnote{In fact, it turns out that $\cU,\cU^\dag,\CTRL$-$(U^\dag A U)$ can all be simulated from oracle access to $\CTRL$-$(U^\dag A U)$ for a particular choice of $A$. One such choice is $A = \SWAP^{\ot n}$ if $U$ acts on $n$ qubits (see \cref{lemma:the-single-swap-oracle}). They can also all be simulated from oracle access to $\cU$ and $\cU^\dag$ (see \cref{lemma:U-and-U-inverse-suffice}).}
\end{itemize}
Such a notion of ideal obfuscation says that any efficiently implementable unitary can be implemented in a way that reveals nothing about the program other than its ability to perform $\CTRL$-$(U^\dag A U)$.

Our definition generalizes the classical notion of ideal obfuscation to the setting of unitary functionalities. In particular, when restricted to programs that implement (the unitary of) classical deterministic functions, our definition recovers the standard notion of classical ideal obfuscation.\footnote{For unitary of the form $U: \ket{x,y} \mapsto \ket{x,y\oplus F(x)}$ for some classical function $F$, the unitary $\CTRL$-$(U^\dag A U)$ can be implemented from black-box access to $\CTRL$-$U$ (because $U^\dag = U$), which can in turn be implemented from black-box access to $\cU$ using controlled swap gates and the efficiently preparable $1$-eigenvectors of $U$.} Furthermore, the definition captures quantum features such as the global phase. Although a global phase introduced by a quantum program has no computational significance when the program is executed in isolation, this phase can become detectable when the program is used under quantum-controlled operations. Our definition implicitly requires the global phase to be obfuscated because the simulator is given oracle access only to $\CTRL$-$(U^\dag AU)$, which is oblivious to the choice of $U$'s global phase. As a result, circuits that implement distinct unitary operators but realize the same unitary transformation (e.g., $\Xg \circ \Zg$ versus $\Yg$, which differ by a global phase $i$) would become indistinguishable under obfuscation schemes that satisfy our definition.
Below, we demonstrate that every quantum program that approximately implements a unitary transformation can be ideally obfuscated in the classical oracle model when assuming a random oracle.

\paragraph{Our Construction.} When designing obfuscation schemes, it is crucial to ensure that the same input is used consistently throughout the entire execution of the program. For classical inputs, prior works \cite{bartusek2023obfuscation,bartusek2024quantum} address the issue with a tokenized signature scheme, which includes a classical verification key $\vk$ and a quantum token state $\tau_\token$ . This token can be used to produce a classical signature on a single \textit{classical} message, but prevents signing two different messages simultaneously, thereby enforcing input consistency. We extend this approach by showing that signature tokens can also be adapted to ensure consistency of \textit{quantum} inputs.
Together with the tools introduced in the previous subsection, we can obfuscate any quantum program that approximately implements a unitary as follows.
\begin{itemize}
    \item Prepare EPR pairs $(\epr_{\intext,\pub},\epr_{\intext,\priv})$ and $(\epr_{\outtext,\priv},\epr_{\outtext,\pub})$.
    \item Consider the extended computation that takes a classical string $\inp$ as input, applies the Pauli correction $P_\inp$ on $\epr_{\intext,\priv}$, computes the quantum program of interest, teleports the output through $\epr_{\outtext,\priv}$, and outputs the teleportation result $y$.
    \item Transform the extended computation into a PLM quantum program consisting of a quantum state $\psi_+$ and instructions $\{(\theta_j, G_j, f_j)\}$ of the adaptive measurements. Specifically, $\psi_+$ contains the description $\psi$ of the original program, the states $\epr_{\intext,\priv}, \epr_{\outtext,\priv}$, and some magic states. The $j$-th instruction $(\theta_j, G_j, f_j)$ suggests an adaptive measurement specified by the classical function $f_j$ and the column basis of $\Hg^{\theta_j} G_j$.
    \item Encode the state $\psi_+$ into a state $\psi_\auth = \Auth.\Enc_k(\psi_+)$ using functional authentication.
    \item Generate a one-time signature token $(\vk, \tau_{\token})$ and take a random function $H$.
    \item Construct the following classical function $\sF\left(j,\widetilde{v}, \inp,s,\ell_1,\dots,\ell_{\nstep} \right)$, which enables homomorphic adaptive measurements without revealing intermediate measurement results.
    \begin{enumerate}
    \setlength{\itemsep}{0pt}
        \item Compute $v \from \Auth.\Dec_{k,\theta_j,G_j}(\widetilde{v})$. If it rejects, output $\bot$.
        \item Compute $\Token.\Verify_\vk(\inp,s)$. If it rejects, output $\bot$.
        \item For $i = 1,\dots, j-1$, reconstruct $r_i$ as follows:
        \begin{itemize}
            \item Set $r_i \in \zo$ so that $\ell_i = H(i,r_i,\inp,s)$.
            \item If there are two or no solutions for $r_i$, output $\bot$. 
        \end{itemize}
        
        \item Compute $r_j = f_j^{\inp,r_1,\dots,r_{j-1}}(v)$.
        \item Output $\begin{cases}
            \text{ the label } \ell_j = H(j,r_j,\inp, s) & \text{ if } j < \nstep.\\
            \text{ the outcome } g(\inp,r_1,\dots,r_\nstep) & \text{ if } j = \nstep.
        \end{cases}$
    \end{enumerate}
\end{itemize}
The obfuscated program consists of the authenticated quantum state $\psi_\auth$, the half EPR states $\epr_{\intext,\pub}, \epr_{\outtext,\pub}$, the signature token $\tau_{\token}$, and the classical oracle $\sF$. The oracle $\sF$ is designed to output random labels $\ell_j$ instead of directly revealing the measurement outcomes $r_j$. To explain, the oracle $\sF$ at step $j$ would first verify and decode the authenticated quantum state in the corresponding basis to ensure that the state is not tampered with. After verifying the tokenized signature, $\sF$ then tries to reconstruct prior measurement results $r_i$ from $\ell_i$. Next, $\sF$ computes the outcome $r_j$ of the function $f_j^{r_1,\dots,r_{j-1}}$, and outputs a random label $\ell_j$ derived from $r_j$ when $j < \nstep$ is an intermediate step. When $j = \nstep$ is the last step, $\sF$ outputs $g(\inp,r_1,\dots,r_\nstep)$ which represents the teleportation result of the quantum output.

\paragraph{Security Proof Sketch.}

We can first apply the simulation-based security of the functional authentication scheme, which shows that it is indistinguishable if the adversary gets a dummy authenticated state instead and the oracle is now simulated by a simulator who holds the state $\psi_+$ on some register $\reg{V}$. Recall that $\psi_+$ contains half EPR states $\epr_{\intext,\priv}, \epr_{\outtext,\priv}$ and the encoding $\psi$ of the original quantum program. We would like to show that the simulator never reveals unnecessary information about $\psi$, even under malicious queries.

When $j < \nstep$, the simulated oracle only outputs either $H(j,r_j,\inp,s)$ or $\bot$. The simulator can apply query-recording techniques (\eg \cite{zhandry2019record}) of random oracles to observe whether any information of the form $H(\cdot,\cdot,\inp,\cdot)$ has been learned by the adversary. At any time, the tokenized signature scheme ensures that the adversary can only learn information of the form $H(\cdot,\cdot,\inp,\cdot)$ for at most one such $\inp$ simultaneously. This mechanism forces the adversary to use a consistent classical string $\inp$ when it performs each evaluation. Since the simulator can extract the adversary's quantum input from $\inp$ and $\epr_{\intext,\priv}$, the mechanism also ensures the consistency of the adversary's quantum input throughout the evaluation.

Controlled that some information of the form $H(\cdot,\cdot,\inp,\cdot)$ has been learned by the adversary, the simulator can expand the current state on $\reg{V}$ under the $\inp$-th orthonormal basis $\set{\ket{\Phi_{\inp,r}}}_r$ of the PLM quantum program. By the design of the oracle and that $\ket{\Phi_{\inp,r}}$ yields deterministic adaptive measurement results, querying the oracle cannot affect the amplitudes of $\reg{V}$ on the basis vectors, except with negligibly small probability. 
Hence, the power of the adversary is essentially limited to passively gathering the outputs of the oracle.

There are three kinds of output that the oracle may produce: $\bot$, $H(\cdot,\cdot,\inp,\cdot)$, or $g(\inp,\cdot)$. The values of $H$ do not provide any useful information because they are random, so the adversary only acquires useful information through $g(\inp,r)$. By the guarantee of the PLM quantum program and that the program approximately implements a unitary $U$, the simulated oracle that conditionally reveals the final output $g(\inp,r)$ can be simulated from $\CTRL$-$(U^\dag A U)$ for some efficient $\CTRL$-$A$, showing that the construction is an ideal obfuscation. We refer the readers to \cref{sec:security} for a detailed security proof.

%\TODO{Circuits that differ only by a global phase (e.g., $I$ vs.\ $XZX$) are functionally equivalent with respect to their black-box input-output behavior. Nonetheless, such circuits can be distinguished when given black-box access to their controlled operations, which are implementable given the white-box program. In contrast, our obfuscation scheme simulates these circuits in the same way, regardless of any global phase difference. Consequently, the resulting obfuscated programs turn these circuits into indistinguishable programs.}

\section{Preliminary}\label{Sec:preliminary}

Let $[n] $ be $\{1, \cdots, n\}$. Sampling uniformly from a set $S$ is denoted by $s\from S$.
A function $f \colon \bbN \to [0,1]$ is called negligible, if for every polynomial $\poly(\cdot)$ and all sufficiently large $n$, it holds that $f(n)<|{1}/{\poly(n)}|$. We use $\negl(\cdot)$ to denote an unspecified negligible function. QPT stands for quantum polynomial time.

An $n$-qubit pure state $\ket{\phi}$ is a unit vector in the Hilbert space $\bbC^{2^n}$, and is identified with the density matrix $\mixed{\ket{\phi}} = \phi = \ketbra{\phi} \in \bbC^{2^n \times 2^n}$. The set of $n$-qubit (mixed) states, denoted $\Den{n}$, consists of positive semi-definite matrices in $\bbC^{2^n \times 2^n}$ with trace $1$. 
Sometimes we identify a mixed state $\rho$ with its purification (denoted $\ket{\rho}$ when we abuse notation), which is a pure state $\ket{\phi}$ such that $\mixednospace{\ket{\phi}}$ has partial trace $\rho$.
A quantum operation $\mathcal{F}$ is a completely-positive trace-preserving (CPTP) map from $\Den{n}$ to $\Den{n'}$. The quantum operation of applying an operator $V$ refers to the map $\rho \mapsto V \rho V^\dag$. We denote a state $\rho$ on register $\reg{R}$ as $\onreg{\rho}{R}$ and a quantum operation $\mathcal{F}$ on register $\reg{R}$ as  $\onreg{\mathcal{F}}{R}$. The notation $(\rho, \sigma)$ denotes a possibly entangled state (\ie need not be a product state) on two registers. Let \text{Tr} denote the trace. The partial trace $\text{Tr}_{\reg{R}}$ is the unique linear map from registers $\reg{R,S}$ to $\reg{S}$ such that $\text{Tr}_{\reg{R}}(\onreg{\rho}{R},\onreg{\sigma}{S}) = \text{Tr}(\rho) \sigma$ for every state $(\rho, \sigma)$.
The trace norm of a linear operator $ A $ is defined as $\norm{A}_1 := \text{Tr}(\sqrt{A^\dagger A})$ and the operator norm is defined as $\norm{A}_{\mathsf{op}} := \max_{\ket{v}} \frac{\norm{Av}}{\norm{v}}$. The operator norm of a linear map $\cF:\Den{n} \to \Den{n'}$ is defined as $\norm{\cF}_{\mathsf{op}} := \max_{\rho \in \Den{n}} \norm{\cF(\rho)}_1$. For quantum operations $\cF$ and $\cG$ from $\Den{n}$ to $\Den{n'}$, their diamond distance is defined as
$\| \cF- \cG \|_\diamond := \max_{\rho \in \Den{2n}} \| (\cF \otimes \identity_n) (\rho) - (\cG \otimes \identity_n) (\rho) \|_{1}$. Given two states $\rho, \sigma$, we write $\rho \approx_\eps \sigma$ if they have trace distance $\frac{1}{2}\norm{\rho-\sigma}_1 \le \eps$.

%We also consider sub-normalized mixed states, which are positive semi-definite operators with trace at most $1$. We identify a distribution $\{\rho_j\}$ of sub-normalized states with the state $\E\left[ \rho_j \right]$. The trace norm of a matrix $A$ is $\| A \|_{tr} = \Tr (\sqrt{A^\dagger A})$. The trace distance between sub-normalized mixed states $\rho, \sigma$ is defined as $\TD{\rho, \sigma} = \frac{1}{2} \norm{\rho-\sigma}_{tr} + \frac{1}{2} \abs{\Tr(\rho - \sigma)}$.

%Two sequences of sub-normalized states $\rho_n, \sigma_n \in \Den{\poly(n)}$ are said to be statistically indistinguishable, denoted $\rho \approx \sigma$, if they have trace distance $\Tr|\rho(n)-\sigma(n)| = \negl(n) \Tr(\rho(n))$. Let $ A $ be a linear operator on a Hilbert space $ H $. The trace norm of $ A $, denoted $ \| A \|_{tr} $, is defined as $\| A \|_{tr} = \text{Tr}(\sqrt{A^\dagger A})$. 

For every classical string, we append an additional status bit that defaults to $0$. When we say that a function outputs $\bot$, we mean that it returns the all-zero string (of the appropriate length) with the status bit set to $1$. All functions will be defined under the convention that if the input contains a status bit set to $1$, then the function outputs $\bot$. As a result, when viewed as a unitary operator, every function has a known $1$-eigenvector by taking the input part as $\ket{\bot}$ and the output part as $\frac{1}{\sqrt{2}}(\ket{0}+\ket{\bot})$. It is well known that given a $1$-eigenvector $\ket{v}$ of a unitary operator $U$, the controlled unitary
$\CTRL\text{-}U := \ketbra{0} \ot I + \ketbra{1} \ot U$ can be efficiently computed using controlled swap gates and a single oracle query to $U$.

\subsection{Quantum Circuits and Programs}

A quantum circuit $Q$ is described by a sequence of unitary gates and standard-basis measurements followed by selecting the output bits or qubits. The width of $Q$ is the number of qubits it acts on, and the size of $Q$ is the number of gates and measurements it has. It is well-known that any quantum computation can be implemented by a quantum circuit whose unitary gates are chosen from the gate set $ \CliffordGroup_2 \cup \{T\} $ defined as follows.

\begin{itemize}

\item\textbf{Pauli Group.} The single-qubit Pauli group $\PauliGroup_1$ consists of the group generated by Pauli gates $\Xg = \begin{pmatrix} 0 & 1 \\ 1 & 0 \end{pmatrix}, \;
\Yg = \begin{pmatrix} 0 & -i \\ i & 0 \end{pmatrix}, \;
\Zg = \begin{pmatrix} 1 & 0 \\ 0 & -1 \end{pmatrix}$.
The  $n$-qubit Pauli group $\PauliGroup_n$ is the $n$-fold tensor product of $\PauliGroup_1$.

\item\textbf{Clifford Group.} The $n$-qubit Clifford group $\CliffordGroup_n$ consists of unitaries $C$ such that $C\PauliGroup_n C^\dagger = \PauliGroup_n$. The group is generated by Hadamard $ \Hg = \frac{1}{\sqrt{2}} \begin{pmatrix} 1 & -1 \\ 1 & 1 \end{pmatrix}$, Phase $ \Sg = \begin{pmatrix} 1 & 0 \\ 0 & i \end{pmatrix}$, and $\CNOT = \footnotesize{\begin{pmatrix} 
1 & 0 & 0 & 0 \\
0 & 1 & 0 & 0 \\
0 & 0 & 0 & 1 \\
0 & 0 & 1 & 0 \\
\end{pmatrix}}$ gates, and contains $\SWAP = \footnotesize{\begin{pmatrix} 
1 & 0 & 0 & 0 \\
0 & 0 & 1 & 0 \\
0 & 1 & 0 & 0 \\
0 & 0 & 0 & 1 \\
\end{pmatrix}}$ gate.
\item {\bf $\Tg$ gate.} The $\Tg$ gate is defined as
$
\Tg = \begin{pmatrix} 
1 & 0 \\
0 & e^{i\pi/4}
\end{pmatrix}
$.
\end{itemize}

A quantum program $(\psi,Q)$ is specified by a quantum circuit $Q$ and an auxiliary state $\psi$. The size of the quantum program is the size of $Q$ plus the width of $Q$. The input length of the quantum program is the width of $Q$ minus the length of $\psi$. Running the quantum program on an input state $\rho$ is to run the quantum circuit $Q$ on the state $(\rho,\psi)$. It is well-known that every quantum program $(\psi,Q)$ can be efficiently transformed into an equivalent quantum program $(\psi',Q^{\mathsf{univ}})$ where $Q^{\mathsf{univ}}$ is a universal quantum circuit that depends only on the size of the original quantum program and the output length, whereas the length of $\psi'$ depends only on the size of the original quantum program. Without loss of generality, we will consider quantum programs of the form $(\psi',Q^{\mathsf{univ}})$ throughout this paper.

\subsection{Quantum Teleportation}
Quantum teleportation facilitates the transmission of quantum information with classical communication and pre-shared quantum states. Define the state $\mathsf{EPR}$ as $\frac{1}{\sqrt{2}} \sum_{b\in\zo} \ket{b_j, b_j}$. An EPR pair of length $n$ is a state $(\epr_L,\epr_R) \gets \mathsf{EPR}^{\ot n}$ where their $i$-th qubits forms an $\mathsf{EPR}$ state for every $i \in [n]$. For $(z,x) \in \zo^{2n}$, we define $P_{(z,x)} := \Xg^x \Zg^z \in \PauliGroup_n$.

\begin{dfn}[Quantum Teleportation]
Let $(\epr_L, \epr_R)$ be an EPR pair of length $n$ pre-shared between a sender holding input $\phi \in \Den{n}$ and a receiver. Quantum teleportation consists of two algorithms. 
\begin{itemize}
    \item  $\TPSend(\onreg{\phi}{M}, \onreg{\epr_L}{L})$: applies $\onreg{\Hg}{M} \onreg{\CNOT}{M,L}$ on $ (\onreg{\phi}{M}, \onreg{\epr_L}{L})$, performs a standard-basis measurement on registers $\reg{(M,L)}$, and outputs the measurement outcome $(z,x)$.
    \item  $\TPRecv((z,x),\onreg{\epr_R}{R})$: applies $P_{(z,x)}{}^\dag$ to $\onreg{\epr_R}{R}$ and outputs register $\gray{R}$.
\end{itemize}
\end{dfn}

We also define the teleportation unitary operator $\onreg{\TP}{M,L} := \onreg{\Hg}{M} \onreg{\CNOT}{M,L}$.
%When we refer to teleporting a state $\onreg{\phi}{M}$ via register $\gray{L}$, we mean performing $\onreg{\TPSend}{M,L}$, measuring $(\gray{M}, \gray{L})$ in the computational basis, and interpreting the measurement outcome $(z,x)$ as the Pauli operator $\Xg^x \Zg^z$. 
The following lemma characterizes the correctness of quantum teleportation.

\begin{lemma}
\label{lemma:teleport}
Let $\ket{\phi, \tau}$ be a purified state separable from $(\epr_L,\epr_R)$. Then
$$
\onreg{\TP}{M,L}\ket{\onreg{\phi}{M},\onreg{\epr_L}{L}, \onreg{\epr_R}{R}, \onreg{\tau}{N}}
= \frac{1}{2^{n}}~\sum_{z,x \in \zo^n}~ \ket{\onreg{z}{M}, \onreg{x}{L}} \ot \onreg{P_{(z,x)}}{R} \ket{\onreg{\phi}{R}, \onreg{\tau}{N}}
$$
\end{lemma}

\subsection{Purified Random Oracle}
\label{sec:purified-ro}

A random oracle is the classical oracle of a random function. One can perfectly simulate a random oracle mapping $\cX$ to $\zo^\kappa$ by purifying the randomness of the random function. To do so, we introduce a database register $\reg{D} := \reg{\{D_x\}_{x \in \cX}}$ that stores the truth table of the random function and initialized it with the uniform superposition state $$\onreg{\ket{\cL_\emptyset}}{D} := \bigotimes_{x \in \cX} \onreg{\ket{+^\kappa}}{D_x}$$
Querying the random oracle with classical input $x$ and output register $\reg{Y}$ can be carried out by applying the unitary
$$\onreg{\WriteH(x)}{Y,D} := \onreg{\CNOT}{D_x, Y}.$$
For comparison, we also define $$\onreg{\Write(x)}{Y} := \onreg{\Xg^{x}}{Y}$$ which writes the value $x$ to register $\reg{Y}$ in the standard basis. 
For a family of unitary operators $U(x)$ parameterized by a classical string $x$, we denote
$$\onreg{U(\reg{A})}{B} := \sum_x \onreg{\ketbra{x}}{A} \ot \onreg{U(x)}{B}.$$
With these notations, the random oracle can be simulated as follows:
\begin{itemize}
    \item Initialize a database register $\reg{D} \gets \ket{\cL_{\emptyset}}$.
    \item Apply $\onreg{\WriteH(\reg{X})}{Y,D}$ for each query, given input register $\reg{X}$ and output register $\reg{Y}$.
\end{itemize}
Define the projector $\onreg{\EqualH(x)}{Y,D} := \onreg{\WriteH(x)}{Y,D} \onreg{\ketbra{0^\kappa}}{Y} \onreg{\WriteH(x)}{Y,D}$ that indicates if the value on $\reg{Y}$ matches the $x$-th entry of the database in the standard basis. A straightforward calculation shows $\onreg{\EqualH(x)}{Y,D} = \sum_{y \in \zo^\kappa} \onreg{\ketbra{y}}{Y} \ot \onreg{\ketbra{y}}{D_x}$ and that $\EqualH(x)$ commutes with $\EqualH(x')$ for every $x \neq x'$. Given a non-empty subset $S \subseteq \cX$, we define the following projectors on $\reg{D}$ that informs if certain function values have been provided to other parties.
\begin{itemize}
    \item $\onreg{\Has_S}{D} := I - \bigotimes_{x \in S} \onreg{\ketbra{+^\kappa}}{D_x}$ indicates that some entries in $S$ have been given out.
    \item $\onreg{\Only_S}{D} := \bigotimes_{x \not \in S} \onreg{\ketbra{+^\kappa}}{D_x} - \onreg{\ketbra{\cL_\emptyset}}{D}$ indicates that some entries in $S$, and only entries in $S$, have been given out.
\end{itemize}

\begin{lemma} \label{lemma:ro-unpredictable}
    $\norm{\onreg{\EqualH(x)}{L,D} \onreg{(I-\Has_{\{x\}})}{D}}_{\mathsf{op}} = 2^{-\kappa/2}$ for every $x \in \cX$.
\end{lemma}
\begin{proof}This follows from
    \begin{align*}
        &\norm{\onreg{\EqualH(x)}{L,D} \onreg{(I-\Has_{\{x\}})}{D}}_{\mathsf{op}}^2\\
        =& \norm{\onreg{(I-\Has_{\{x\}})}{D} \onreg{\EqualH(x)}{L,D} \onreg{(I-\Has_{\{x\}})}{D}}_{\mathsf{op}}\\
        =& \norm{\onreg{\ketbra{+^{\kappa}}}{D_x} \onreg{\left(\sum_{y \in \zo^{\kappa}} \ketbra{y} \ot \ketbra{y}\right)}{L,D_x} \onreg{\ketbra{+^{\kappa}}}{D_x} }_{\mathsf{op}}\\
        =& \norm{\frac{1}{2^\kappa} \left(\sum_{y \in \zo^{\kappa}} \ketbra{y} \ot \ketbra{+^{\kappa}}\right)}_{\mathsf{op}} = \norm{\frac{1}{2^\kappa} \left(I \ot \ketbra{+^{\kappa}}\right)}_{\mathsf{op}} = \frac{1}{2^{\kappa}}.
    \end{align*}
\end{proof}

\subsection{Tokenized Signature}
\label{sec:token-sig}
\begin{dfn}
\label{dfn:token}
A \textit{tokenized signature scheme} $\Token$ consists of algorithms $(\Gen, \Sign, \Verify)$ with the following syntax:

\begin{itemize}
    \item
    $\Gen(1^\secparam, 1^n) \to (\vk, \tau_{\token})$ takes as input the security parameter $\secparam$ and the message length $n$, and outputs a classical verification key $\vk$ and a quantum signing key $\tau_{\token}$.

    \item 
    
    $\Sign(m, \tau_{\token}) \to s$ takes as input a message $m \in \{0,1\}^n$ and the quantum signing key $\tau_{\token}$, and outputs a classical signature $s$.

    \item $\Verify_\vk(m, s) \to \top / \bot$ takes as input the verification key $\vk$, a message $m$, and a signature $s$, and outputs $\top$ (accept) or $\bot$ (reject).
\end{itemize}    
A tokenized signature scheme should satisfy correctness and unforgeability as follows.
\begin{itemize}
    \item Correctness: For any message $r \in \{0,1\}^n$,
\[
\Pr\left[
  \Verify_\vk(m,s) = \top 
  \,\middle|\,
  \begin{array}{l}
    (\vk, \tau_{\token}) \leftarrow \Gen(1^\secparam, 1^n) \\
    s \leftarrow \Sign(m, \tau_{\token})
  \end{array}
\right] = 1.
\]

\item Unforgeability: For any $n = \poly(\secparam)$ and polynomial-query quantum adversary $\adv$,
\[
\Pr\left[
\begin{array}{c}
m_0 \neq m_1 \\
\land \; \Verify_\vk(m_0, s_0) = \top \\
\land \; \Verify_\vk(m_1, s_1) = \top
\end{array}
\,\middle|\,
\begin{array}{l}
(\vk, \tau_{\token}) \leftarrow \Gen(1^\secparam, 1^n) \\
(m_0, m_1, s_0, s_1) \leftarrow \adv^\mathcal{\Verify_\vk(\cdot,\cdot)}(\tau_{\token})
\end{array}
\right] = 2^{-\Omega(\secparam)}
\]
\end{itemize}
\end{dfn}

The definition requires that honestly generated signatures will always be accepted by the verification algorithm. Moreover, it asserts that any adversary with polynomially-many quantum queries to the verification oracle cannot produce valid signatures for two distinct messages simultaneously, except with exponentially small probability.

\begin{theorem}[\cite{ben2023quantum}]\label{thm:tokenized-signature}
There exists a tokenized signature scheme $\Token$ unconditionally with signature length $O(\secparam n)$.
\end{theorem}

\subsection{Post-quantum PRF}
\label{sec:prf}
A pseudorandom function is a family of efficiently computable functions that are computationally indistinguishable to truly random functions under query access. Here, we consider post-quantum pseudorandom functions, where indistinguishability holds even for adversaries who can query the function in superposition.

\begin{dfn}[Post-quantum PRF] \label{dfn:PRF}
A post-quantum PRF is a family $\set{\PRF_k: \cX \to \cY}_{k \in \cK}$ of efficiently computable functions, where the key space $\cK$, the domain $\cX$, and the range $\cY$ implicitly depend on the security parameter $\secparam$. We require computational indistinguishability between $\PRF$ and the truly random function, in the sense that the following should hold for every QPT adversary $\adv$ with quantum queries to the function.
$$\abs{\Pr_{k\gets\mathcal{K}}[\adv^{\text{PRF}_k}(\cdot)=1]-\Pr_{F\gets\cY^\cX}[\adv^{F}(\cdot)=1]} \le \negl(\secparam)$$
\end{dfn}

\begin{theorem}[\cite{zhandry2021construct}]\label{thm:PRF}
Post-quantum \text{PRF} can be constructed from post-quantum one-way functions.    
\end{theorem}

\subsection{Useful Lemmas}
\begin{lemma}[\cite{watrous2018theory}] \label{lemma:diamond-norm}
    Let $\Phi_0, \Phi_1: \Den{n} \to \Den{m}$ be quantum operations.
    \begin{itemize}
        \item If $\Phi_1$ is a unitary transformation, then $\norm{\Phi_0 - \Phi_1}_\diamond \le \sqrt{ 2 \max_{\rho \in \Den{n}} \norm{\Phi_0(\rho) - \Phi_1(\rho)}_{1} }$.
        \item If $\Phi_0, \Phi_1$ are both isometries, then $\norm{\Phi_0 - \Phi_1}_\diamond = \norm{\Phi_0 - \Phi_1}_{\mathsf{op}}$.
    \end{itemize}
\end{lemma}

\begin{theorem}[\cite{kretschmann2008information}] \label{thm:continuity-of-dilation}
    Let $\Phi_0, \Phi_1: \Den{n} \to \Den{m}$ be quantum operations and $U_0, U_1: \bbC^n \to \bbC^m \ot \bbC^k$ be isometries such that
    $\Phi_i(\rho) \equiv \Tr_{2} ( U_i \rho U_i^\dag )$ for $i=0,1$. Then we have
    $$ \inf_{\text{unitary }V} \norm{U_0 - (I \ot V) U_1}_{\mathsf{op}} \le \sqrt{\norm{\Phi_0 - \Phi_1}_\diamond} $$
\end{theorem}

\begin{lemma}\label{lemma:easy-norm}
    Let $U, W: \bbC^n \to \bbC^m$ be isometries. Then $\norm{U(\cdot)U^\dag - W(\cdot)W^\dag}_{\diamond} \le 2\norm{U - W}_{\mathsf{op}}$.
\end{lemma}
\begin{proof}
    By \cref{lemma:diamond-norm}, the triangle inequality, and that operator norm is sub-multiplicative,
    \begin{align*}
        & \norm{U(\cdot)U^\dag - W(\cdot)W^\dag}_{\diamond} = \norm{U(\cdot)U^\dag - W(\cdot)W^\dag}_{\mathsf{op}} = \max_{\text{state }\rho} \norm{U\rho U^\dag - W\rho W^\dag}_{1}\\
        \le& \max_{\text{state }\rho} \left(\norm{(U-W)\rho U^\dag}_{1} + \norm{(W\rho (U-W)^\dag}_{1}\right) \le 2 \norm{U-W}_{\mathsf{op}}
    \end{align*}
\end{proof}

\begin{lemma}[Gentle Measurement, \cite{winter1999coding}]\label{lemma:gentle-measurement}
Let $\rho$ be a quantum state, and $\{\Pi,\Ig-\Pi\}$ be a projective measurement such that $\Tr(\Pi \rho) \leq \eps$. Let $\rho' = \frac{(I-\Pi) \rho (I-\Pi)}{\Tr((I-\Pi) \rho)}$ be the state after applying $\{\Pi,\Ig-\Pi\}$ on $\rho$ and post-selecting on obtaining the second outcome. Then the states $\rho$ and $\rho'$ are within trace distance $2 \sqrt{\eps}$.
\end{lemma}

The following lemma allows us to switch between oracles that differ on hard-to-find inputs without perturbing the state too much.

\begin{lemma}[Oracle Hybrid Lemma] \label{lemma:oracleswitch}
Let $\mathcal{K}$ be a distribution. For all possible $k \from \mathcal{K}$, let $\ket{\psi_k}$ be a quantum state, $\Pi_k$ be a projector, and $O_k$ and $O'_k$ be unitary operators such that $O_k (I - \Pi_k) = O'_k (I-\Pi_k)$.  
Suppose for every $q$-query algorithm $\adv$, it holds that 
$
\E_{k \from \cK} \left[ \norm{\Pi_{k} \adv^{O_k} \ket{\psi_k}}^2 \right] \le \eps
$.
Then for every $q$-query algorithm $\adv$, we have $$\E_{k \gets \cK}\left[ \mixed{\adv^{O_k}\ket{\psi_{k}}}\right] \;\;\underset{O( q \sqrt{\eps})}{\approx}\;\; \E_{k \gets \cK}\left[\mixed{\adv^{O'_k}\ket{\psi_{k}}}\right]$$
\end{lemma}
\begin{proof} 
For each possible $k \gets \cK$ and $i = 0,1,\dots,q$, we define the following states
\begin{itemize}
    \item $\ket{\psi_{k,i}}$ is obtained by applying $\adv$ to $\ket{\psi_k}$ where the first $i$ queries are made to $O_k$ and the rest of the queries are made to $O'_k$.
    \item $\ket{\psi'_{k,i}}$ is obtained similarly to $\ket{\psi_{k,i}}$, except that we apply the measurement $\{\Pi_k,I-\Pi_k\}$ right before the $i$-th oracle query and post-selecting on obtaining the second outcome.
    \item $\ket{\varphi_{k,i}}$ is obtained by applying $\adv^{O_k}$ to $\ket{\psi_k}$ up to and not including the $i$-th query.
\end{itemize}
Set $\eps_{k,i} := \norm{\Pi_k \ket{\varphi_{k,i}}}^2$. By assumption, we have $\E_{k \gets \cK}\left[\eps_{k,i}\right] \le \eps$. Applying the triangle inequality, the gentle measurement lemma, and the Cauchy-Schwartz inequality, we obtain
\begin{align*}
    & \frac{1}{2}\text{Tr}\abs{\E_{k\gets \cK}\mixed{\ket{\psi_{k,i}}} - \E_{k\gets \cK}\mixed{\ket{\psi'_{k,i}}}} \\
    \le & \E_{k\gets \cK} \left[\frac{1}{2}\text{Tr} \abs{\mixed{\ket{\psi_{k,i}}} - \mixed{\ket{\psi'_{k,i}}}}\right] \\
    \le & \E_{k\gets \cK} \left[2 \sqrt{\eps_{k,i}}\right] 
    \le 2 \sqrt{\E_{k\gets \cK} \left[\eps_{k,i}\right]} \le 2\sqrt{\eps}
\end{align*}
By assumption, right after post-selecting on the measurement result of $I-\Pi_k$, applying $O_k$ has the same effect as applying $O'_k$. Hence, the state $\ket{\psi'_{k,i}}$ can also be obtained similarly to $\ket{\psi_{k,i-1}}$, except that we apply the measurement $\{\Pi_k,I-\Pi_k\}$ right before the $i$-th oracle query and post-selecting on obtaining the second outcome. By a similar argument, we have
\begin{align*}
    \frac{1}{2}\text{Tr}\abs{\E_{k\gets \cK}\mixed{\ket{\psi_{k,i-1}}} - \E_{k\gets \cK}\mixed{\ket{\psi'_{k,i}}}} \le 2\sqrt{\eps}
\end{align*}
Putting together, we have 
$$\E_{k \gets \cK}\mixed{\adv^{O_k}\ket{\psi_{k}}} = \E_{k\gets \cK}\mixed{\ket{\psi_{k,q}}} \underset{O(q \sqrt{\eps})}{\approx} \E_{k\gets \cK}\mixed{\ket{\psi_{k,0}}} = \E_{k \gets \cK}\mixed{\adv^{O'_k}\ket{\psi_{k}}}$$
\end{proof}

\section{Functional Authentication}
\label{Sec:auth}
In this section, we introduce functional authentication, which enhances standard quantum authentication by enabling fine-grained access to measurement results of the state. In particular, the functional authentication scheme supports classical decoding oracles, which are interpreted as functional keys. The security of functional authentication will ensure that the adversaries gain no additional advantage beyond the combined functionality allowed by the given functional keys. Below, we call a unitary gate $G$ linear if it is composed of $\CNOT$ gates. We sometimes use the expression $\onreg{\Write(\reg{A})}{B} := \sum_{a} \onreg{\ketbra{a}}{A} \ot \onreg{\Xg^a}{B}$ for $\CNOT$ gates. 

%Viewing the scheme through the lens of functional authentication, we interpret the classical verification and decoding oracles as functional keys that securely reveal only the permitted measurement outcomes, even when the adversary dynamically interacts with the ciphertext state. In contrast to the quantum authentication scheme in \cite{bartusek2024quantum}, which only ensures static distributional guarantees for individual measurement results, functional authentication robustly captures the joint distribution with quantum states and better characterizes adversarial dynamic behavior. This stronger security ensures that, when applied to quantum state obfuscation in subsequent sections, the adversaries gain no additional advantage beyond the allowed functionality.

\subsection{Definition}

\begin{dfn}[Functional Authentication]
\label{dfn:FuncAuth}
A functional authentication scheme consists of algorithms with the following syntax:
\begin{itemize}
    \item $\KeyGen(1^\secparam, 1^n) \to k$ takes as input the security parameter $\secparam$ and the qubit count $n = \poly(\secparam)$, and outputs a classical authentication key $k$.
    \item $\Enc_k(\rho) \to \sigma$ takes as input an $n$-qubit state $\rho$ and the key $k$, and outputs a state $\sigma = (\sigma_1,\dots,\sigma_n)$ where each $\sigma_i$ has $p(\secparam)$ qubits. This algorithm is an isometry parameterized by the authentication key $ k $.

    \item $\Eval_{\theta} \to \wt{\theta}$ and $\Eval_{G} \to \wt{G}$ are classical algorithms that output the descriptions of homomorphic gates. Here, $ \theta \in \zo^n $, $\wt{\theta} \in \zo^{n p(\secparam)}$ stand for measurement bases, and $ G ,\wt{G}$ are linear gates on $n,n p(\secparam)$ qubits respectively.
    
    \item $\Dec_{k,\theta,G}(c) \to m / \bot$ is a classical algorithm that takes a string $c$, an authentication key $ k $, measurement bases $ \theta \in \zo^n $ and a linear gate $ G $ on $n$ qubits, and outputs either a decoded outcome $ m $ or $ \bot $ (failure).

    \item $\Verify_{k,\theta,G}(c) \to \top/\bot$ is a classical algorithm that verifies the string $ c $ in a manner similar to the decoding algorithm, and outputs either  $ \top $ if $m \neq \bot$, or $\bot$ if $m = \bot$.

\end{itemize}

For a classical function $f: \zo^n \to \zo^*$, a string $\theta \in \zo^{n}$, and a linear gate $G$ on $n$-qubits, we define the measurement
    $$\meas{f,\theta,G} := \set{G^\dag \Hg^\theta \left(\sum_{x: f(x)=y}  \ketbra{x}\right) \Hg^\theta G}_y$$

Functional authentication scheme needs to satisfy the following properties:
\begin{itemize}
    \item Correctness: For any classical function $f: \{0,1\}^{n} \to \{0,1\}^*$, bases $\theta \in \zo^{n}$, linear gate $G$ on $n$ qubits, and all possible $k \from \KeyGen(1^\secparam,1^n)$, we have
    \begin{align*}
    \meas{\left(f \circ \Dec_{k,\theta,G}\right),\wt{\theta},\wt{G}} \circ \Enc_k  = \Enc_k \circ \meas{f,\theta,G}.
    \end{align*}
    under the convention $f(\bot) := \bot$. Moreover, for every $n$-qubit state $\rho$, we have
    \begin{align*}
    \Pr\left[\top \gets \meas{\Ver_{k,\theta,G},\wt{\theta},\wt{G}} \left( \Enc_k(\rho)\right) \right]=1.
    \end{align*}
    \item Functional Security: Let $f_1,\dots,f_t: \zo^n \rightarrow \zo^{n'}$ be any functions, $\theta_1,\dots,\theta_t \in \{0,1\}^n$ be any measurement bases, and $G_1,\dots,G_t$ be any linear gates on $n$ qubits.
Define corresponding unitary operators $\fDec_k(j, \wt{v}, \reg{Y})$ and $\fVer_k(j, \wt{v}, \reg{Y}, \reg{V})$ as follows. The ``If-Then'' expressions should be interpreted as controlled operators, whereas the measurements are to be applied coherently.

\begin{minipage}{0.41\textwidth}
    \begin{itemize}[leftmargin=*]
        \item $\fDec_k(j, \wt{v}, \reg{Y})$:
    \begin{enumerate}[leftmargin=*]
        \item $v \gets \Dec_{k,\theta_j,G_j}(\wt{v})$
        \item If $v=\bot$, apply $\onreg{\Write(\bot)}{Y}$
        \item Otherwise, apply $\onreg{\Write(f_j(v))}{Y}$
    \end{enumerate}
    \end{itemize}
\end{minipage}
\hfill
\begin{minipage}{0.6\textwidth}
    \begin{itemize}[leftmargin=*]
        \item $\fVer_k(j, \wt{v}, \reg{Y}, \reg{V})$:
    \begin{enumerate}[leftmargin=*]
        \item $b \gets \Ver_{k,\theta_j,G_j}(\wt{v})$
        \item If $b=\bot$, apply $\onreg{\Write(\bot)}{Y}$
        \item Otherwise, apply $\onreg{\Write\left(\meas{f_j,\theta_j,G_j}(\reg{V})\right)}{Y}$
    \end{enumerate}
    \end{itemize}
\end{minipage}\\

Let $t,n' = \poly(\secparam)$. Then for every $\poly(\secparam)$-query adversary $\adv$ and state $\onreg{\ket{\psi}}{V,W}$,
\[
\left\{(\reg{\wt{V}},\reg{W}) \,\middle|\, 
\begin{array}{l}
    \reg{V}, \reg{W} \gets \ket{\psi}\\
    k \gets \KeyGen(1^\secparam,1^n) \\
    \reg{\wt{V}} \gets \Enc_k(\reg{V}) \\
    \reg{\wt{V}},\reg{W} \gets \adv^{\fDec_k(\cdot,\cdot,\cdot)}(\reg{\wt{V}},\reg{W})
\end{array}
\right\} \underset{2^{-\Omega(\secparam)}}{\approx} \left\{(\reg{\wt{V}},\reg{W}) \,\middle|\, 
\begin{array}{l}
    \reg{V}, \reg{W} \gets \ket{\psi}\\
    k \gets \KeyGen(1^\secparam,1^n) \\
    \reg{\wt{V}} \gets \Enc_{k}(0^n) \\ \reg{\wt{V}},\reg{W} \gets \adv^{\fVer_{k}(\cdot,\cdot,\cdot,\reg{V})}(\reg{\wt{V}},\reg{W})
\end{array}
\right\} 
\]
\end{itemize}
\end{dfn}
Our notion of functional security provides strong guarantees by ensuring that the authenticated quantum state remains secure under a simulation-based definition, even in the presence of both decryption and verification oracles. While our scheme shares syntactic similarities with the quantum authentication scheme of \cite{bartusek2024quantum}, their security analysis is shown secure only under a property-based definition in the presence of verification oracles. As a result, their scheme falls short of achieving the goals of functional authentication. In contrast, our definition of functional security captures dynamic adversarial behaviors beyond the static settings considered in their work, thereby enabling a wider range of applications.

\begin{theorem}[\cite{bartusek2024quantum}] \label{thm:auth-code-old-soundness}
There exists a quantum authentication scheme with the following properties:
\begin{itemize}
    \item Syntax: Same as functional authentication.
    \item Correctness: Same as functional authentication.
    \item Soundness: 
    For every classical function $f: \{0,1\}^{n} \to \{0,1\}^*$, bases $\theta \in \zo^{n}$, linear gate $G$, and polynomial-query adversary $\adv$, there exists an $\epsilon(\secparam) \in [0,1]$ such that for every $n$-qubit state $\rho$, 
    \[\left\{y : \begin{array}{c} k \gets \KeyGen(1^\secparam,1^n) \\ \sigma \from \adv^{\Verify_{k,\cdot,\cdot}(\cdot)} \left( \Enc_k(\rho) \right) \\ y \gets \meas{\left(f \circ \Dec_{k,\theta,G}\right),\wt{\theta},\wt{G}}(\sigma) \end{array}\right\} \underset{2^{-\Omega(\secparam)}}{\approx} \begin{array}{c}  (1-\epsilon(\secparam))\left\{y : y \gets \meas{f,\theta,G} (\rho)\right\} \\
    + \epsilon(\secparam)\{\bot\} 
    \end{array}.\]
    \item Privacy: For every quantum state $\rho$ with $n = \poly(\secparam)$ qubits and every $\poly(\secparam)$-query quantum adversary $\adv$, we have
$$\E_{k \from \KeyGen(1^\secparam, 1^n)}\left[ \adv^{\Verify_{k,\cdot,\cdot}(\cdot)}(\Enc_k(\rho)) \right] \;\underset{2^{-\Omega(\secparam)}}{\approx}\;
\E_{k \from \KeyGen(1^\secparam, 1^n)}\left[ \adv^{\Verify_{k,\cdot,\cdot}(\cdot)}(\Enc_k(0^n))\right].$$

\end{itemize}
\end{theorem}

\subsection{Construction and Security Proof}

Interestingly, we show that the construction of the quantum authentication scheme proposed in \cite{bartusek2024quantum} can, in fact, be an instance of our functional authentication. Although the authors of \cite{bartusek2024quantum} established security under a weaker notion, our analysis in this section demonstrates that their construction indeed satisfies the stronger definition of functional authentication (\cref{dfn:FuncAuth}). 

For clarity, before presenting our security analysis, we briefly recall their construction, which we refer to as the coset authentication code. 
For a linear subspace $S \subset \bbF_2^d$ and a vector $v \in \bbF_2^d$, the coset state $\ket{S+v}$ is defined as
$$
\ket{S+v} := \frac{1}{\sqrt{|S|}} \sum_{s \in S} \ket{s+v}
$$

\begin{Construction}{(\cite{bartusek2024quantum}) Coset Authentication Code}
\label{con:CosetAuth}
{\begin{itemize}
    \item $\KeyGen(1^\secparam, 1^n)$:  
    Sample a uniformly random subspace $S \subset \mathbb{F}_2^{2\secparam+1}$ of dimension $\secparam$. Sample a uniformly random vector $\Delta \in \mathbb{F}_2^{2\secparam+1} \setminus S$, and for each $i \in [n]$, sample uniformly random vectors $x_i, z_i \in \mathbb{F}_2^{2\secparam+1}$. The output is the authentication key $k := (S, \Delta, x, z)$.
    
    \item $\Enc_k(\rho)$:
    First applies the isometry $\sum_{b\in\zo} \alpha_b\ket{b} \mapsto \sum_{b\in\zo} \alpha_b\ket{S + b \Delta}$ to each qubit of $\rho$, and then applies the quantum one-time pad $\Xg^x \Zg^z$.

    \item $\Eval_\theta$ and $\Eval_G$:
    For $\theta = (\theta_1,\cdots,\theta_n)$, set $\wt{\theta} = (\theta_1^{p(\secparam)} || \cdots || \theta_n^{p(\secparam)})$. Set $\wt{G} = G^{\ot n}$.

    \item $\Dec_{k,\theta,G}(c)$: Parse $k = (S,\Delta,x,z)$. Let $S_\Delta = S \cup (S+\Delta),\; \wh{S} = (S_{\Delta})^\perp,\; \wh{\Delta} \in S^\perp \setminus (S_{\Delta})^{\perp},$ and $\wh{S}_{\wh{\Delta}} = \wh{S}\cup (\wh{S}+\wh{\Delta})$.
    Let $(z_G, x_G)$ be the result of starting with $(z,x)$ and applying the following update rule sequentially for each $\CNOT$ gate (say acting on qubits $i,j$) in $G$: $((z_i, x_i), (z_j, x_j)) \mapsto ((z_i \oplus z_j, x_i), (z_j, x_i \oplus x_j))$. For $i \in [n]$, compute
    $$m_i = \begin{cases}0 &  \text{ if } (\theta_i = 0 \text{ and } c_i \in S + x_{G,i}) \text{ or } (\theta_i = 1 \text{ and } c_i \in \wh{S} + z_{G,i}) \\ 1 & \text{ if } (\theta_i = 0 \text{ and } c_i \in S + \Delta + x_{G,i}) \text{ or } (\theta_i = 1 \text{ and } c_i \in \wh{S} + \wh{\Delta} + z_{G,i}) \\ \bot & \text{ otherwise}\end{cases}$$
    If any $m_i = \bot$, then output $\bot$. Otherwise output $m = (m_1,\dots,m_n)$.
    \item $\Verify_{k,\theta,G}(c)$: 
    For $i \in [n]$, check whether $c_i \in S_\Delta + x_{G,i}$ whenever $\theta_i = 0$ and $c_i \in \wh{S}_{\wh{\Delta}} + z_{G,i}$ whenever $\theta_i = 1$. If all checks pass, output $\top$. Otherwise, output $\bot$.
\end{itemize}}
\end{Construction}

\begin{theorem}
\label{thm:FuncAuth}
    Construction \ref{con:CosetAuth} is a functional authentication scheme for a regime $\secparam \ge \Omega(n)$.
\end{theorem}

By \cref{thm:auth-code-old-soundness}, we are left with proving that construction \ref{con:CosetAuth} satisfies functional security (\cref{dfn:FuncAuth}). Consider an alternate key generation algorithm $\KeyGen'$ that, in addition to generating the key $k = (S, \Delta, x, z)$ as in $\KeyGen$, also samples larger subspaces $R$ and $\widehat{R}$, and a modified verification algorithm $\Verify'$ that checks memberships in cosets of $R$ and $\wh{R}$.

\begin{itemize}
\setlength{\itemsep}{1pt}
    \item $\KeyGen'(1^\secparam,1^n)$:
    \begin{itemize}[leftmargin=*]
    \setlength{\itemsep}{0pt}
        \item Sample $k=(S,\Delta,x,z) \from \KeyGen(1^\secparam, 1^n)$.
        \item Sample a random subspace $R \supset S_\Delta = S \cup (S + \Delta)$ such that $\dim(R) = \tfrac{3\secparam}{2} + 1$.
        \item Sample a random subspace  $\wh{R} \supset \wh{S}_{\widehat{\Delta}} = S^\perp$
        such that $\dim(\widehat{R}) = \tfrac{3\secparam}{2} + 1$.
        \item Sample a random $x^+ \gets x + R^n$ and a random $z^+ \gets z + \wh{R}^n$. \item Output $(k,k^+)$ where we define the extended key $k^+ := (R, \wh{R}, x^+, z^+)$.
    \end{itemize}
    \item $\Verify'_{k^+,\theta,G}(c)$: For $i \in [n]$, check whether $c_i - x^+_{G,i} \in R$ whenever $\theta_i = 0$ and $c_i - z^+_{G,i} \in \wh{R}$ whenever $\theta_i = 1$. If all checks pass, output $\top$. Otherwise, output $\bot$.
\end{itemize}
We show that oracle access to $\Ver_k$ is indistinguishable from oracle access to $\Ver'_{k^+}$.

\begin{lemma}\label{lemma:Ver-to-Ver'}
For every $\poly(\secparam)$-query adversary $\adv$ and quantum state $\onreg{\ket{\psi}}{V,W}$,
{\small
\[
\left\{(k,\reg{\wt{V}},\reg{W}) \,\middle|\, 
\begin{array}{l}
    \reg{V}, \reg{W} \gets \ket{\psi}\\
    (k,k^+) \gets \KeyGen'(1^\secparam,1^n) \\
    \reg{\wt{V}} \gets \Enc_k(\reg{V}) \\
    \reg{\wt{V}},\reg{W} \gets \adv^{\Ver_{k}}(\reg{\wt{V}},\reg{W})
\end{array}
\right\}  \underset{2^{-\Omega(\secparam)}}{\approx} 
\left\{(k,\reg{\wt{V}},\reg{W}) \,\middle|\, 
\begin{array}{l}
    \reg{V}, \reg{W} \gets \ket{\psi}\\
    (k,k^+) \gets \KeyGen'(1^\secparam,1^n) \\
    \reg{\wt{V}} \gets \Enc_k(\reg{V}) \\
    \reg{\wt{V}},\reg{W} \gets \adv^{\Ver'_{k^+}}(\reg{\wt{V}},\reg{W})
\end{array}
\right\}
\]
}\end{lemma}
\begin{proof}
To highlight the difference, $\Ver_k$ checks memberships in cosets $x_{G,i} + S_\Delta$ and $z_{G,i} + \wh{S}_{\wh{\Delta}}$ whereas $\Ver'_{k^+}$ checks memberships in larger cosets $x_{G,i} + R$ and $z_{G,i} + \wh{R}$. For every $(\theta,G,\wt{v})$ output by a polynomial-query adversary on the left-hand side, the two verification oracles behave differently when either $\wt{v}_i - x_{G,i} \in R \setminus S_\Delta$  or $\wt{v}_i - z^{+}_{G,i} \in \wh{R} \setminus \wh{S}_{\wh{\Delta}}$ for some $i \in [n]$. Since the adversary on the left-hand side has no information about $(R,\wh{R})$, the first type of event occurs with probability at most
\begin{align*}
    \Pr_{R}\left[ \wt{v}_i - x^{+}_{G,i} \in R \setminus S_\Delta\right] =\frac{|R \setminus S_\Delta|}{|\mathbb{F}_2^{2\secparam+1} \setminus S_\Delta|} = \frac{2^{(3\secparam/2)+1} - 2^{\secparam+1}}{2^{2\secparam+1} - 2^{\secparam+1}} = 2^{-\Omega(\secparam)}.
\end{align*}
Similarly, the second type of event occurs with probability at most $2^{-\Omega(\secparam)}$. By a union bound, the two verification oracles behave differently on $(\theta, G,\wt{v})$ with probability at most $2^{-\Omega(\secparam)}$. The lemma now follows by applying the oracle hybrid argument (\cref{lemma:oracleswitch}). 
\end{proof}

An additional feature of Construction \ref{con:CosetAuth} is that homomorphic evaluation of Pauli gates can be performed solely via key updates.

\begin{lemma}\label{lemma:switch-key}
    For every $k = (S\Delta,x,z) \gets \KeyGen(1^\secparam, 1^n)$ and every Pauli $P = \Xg^u \Zg^v \in \PauliGroup_n$, we define the key $k_P := (S,\Delta,x',z')$ where $x'_i = x_i \oplus u_i \Delta$ and $z'_i = z_i \oplus v_i \wh{\Delta}$ for every $i\in [n]$. 
    Then $\Ver_{k_P,\cdot,\cdot}(\cdot) = \Ver_{k,\cdot,\cdot}(\cdot)$ and $\Enc_{k_P} = \Enc_k \circ P$ up to a global phase.
\end{lemma}
\begin{proof}
    Since $x'_i - x_i = u_i \Delta \in S_{\Delta}$ for every $i\in [n]$, we have $x'_{G,i} - x_{G,i} \in S_{\Delta}$ and similarly we have $z'_{G,i} - z_{G,i} \in \wh{S}_{\wh{\Delta}}$ for every $i \in [n]$. Hence, $\Ver_{k_P,\cdot,\cdot}(\cdot)$ and $\Ver_{k,\cdot,\cdot}(\cdot)$ computes the same function because they verify memberships of the same cosets. Next, for every $w \in \zo^{n}$,
    \begin{align*}
        \Enc_{k_P} \ket{w}
        &= \bigotimes_{i=1}^n \Xg^{x_i \oplus u_i \Delta} \Zg^{z_i \oplus v_i \wh{\Delta}} \ket{S+w_i\Delta}
        = \bigotimes_{i=1}^n \Xg^{x_i \oplus u_i \Delta} \Zg^{z_i \oplus v_i \wh{\Delta}} \Xg^{w_i\Delta} \ket{S}\\
        &= \bigotimes_{i=1}^n (-1)^{u_i\inner{\Delta, z_i} + v_i w_i\inner{\wh{\Delta}, \Delta}} \Xg^{x_i} \Zg^{z_i} \Xg^{(u_i \oplus w_i) \Delta} \Zg^{v_i \wh{\Delta}} \ket{S}\\
        &= (-1)^{\sum_i u_i\inner{\Delta, z_i}} (-1)^{\inner{v,w}} \bigotimes_{i=1}^n \Xg^{x_i} \Zg^{z_i} \ket{S + (u_i \oplus w_i) \Delta}\\
        &= (-1)^{\sum_i u_i\inner{\Delta, z_i}} (-1)^{\inner{v,w}} \Enc_k \ket{u \oplus w}\\
        &= (-1)^{\sum_i u_i\inner{\Delta, z_i}} \Enc_k \; \Xg^u \Zg^v \ket{w} = (-1)^{\sum_i u_i\inner{\Delta, z_i}} \Enc_k \circ P \ket{w}
    \end{align*}
    where the second line uses the anti-commutativity of $\Xg$ and $\Zg$, and the third line follows because we have chosen $\Delta, \wh{\Delta}$ such that $\inner{\wh{\Delta}, \Delta} = 1$ and $\wh{\Delta} \in S^{\perp}$. Hence, we observe that $\Enc_{k_P}$ and $\Enc_k \circ P$ differs only by a global phase $(-1)^{\sum_i u_i \inner{\Delta,z_i}}$ independent of $w$.    
\end{proof}

In the next lemma, we focus on the case $t=1$ and $n'=1$ (\ie $f$ outputs $1$ bit), therefore omitting reference to the index $j$ in $\fDec$ and $\fVer$. 

\begin{lemma}\label{lemma:Dec-to-Ver}
The following holds for a regime $\secparam \ge \Omega(n)$.
For every quantum adversary $\adv$ and quantum state $\onreg{\ket{\psi}}{V,W}$ where we parse $\reg{W} = (\reg{Y}, \reg{Z})$,

{\small
\[
\left\{(k_P,\reg{\wt{V}},\reg{W}) \,\middle|\, 
\begin{array}{l}
    \reg{V}, \reg{W} \gets \ket{\psi}\\
    (k,k^+) \gets \KeyGen'(1^\secparam,1^n) \\
    (\epr_{L},\epr_R) \gets \mathsf{EPR}^{\ot n}\\
    \reg{\wt{V}} \gets \Enc_{k}(\epr_R) \\
    \reg{\wt{V}},\reg{W}, \reg{V} \gets \adv^{\Ver_{k}}(\reg{\wt{V}},\reg{W}, \reg{V})\\
    P \gets \TPSend(\reg{V},\epr_{L})\\
    \reg{\wt{V}},\reg{Y} \gets \fDec_{k_P}(\reg{\wt{V}},\reg{Y})
\end{array}
\right\} \underset{2^{-\Omega(\secparam)}}{\approx} \left\{(k_P,\reg{\wt{V}},\reg{W}) \,\middle|\, 
\begin{array}{l}
    \reg{V}, \reg{W} \gets \ket{\psi}\\
    (k,k^+) \gets \KeyGen'(1^\secparam,1^n) \\
    (\epr_{L},\epr_R) \gets \mathsf{EPR}^{\ot n}\\
    \reg{\wt{V}} \gets \Enc_{k}(\epr_R) \\
    \reg{\wt{V}},\reg{W}, \reg{V} \gets \adv^{\Ver_{k}}(\reg{\wt{V}},\reg{W}, \reg{V})\\
    \reg{\wt{V}},\reg{Y},\reg{V} \gets \fVer_{k}(\reg{\wt{V}},\reg{Y},\reg{V})\\
    P \gets \TPSend(\reg{V},\epr_{L})
\end{array}
\right\} 
\]
}
\end{lemma}
\begin{proof}
By \cref{lemma:Ver-to-Ver'}, it suffices to prove the following where we switched $\Ver_k$ to $\Ver'_{k^+}$:
{\small
\[
\left\{(k_P,\reg{\wt{V}},\reg{W}) \,\middle|\, 
\begin{array}{l}
    \reg{V}, \reg{W} \gets \ket{\psi}\\
    (k,k^+) \gets \KeyGen'(1^\secparam,1^n) \\
    (\epr_{L},\epr_R) \gets \mathsf{EPR}^{\ot n}\\
    \reg{\wt{V}} \gets \Enc_{k}(\epr_R) \\
    \reg{\wt{V}},\reg{W}, \reg{V} \gets \adv^{\Ver'_{k^+}}(\reg{\wt{V}},\reg{W}, \reg{V})\\
    P \gets \TPSend(\reg{V},\epr_{L})\\
    \reg{\wt{V}},\reg{Y} \gets \fDec_{k_P}(\reg{\wt{V}},\reg{Y})
\end{array}
\right\} \underset{2^{-\Omega(\secparam)}}{\approx} \left\{(k_P,\reg{\wt{V}},\reg{W}) \,\middle|\, 
\begin{array}{l}
    \reg{V}, \reg{W} \gets \ket{\psi}\\
    (k,k^+) \gets \KeyGen'(1^\secparam,1^n) \\
    (\epr_{L},\epr_R) \gets \mathsf{EPR}^{\ot n}\\
    \reg{\wt{V}} \gets \Enc_{k}(\epr_R) \\
    \reg{\wt{V}},\reg{W}, \reg{V} \gets \adv^{\Ver'_{k^+}}(\reg{\wt{V}},\reg{W}, \reg{V})\\
    \reg{\wt{V}},\reg{Y},\reg{V} \gets \fVer_{k}(\reg{\wt{V}},\reg{Y},\reg{V})\\
    P \gets \TPSend(\reg{V},\epr_{L})
\end{array}
\right\} 
\]
}

We will prove the lemma even when $k^+$ is known to the adversary. Hence, it suffices to consider adversaries that can depend on $k^+$, and prove indistinguishability when $k$ is generated from the output distribution of $\KeyGen'(1^\secparam,1^n)$ conditioned on $k^+$, denoted as $\cK$. More concretely, the distribution $\cK$ can be sampled as follows:
\begin{itemize}
    \item Parse $k^+ = (R,\wh{R},x^+,z^+)$.
    \item Sample a uniformly random subspace $S$ conditioned on $\wh{R}^\perp \subset S \subset R$.
    \item Sample a uniformly random $\Delta \gets R \setminus S$.
    \item Sample uniformly random $x^R_i \gets R$ and $z^{\wh{R}}_i \gets \wh{R}$ for each $i \in [n]$.
    \item Output $k = (S,\Delta, x, z)$ where $x = x^+ + x^R$ and $z = z^+ + z^R$.
\end{itemize}

Without loss of generality, $\adv$ performs a unitary operator $\onreg{A}{\wt{V},W,V}$, which we decompose into $\onreg{A}{\wt{V},W,V} = \sum_{Q \in \PauliGroup_{n}} \onreg{A_Q}{\wt{V},W} \ot \onreg{Q}{V}$.
We define the following projectors.
\begin{itemize}
    \item $\Pi_{k,b} := \sum_{\wt{v}: \Ver_{k,\theta,G}(\wt{v}) = b}$ for $b \in \set{\top,\bot}$.
    \item $\Pi_{k,y} := \sum_{\wt{v}: f(\Dec_{k,\theta,G}(\wt{v})) = y}$ for $j \in \zo$. 
    \item $\hat{\Pi}_y := \measureto{=y}{f,\theta,G} = G^\dag \Hg^\theta \left(\sum_{x: f(x)=y} \ketbra{x} \right) \Hg^\theta G$ 
\end{itemize}
The state on the left hand side of the lemma is
{\scriptsize
\begin{align*}
    & \E_{k \gets \cK} \left[ \sum_{P \in \PauliGroup_n} k_P \ot \mixed{ \left(\onreg{\Xg^\bot}{Y}\onreg{\Pi_{k_P,\bot}}{\wt{V}} + \sum_{y\in\zo} \onreg{\Xg^y}{Y} \onreg{\Pi_{k_P,y}}{\wt{V}}\right) \onreg{(\bra{P} \TP)}{V,M_0} \onreg{A}{\wt{V},W, V} \onreg{\Enc_k}{M_1 \to \wt{V}} \onreg{\ket{\epr}}{M_0,M_1} \onreg{\ket{\psi}}{V,W} } \right]\\
    =& \E_{k \gets \cK} \left[ \sum_{P \in \PauliGroup_n} k_P \ot \mixed{ \left(\onreg{\Xg^\bot}{Y}\onreg{\Pi_{k_P,\bot}}{\wt{V}} + \sum_{y\in\zo} \onreg{\Xg^y}{Y} \onreg{\Pi_{k_P,y}}{\wt{V}}\right) \sum_{Q \in \PauliGroup_n} \onreg{A_Q}{\wt{V},W} \onreg{\Enc_k}{M_1 \to \wt{V}} \onreg{(\bra{P} \TP)}{V,M_0} \onreg{\ket{\epr}}{M_0,M_1} \onreg{Q}{V} \onreg{\ket{\psi}}{V,W} } \right]\\
    =& \E_{\begin{smallmatrix} k \gets \cK \\ P \gets \PauliGroup_n \end{smallmatrix}} \left[ k_P \ot \mixed{ \sum_{Q \in \PauliGroup_n} \left(\onreg{\Xg^\bot}{Y}\onreg{\Pi_{k_P,\bot}}{\wt{V}} + \sum_{y\in\zo} \onreg{\Xg^y}{Y} \onreg{\Pi_{k_P,y}}{\wt{V}}\right) \onreg{A_Q}{\wt{V},W} \onreg{\Enc_k}{V \to \wt{V}} \onreg{P}{V} \onreg{Q}{V} \onreg{\ket{\psi}}{V,W} } \right]\\
    =& \E_{\begin{smallmatrix} k \gets \cK \\ P \gets \PauliGroup_n \end{smallmatrix}} \left[ k_P \ot \mixed{ \sum_{Q \in \PauliGroup_n} \left(\onreg{\Xg^\bot}{Y}\onreg{\Pi_{k,\bot}}{\wt{V}} + \sum_{y\in\zo} \onreg{\Xg^y}{Y} \onreg{\Pi_{k_P,y}}{\wt{V}}\right) \onreg{A_Q}{\wt{V},W} \onreg{\Enc_{k_P}}{V \to \wt{V}} \onreg{Q}{V} \onreg{\ket{\psi}}{V,W} } \right]
\end{align*}
}

\noindent where the first equality follows from the commutativity between several operators, the second equality follows from quantum teleportation, and the last equality follows from \cref{lemma:switch-key}.
The state on the right-hand side of the lemma is

{\scriptsize
\begin{align*}
    & \E_{k \gets \cK} \left[ \sum_{P \in \PauliGroup_n} k_P \ot \mixed{ \onreg{(\bra{P} \TP)}{V,M_0} \left( \onreg{\Xg^\bot}{Y}\onreg{\Pi_{k,\bot}}{\wt{V}} + \sum_{y\in\zo} \onreg{\Xg^y}{Y} \onreg{\hat{\Pi}_{y}}{V} \onreg{\Pi_{k,\top}}{\wt{V}}\right) \onreg{A}{\wt{V},W,V} \onreg{\Enc_k}{M_1 \to \wt{V}} \onreg{\ket{\epr}}{M_0,M_1} \onreg{\ket{\psi}}{V,W} } \right]\\
    =& \E_{k \gets \cK} \left[ \sum_{P \in \PauliGroup_n} k_P \ot \mixed{\begin{array}{l} 
        \sum_{Q \in \PauliGroup_n} \onreg{\Xg^\bot}{Y} \onreg{\Pi_{k,\bot}}{\wt{V}} \onreg{A_Q}{\wt{V},W} \onreg{\Enc_k}{M_1 \to \wt{V}} \onreg{(\bra{P} \TP)}{V,M_0} \onreg{\ket{\epr}}{M_0,M_1} \onreg{Q}{V} \onreg{\ket{\psi}}{V,W} \\
        + \sum_{Q \in \PauliGroup_n} \sum_{y\in\zo} \onreg{\Xg^y}{Y} \onreg{\Pi_{k,\top}}{\wt{V}} \onreg{A_Q}{\wt{V},W} \onreg{\Enc_k}{M_1 \to \wt{V}} \onreg{(\bra{P} \TP)}{V,M_0} \onreg{\ket{\epr}}{M_0,M_1} \onreg{\hat{\Pi}_{y}}{V} \onreg{Q}{V} \onreg{\ket{\psi}}{V,W} 
    \end{array}} \right]\\
    =& \E_{\begin{smallmatrix} k \gets \cK \\ P \gets \PauliGroup_n \end{smallmatrix}} \left[ k_P \ot \mixed{\begin{array}{l} 
        \sum_{Q \in \PauliGroup_n} \onreg{\Xg^\bot}{Y} \onreg{\Pi_{k,\bot}}{\wt{V}} \onreg{A_Q}{\wt{V},W} \onreg{\Enc_{k_P}}{V \to \wt{V}} \onreg{Q}{V} \onreg{\ket{\psi}}{V,W} \\
        + \sum_{Q \in \PauliGroup_n} \sum_{y\in\zo} \onreg{\Xg^y}{Y} \onreg{\Pi_{k,\top}}{\wt{V}} \onreg{A_Q}{\wt{V},W} \onreg{\Enc_{k_P}}{V \to \wt{V}} \onreg{\hat{\Pi}_{y}}{V} \onreg{Q}{V} \onreg{\ket{\psi}}{V,W} 
    \end{array}} \right]
\end{align*}
}where the first equality follows from the commutativity between several operators, and the second equality follows from quantum teleportation and \cref{lemma:switch-key}.
Since we always have $\Tr \big| \mixed{\ket{\phi_0}} - \mixed{\ket{\phi_1}} \big| \le 2 \norm{\ket{\phi_0} - \ket{\phi_1}}$, the two states we considered above have trace distance at most

{\scriptsize
\begin{align*}
    & \;2 \E_{\begin{smallmatrix} k \gets \cK \\ P \gets \PauliGroup_n \end{smallmatrix}} \norm{
        \sum_{Q \in \PauliGroup_n} \sum_{y\in\zo} \onreg{\Xg^y}{Y} \left(\onreg{\Pi_{k_P, y}}{\wt{V}} \onreg{A_Q}{\wt{V},W} \onreg{\Enc_{k_P}}{V\to\wt{V}} - \onreg{\Pi_{k, \top}}{\wt{V}} \onreg{A_Q}{\wt{V},W} \onreg{\Enc_{k_P}}{V\to\wt{V}} \onreg{\hat{\Pi}_y}{V}\right) \onreg{Q}{V} \onreg{\ket{\psi}}{V,W}
    }\\
    =& \;2 \E_{\begin{smallmatrix} k \gets \cK \\ P \gets \PauliGroup_n \end{smallmatrix}} \norm{
        \sum_{Q \in \PauliGroup_n} \sum_{y\in\zo} \onreg{\Xg^y}{Y} \left(\begin{array}{l}
            \onreg{\Pi_{k_P, y}}{\wt{V}} \onreg{A_Q}{\wt{V},W} \onreg{\Enc_{k_P}}{V\to\wt{V}} \onreg{\hat{\Pi}_y}{V} + \onreg{\Pi_{k_P, y}}{\wt{V}} \onreg{A_Q}{\wt{V},W} \onreg{\Enc_{k_P}}{V\to\wt{V}} \onreg{\hat{\Pi}_{y \oplus 1}}{V} \\
            - \onreg{\Pi_{k_P, y}}{\wt{V}} \onreg{A_Q}{\wt{V},W} \onreg{\Enc_{k_P}}{V\to\wt{V}} \onreg{\hat{\Pi}_y}{V} - \onreg{\Pi_{k_P, y \oplus 1}}{\wt{V}} \onreg{A_Q}{\wt{V},W} \onreg{\Enc_{k_P}}{V\to\wt{V}} \onreg{\hat{\Pi}_y}{V}
        \end{array}\right)\onreg{Q}{V} \onreg{\ket{\psi}}{V,W}
    }\\
    \le& \;4 \sum_{y \in \zo} \sum_{Q \in \PauliGroup_n} \E_{\begin{smallmatrix} k \gets \cK \\ P \gets \PauliGroup_n \end{smallmatrix}} \norm{
        \onreg{\Pi_{k_P, y \oplus 1}}{\wt{V}} \onreg{A_Q}{\wt{V},W} \onreg{\Enc_{k_P}}{V\to\wt{V}} \onreg{\hat{\Pi}_y}{V} \onreg{Q}{V} \onreg{\ket{\psi}}{V,W}
    }\\
    \le& \;4 \sum_{y \in \zo} \sum_{Q \in \PauliGroup_n} \left(\E_{k \gets \cK} \norm{
        \onreg{\Pi_{k, y \oplus 1}}{\wt{V}} \onreg{A_Q}{\wt{V},W} \onreg{\Enc_{k}}{V\to\wt{V}} \onreg{\hat{\Pi}_y}{V} \onreg{Q}{V} \onreg{\ket{\psi}}{V,W}
    }^2\right)^{\frac{1}{2}}\\
    \le& \;4 \sum_{y \in \zo} \sum_{Q \in \PauliGroup_n} \left(\E_{P' \gets \PauliGroup_n} \E_{k \gets \cK} \norm{
        \onreg{\Pi_{k, y \oplus 1}}{\wt{V}} \left(\onreg{P'^{\dag}}{V'} \onreg{A}{\wt{V},W,V'} \onreg{P'}{V'}\right) \onreg{\Enc_{k}}{V\to\wt{V}} \onreg{\hat{\Pi}_y}{V} \left(\onreg{Q}{V} \onreg{\ket{\psi}}{V,W} \onreg{\ket{0^n}}{V'} \right)
    }^2\right)^{\frac{1}{2}}
\end{align*}
}where the second line uses $I = \hat{\Pi}_y + \hat{\Pi}_{y \oplus 1}$ and $\Pi_{k,\top} = \Pi_{k_P,\top} = \Pi_{k_P,y} + \Pi_{k_P,y \oplus 1}$, the third line follows from cancellation and the triangle inequality, the fourth line makes the observation that the distribution of $k_P$ is $\cK$ and applies the Cauchy-Schwartz inequality, and the last line follows from a standard Pauli twirling on $A = \sum_{Q \in \PauliGroup_n} A_Q \ot Q$. At this point, recall that by definition $\Pi_{k,y \oplus 1} = \measureto{=y \oplus 1}{(f \circ \Dec_{k,\theta,G}), \wt{\theta}, \wt{G}}$ whereas $\hat{\Pi}_y = \measureto{=y}{f,\theta,G}$. To conclude the proof, we incorporate the following lemma from \cite{bartusek2024quantum} stating that the coset authentication code satisfies soundness even given $k^+$, which implies that the trace distance is upper bounded by $2^{O(n)} \cdot 2^{-\Omega(\secparam)} \le 2^{-\Omega(\secparam)}$ for a regime where $\secparam = \Omega(n)$.
\end{proof}

\begin{lemma}[Soundness given $k^+$, \cite{bartusek2024quantum}]
Let $\cK$ be the conditional distribution of $k$ from $\KeyGen'(1^\secparam,1^n)$ given $k^+$. For every $k^+, f, \theta, G$, and quantum adversary $\adv$, 
there exists an $\epsilon = \epsilon(\secparam) \in [0,1]$ such that for any state $\ket{\phi}$,

{\footnotesize
\begin{align*}
    \left\{y \gets \meas{(f \circ \Dec_{k,\theta,G}), \wt{\theta}, \wt{G}} (\reg{\wt{V}}) \middle| \begin{array}{l} 
        \reg{V,W} \gets \ket{\phi}\\
        k \gets \cK \\
        \reg{\wt{V}} \gets \Enc_k({\reg{V}})\\
        \reg{\wt{V}},\reg{W} \gets \adv(\reg{\wt{V}},\reg{W})
    \end{array}\right\} \underset{2^{-\Omega(\secparam)}}{\approx} (1-\epsilon)\left\{y \gets \meas{f,\theta,G} (\phi) \right\} + \epsilon \left\{\bot\right\}.
\end{align*}
}In particular, if $\Pr\left[y_1 \gets \meas{f,\theta,G} (\phi) \right] = 0$ for some $y_1$, then for every adversary $\adv$,
$$
\Pr\left[y_1 \gets \meas{(f \circ \Dec_{k,\theta,G}), \wt{\theta}, \wt{G}} (\reg{\wt{V}}) \middle| \begin{array}{l} 
        \reg{V,W} \gets \ket{\phi}\\
        k \gets \cK \\
        \reg{\wt{V}} \gets \Enc_k({\reg{V}})\\
        \reg{\wt{V}},\reg{W} \gets \adv(\reg{\wt{V}},\reg{W})
    \end{array}\right] \le 2^{-\Omega(\secparam)}.
$$
\end{lemma}
We are ready to prove that the coset authentication code satisfies functional security.
\begin{proof}[Proof of \cref{thm:FuncAuth}]
    Consider the function $f_{j,i}(x)$ that outputs the $i$-th bit of $f_j(x)$. We reserve the notations $\fDec, \fVer$ for the unitary operators defined with respect to the tuple $(f_1,\dots,f_t,\theta_1,\dots,\theta_t,G_1,\dots,G_t)$ and use $\fDec^{(j,i)}, \fVer^{(j,i)}$ to denote the unitary operators defined with respect to $(f_{j,i},\theta_j,G_j)$. By our convention in \cref{Sec:preliminary}, the classical oracles $\{\fDec^{(j,i)}\}_{j,i}, \{\fVer^{(j,i)}\}_{j,i}$ can be used to construct their controlled versions. It is clear that $\fDec, \fVer$ can be efficiently simulated using the same oracle algorithm from the controlled versions of $\{\fDec^{(j,i)}\}_{j,i}, \{\fVer^{(j,i)}\}_{j,i}$ respectively.
    Hence, it suffices to show the following indistinguishability for every adversary $\adv$ making $\poly(t,n',\secparam) = \poly(\secparam)$ queries.
    {\footnotesize
    \[
\left\{(\reg{\wt{V}},\reg{W}) \,\middle|\, 
\begin{array}{l}
    \reg{V}, \reg{W} \gets \ket{\psi}\\
    k \gets \KeyGen(1^\secparam,1^n) \\
    \reg{\wt{V}} \gets \Enc_k(\reg{V}) \\
    \reg{\wt{V}},\reg{W} \gets \adv^{\{\fDec_{k,\cdot,\cdot}^{(j,i)}(\cdot,\reg{V})\}_{j,i}}(\reg{\wt{V}},\reg{W})
\end{array}
\right\} \underset{2^{-\Omega(\secparam)}}{\approx} \left\{(\reg{\wt{V}},\reg{W}) \,\middle|\, 
\begin{array}{l}
    \reg{V}, \reg{W} \gets \ket{\psi}\\
    k \gets \KeyGen(1^\secparam,1^n) \\
    \reg{\wt{V}} \gets \Enc_{k}(0^n) \\ \reg{\wt{V}},\reg{W} \gets \adv^{\{\fVer^{(j,i)}_{k,\cdot,\cdot}(\cdot,\cdot,\reg{V})\}_{j,i}}(\reg{\wt{V}},\reg{W})
\end{array}
\right\} 
\]}

We can express the adversary as $(U_q\; \fDec^{(q)} \cdots U_1\; \fDec^{(1)} U_0)$ on the left hand side and as $(U_q\; \fVer^{(q)} \cdots U_1\; \fVer^{(1)} U_0)$ on the right-hand side for some unitaries $U_0,\dots,U_q$, where $q = \poly(\secparam)$ is the number of queries and each $\fDec^{(a)}, \fVer^{(a)}$ refers to some $\fDec^{(j,i)}, \fVer^{(j,i)}$. Let us start with the state on the left where we also keep track of the key $k$ in the output.
$$\left\{(k,\reg{\wt{V}},\reg{W}) \,\middle|\, 
\begin{array}{l}
    \reg{V}, \reg{W} \gets \ket{\psi}\\
    k \gets \KeyGen(1^\secparam,1^n) \\
    \reg{\wt{V}} \gets \Enc_k(\reg{V}) \\
    \reg{\wt{V}},\reg{W} \gets \left(U_q\; \fDec_{k}^{(q)}(\cdot,\reg{V}) \cdots U_1\; \fDec_{k}^{(1)}(\cdot,\reg{V})\; U_0\right) (\reg{\wt{V}},\reg{W})
\end{array}
\right\}$$
By quantum teleportation (\cref{lemma:teleport}) and Pauli key update (\cref{lemma:switch-key}), the state equals to
$$\left\{(k_P,\reg{\wt{V}},\reg{W}) \,\middle|\, 
\begin{array}{l}
    \reg{V}, \reg{W} \gets \ket{\psi}\\
    k \gets \KeyGen(1^\secparam,1^n) \\
    (\epr_{L},\epr_R) \gets \mathsf{EPR}^{\ot n}\\
    \reg{\wt{V}} \gets \Enc_k(\epr_R) \\
    P \gets \TPSend(\reg{V}, \epr_{L})\\
    \reg{\wt{V}},\reg{W} \gets \left(U_q\; \fDec_{k_P}^{(q)}(\cdot,\reg{V}) \cdots U_1\; \fDec_{k_P}^{(1)}(\cdot,\reg{V})\; U_0\right) (\reg{\wt{V}},\reg{W})
\end{array}
\right\}$$
By \cref{lemma:Dec-to-Ver}, the state is within $2^{-\Omega(\secparam)}$ trace distance to
$$\left\{(k_P,\reg{\wt{V}},\reg{W}) \,\middle|\, 
\begin{array}{l}
    \reg{V}, \reg{W} \gets \ket{\psi}\\
    (k,k^+) \gets \KeyGen'(1^\secparam,1^n) \\
    (\epr_{L},\epr_R) \gets \mathsf{EPR}^{\ot n}\\
    \reg{\wt{V}} \gets \Enc_k(\epr_R) \\
    \reg{\wt{V}},\reg{W} \gets \left(\fVer_{k}^{(1)}(\cdot, \cdot,\reg{V})\; U_0\right)(\reg{\wt{V}},\reg{W})\\
    P \gets \TPSend(\reg{V}, \epr_{L})\\
    \reg{\wt{V}},\reg{W} \gets \left(U_q\; \fDec_{k_P}^{(q)}(\cdot,\reg{V}) \cdots U_2\; \fDec_{k_P}^{(2)}(\cdot,\reg{V})\; U_1\right) (\reg{\wt{V}},\reg{W})
\end{array}
\right\}$$
By repeatedly applying \cref{lemma:Dec-to-Ver}, an induction proof would show that the starting state is within $O(q \cdot 2^{-\Omega(\secparam)}) = 2^{-\Omega(\secparam)}$ trace distance to the following state.
$$\left\{(k_P,\reg{\wt{V}},\reg{W}) \,\middle|\, 
\begin{array}{l}
    \reg{V}, \reg{W} \gets \ket{\psi}\\
    k \gets \KeyGen(1^\secparam,1^n) \\
    (\epr_{L},\epr_R) \gets \mathsf{EPR}^{\ot n}\\
    \reg{\wt{V}} \gets \Enc_k(\epr_R) \\
    \reg{\wt{V}},\reg{W} \gets \left(U_q\; \fVer_{k}^{(q)}(\cdot, \cdot,\reg{V}) \cdots U_1\;\fVer_{k}^{(1)}(\cdot, \cdot,\reg{V})\; U_0\right)(\reg{\wt{V}},\reg{W})\\
    P \gets \TPSend(\reg{V}, \epr_{L})
\end{array}
\right\}$$
By tracing out the register that holds the key, we obtain
{\footnotesize
\[
\left\{(\reg{\wt{V}},\reg{W}) \,\middle|\, 
\begin{array}{l}
    \reg{V}, \reg{W} \gets \ket{\psi}\\
    k \gets \KeyGen(1^\secparam,1^n) \\
    \reg{\wt{V}} \gets \Enc_k(\reg{V}) \\
    \reg{\wt{V}},\reg{W} \gets \adv^{\{\fDec_{k,\cdot,\cdot}^{(j,i)}(\cdot,\reg{V})\}_{j,i}}(\reg{\wt{V}},\reg{W})
\end{array}
\right\} \underset{2^{-\Omega(\secparam)}}{\approx} \left\{(\reg{\wt{V}},\reg{W}) \,\middle|\, 
\begin{array}{l}
    \reg{V}, \reg{W} \gets \ket{\psi}\\
    k \gets \KeyGen(1^\secparam,1^n) \\
    (\epr_{L},\epr_R) \gets \mathsf{EPR}^{\ot n}\\
    \reg{\wt{V}} \gets \Enc_k(\epr_R) \\ \reg{\wt{V}},\reg{W} \gets \adv^{\{\fVer^{(j,i)}_{k,\cdot,\cdot}(\cdot,\cdot,\reg{V})\}_{j,i}}(\reg{\wt{V}},\reg{W})\\
    P\gets \TPSend(\reg{V},\epr_{L})
\end{array}
\right\}
\]
}On the right-hand side, note that the information of $(P, \reg{V}, \epr_{L})$ is traced out in the end, so removing the teleportation does not change the final state. Moreover, since $\epr_{L}$ is never used after initialization, we can apply the teleportation $P \gets \TPSend(0^n, \epr_{L})$ right after the initialization without changing the final state. By teleportation and \cref{lemma:switch-key}, we then have $\reg{\wt{V}} \gets \Enc_{k}(\epr_R) = \Enc_k \circ P (0^n) = \Enc_{k_P}(0^n)$ and that replacing $\fVer_{k}^{(j,i)}$ with $\fVer_{k_P}^{(j,i)}$ does not change the final state. Now the key being used has been changed to $k_P$, and since $k_P$ has the same distribution as $k$, the state on the right equals to
\[
\left\{(\reg{\wt{V}},\reg{W}) \,\middle|\, 
\begin{array}{l}
    \reg{V}, \reg{W} \gets \ket{\psi}\\
    k \gets \KeyGen(1^\secparam,1^n) \\
    \reg{\wt{V}} \gets \Enc_{k}(0^n) \\ \reg{\wt{V}},\reg{W} \gets \adv^{\{\fVer^{(j,i)}_{k,\cdot,\cdot}(\cdot,\cdot,\reg{V})\}_{j,i}}(\reg{\wt{V}},\reg{W})
\end{array}
\right\}
\]
\end{proof}

\section{Projective LM Quantum Program}
Finding a good representation of quantum circuits is crucial to program obfuscation. 
Recently,  \cite{bartusek2024quantum} proposed LM quantum programs and showed how to obfuscate these programs as long as they have classical inputs and pseudo-deterministic classical outputs. In this section, we transform any quantum circuit into a ``projective'' LM (PLM) quantum program whose structure will facilitate the obfuscation of general quantum circuits.

First, we refine the definition of LM quantum program of \cite{bartusek2024quantum}. A unitary gate $G$ is said to be linear if it is composed of $\CNOT$ gates. For a classical function $f: \zo^n \to \zo^*$, a string $\theta \in \zo^n$, and a linear gate $G$ on $n$-qubits, we define the measurement
$$\meas{f,\theta,G} := \set{G^\dag \Hg^\theta \left(\sum_{x: f(x)=y}  \ketbra{x}\right) \Hg^\theta G}_y$$
Moreover, for $S \subseteq \zo^*$, we define the projector
$$
\measureto{\in S}{f,\theta,G} := G^\dag \Hg^\theta \left(\sum_{x: f(x) \in S}  \ketbra{x}\right) \Hg^\theta G
$$
We sometimes write $\measureto{=y}{f,\theta,G}$ as a shorthand for $\measureto{\in \set{y}}{f,\theta,G}$.

\begin{dfn}[LM quantum program] \label{dfn:LM}
An LM quantum program $P$ is described by a quantum state $\psi$, a sequence of instructions $((f_j, \theta_j, G_j)_{j \in [\nstep]}, g)$, and parameters $(n_q,n_c,n')$. Here, $f_j = \set{f_j^{\inp,r_1,\cdots,r_{j-1}}}_{\inp \in \zo^{n_c},\; r_1,\cdots,r_{j-1} \in \zo}$ is a set of classical functions, $\theta_j$ is a bit string, $G_j$ is a linear gate, and $g$ is a classical function. The program $P(\inp,\rho)$ is executed as follows:
    \begin{enumerate}
    \setlength{\itemsep}{0pt}
        \item Input a classical string $\inp \in \zo^{n_c}$ and a quantum state $\rho \in \Den{n_q}$.
        \item Initialize a register $\reg{V}$ with the state $(\rho,\psi)$.
        \item Iteratively apply $r_j \from \onreg{\meas{f_{j}^{\inp,r_1,\cdots,r_{j-1}}, \theta_j, G_j}}{V}$ for $j=1,\cdots,t$.
        \item Output the classical outcome $g(\inp,r_1,\dots, r_\nstep)  \in \zo^{n'}$.
    \end{enumerate}
\end{dfn}

Now, we introduce our notion of PLM quantum programs, which will be a useful representation of quantum circuits.
\begin{dfn}[PLM quantum program] \label{dfn:projective-LM}
An LM quantum program with instructions $((f_j,\theta_j,G_j)_{j\in [\nstep]}, g)$ is said to be projective if for every $\inp$ there exists an orthonormal basis $\set{\Phi_{\inp,(r_1,\dots,r_\nstep)}}$ on $\reg{V}$ such that the product 
$$\measureto{=r_\nstep}{f_\nstep^{\inp,r_1,\cdots,r_{\nstep-1}}, \theta_\nstep, G_\nstep} \;\cdots\; \measureto{=r_2}{f_2^{\inp,r_1}, \theta_2, G_2} \; \measureto{=r_1}{f_1^{\inp}, \theta_1, G_1}
$$
is the projector $\ketbra{\Phi_{{\inp},(r_1,\dots,r_\nstep)}}$ for every $r_1,\dots,r_\nstep$.
\end{dfn}

In other words, applying adaptive measurements on these basis elements always yield deterministic results. These bases will become useful in the security proof of our obfuscation scheme in \cref{sec:obfuscation}. The following theorem states that every quantum circuit with classical outputs can be efficiently compiled into a PLM quantum program.

\begin{thm}[Compilation]
\label{thm:compile}
    There is a polynomial time algorithm that converts any quantum circuit $Q$ with classical output into a PLM quantum program $P = (\psi_\PLM, \set{f_j,\theta_j,G_j}_{j \in [t]})$ with the following properties.
    \begin{itemize}
        \item $|\psi_{\PLM}|, t = O(m)$ where $m$ is the size plus the width of $Q$.
        \item For every classical input $\inp$, quantum state $\rho$, and possibly entangled state $\tau$, running the circuit $Q$ and running the PLM quantum program $P$ yield identical distributions
        $$\bigg\{(y, \tau) \;\bigg|\;  y \from Q(\inp, \rho)\bigg\} \equiv \bigg\{(y, \tau) \;\bigg|\; y \from P(\inp, \rho)\bigg\}$$
        \item By deferring measurements, let $U_{Q,\inp}$ be a unitary operator where $Q(\inp,\rho)$ is equivalent to first applying $U_{Q,\inp}$ to $\rho$ then measuring $n'$ qubits in the standard basis as the output. Let $\set{\ket{\Phi_{\inp,r}}}_{r \in \zo^{t}}$ be the orthonormal basis associated with $P$ and $\inp$ as in \cref{dfn:projective-LM}. Then for every $\inp \in \zo^{n_c}$ and $y \in \zo^{n'}$, 
        {\footnotesize
        $$\sum_y X^y \ot \sum_{r: g(\inp,r) = y}\ketbra{\Phi_{\inp,r}} (I^{\ot n_q} \ot \ket{\psi_{\PLM}}) = \sum_y X^y \ot \left(U_{Q,\inp}^\dag \left(\ketbra{y} \ot I^{\ot (n_q - n')}\right) U_{Q,\inp}\right) \ot \ket{\psi_{\PLM}}$$
        }
   \end{itemize}
\end{thm}
We start by describing how to implement each gate in the universal gate set $\set{\CNOT, \Hg, \Tg}$ using magic states, Clifford gates, adaptive measurements, and Pauli corrections.
For each implementation, we also provide an orthonormal basis on which the adaptive measurements yield deterministic outcomes. 
These orthonormal bases will ultimately be tensored into an orthonormal basis over the entire circuit, which facilitates the construction of projective LM quantum programs. The most distinctive aspect of our construction lies in our implementation of $\Tg$ gates.

\paragraph{Implementing the $\Hg$ gate.} We prepare the magic state
$\ket{\phi_\Hg} = \frac{1}{2}(\ket{00}+\ket{01}+\ket{10}-\ket{11})$
and perform the following implementation of $\Hg$ as in \cite{broadbent2013quantum}.

\begin{Mycircuit}\label{circ:H}
\end{Mycircuit}
\begin{center}
    \begin{quantikz}
     \lstick{$\ket{\psi}$} & \ctrl{1} & \gate{\Hg}  & \meter{} & \setwiretype{c}\; c_0\\
     \lstick[2]{$\ket{\phi_\Hg}$}& \targ{} & & & \meter{} & \setwiretype{c}\; c_1 \\
     & \qw & & & & & \gate{\Xg^{c_0} \Zg^{c_1}} & \rstick{$\Hg \ket{\psi}$}
    \end{quantikz}
\end{center}
The correctness of the computation has been shown in both \cite{broadbent2013quantum,bartusek2024quantum}.
We remark that the two measurements would completely collapse the state in the top two wires, and the measurement results would decide the Pauli correction on the last wire. 
Also, it is immediate to see that on the first two wires, every state in the Bell basis will produce deterministic measurement results. More precisely, it is the basis $$\beta^{\Hg} := \set{\ket{\wt{\Phi}^{\Hg}_{c_0,c_1}} \given c_0,c_1 \in \zo} := \set{(I \ot \Xg^{c_1} \Zg^{c_0}) \frac{\ket{00}+\ket{11}}{\sqrt{2}} \given c_0,c_1 \in \zo}$$
where $c_0,c_1$ would be the deterministic measurement results. The correctness of the circuit shows that for some coefficients $\{\alpha^{\Hg}_{c_0,c_1}\}$, we have
$$
    \ket{\psi} \ot \ket{\phi_\Hg} = \sum_{c_0,c_1} \alpha^{\Hg}_{c_0,c_1} \ket{\wt{\Phi}^{\Hg}_{c_0,c_1}} \ot \Xg^{c_0} \Zg^{c_1} \Hg \ket{\psi}
$$

\paragraph{Implementing the $\CNOT$ gate.} We prepare the state 
$\phi_\CNOT = (\epr_L, \epr_R) \gets \mathsf{EPR}^{\ot 2}$.
\begin{Mycircuit}\label{circ:CNOT}
\end{Mycircuit}
\begin{center}
    \begin{quantikz}
     \lstick[2]{$\ket{\psi}$} & \ctrl{1} & \ctrl{2} & & \gate{\Hg} &\meter{} & \setwiretype{c}\; c_0 \\
     & \targ{} & &\ctrl{2} & \gate{\Hg} &\meter{} & \setwiretype{c}\; c_1\\
     \lstick[2]{$\epr_L$}& &\targ{} & & & \meter{} & \setwiretype{c}\; c_2 \\
     & \qw & & \targ{} & &\meter{} & \setwiretype{c}\; c_3\\
     \lstick[2]{$\epr_R$}& & & & & & \gate{\Xg^{c_2} \Zg^{c_0}} & \rstick[2]{$\CNOT \ket{\psi}$} \\
     & \qw & & & & & \gate{\Xg^{c_3} \Zg^{c_1}} & \\
    \end{quantikz}
\end{center}
The correctness of the computation follows from quantum teleportation (\cref{lemma:teleport}). We remark that the four measurements would completely collapse the state in the top four wires, and the measurement results would determine the Pauli corrections on the last two wires. Also, it is immediate to see that on the first four wires, every state in the two-fold tensor of the Bell basis will produce deterministic measurement results. We denote the basis as $$\beta^{\CNOT} := \set{\ket{\wt{\Phi}^{\CNOT}_{c_0,c_1,c_2,c_3}}\given c_0,c_1,c_2,c_3 \in \zo}$$ where $c_0,c_1,c_2,c_3$ would be the deterministic measurement results. The correctness of the circuit shows that for some coefficients $\{\alpha^{\CNOT}_{c_0,c_1,c_2,c_3}\}$, we have
$$\ket{\psi}\ot \ket{\phi_\CNOT} = \sum_{c_0,c_1,c_2,c_3} \alpha^{\CNOT}_{c_0,c_1,c_2,c_3} \ket{\wt{\Phi}^{\CNOT}_{c_0,c_1,c_2,c_3}} \ot (\Xg^{c_2}\Zg^{c_0}\ot\Xg^{c_3}\Zg^{c_1})\CNOT\ket{\psi}$$

\paragraph{Implementing the $\Tg$ gate.} We prepare the magic state $\ket{\phi'_\Tg} = \ket{\phi_\Tg} \ot \ket{\phi_{\Sg^\dag}} \ot \ket{\epr}$, where
\begin{align*}
    \ket{\phi_\Tg} = \frac{1}{\sqrt{2}}(\ket{0} + e^{i\pi/4} \ket{1}),\quad
    \ket{\phi_{\Sg^\dag}} = \frac{1}{\sqrt{2}}(\ket{0} - i \ket{1}),\quad \ket{\epr} = \frac{1}{\sqrt{2}}(\ket{00} + \ket{11}).
\end{align*}
We then define the measurements $\Gamma_0, \Gamma'_0, \Gamma_1, \Gamma'_1$ so that
\begin{align*}
    \Gamma_0 \text{ measures the output bit of the classical function } &(a,b) \mapsto b \\
    \Gamma'_0 \text{ measures the output bit of the classical function } &(a,b) \mapsto a \\
    \Gamma_1, \Gamma'_1 \text{ measure the output bit of the classical function } &(a,b) \mapsto a \oplus b
\end{align*}
In other words, these measurements are 
\begin{align*}
    \Gamma_0 &= \set{\proj{00} + \proj{10},\; \proj{01} + \proj{11}}\\
    \Gamma'_0 &= \set{\proj{00} + \proj{01},\; \proj{10} + \proj{11}}\\
    \Gamma_1 = \Gamma'_1 &= \set{\proj{00} + \proj{11},\; \proj{01} + \proj{10}}
\end{align*}
\begin{lemma}\label{lemma:T-gate-correctness}
    The following circuit implements the $\Tg$ gate.
\end{lemma}
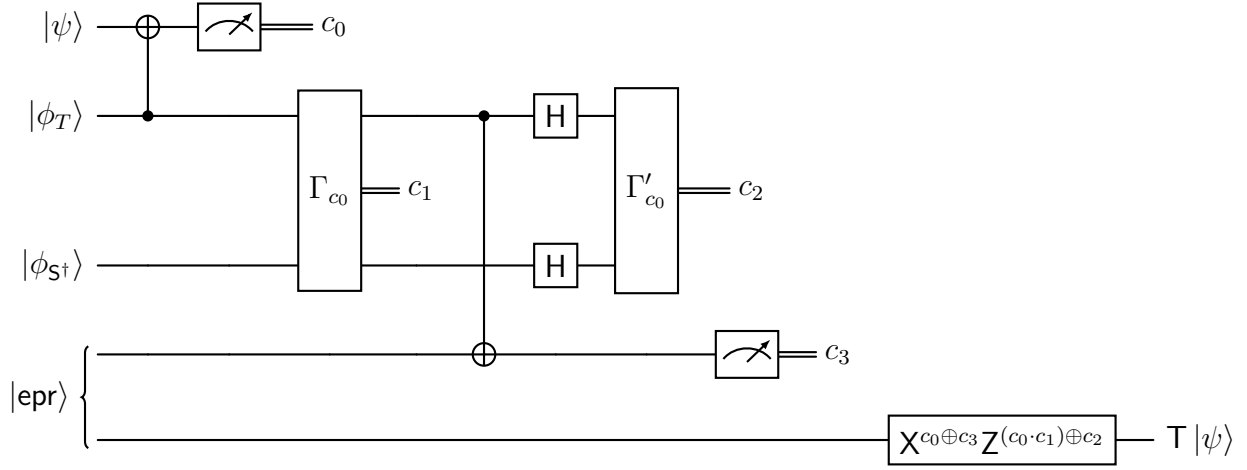
\begin{Mycircuit}\label{circ:T}
\end{Mycircuit}
\begin{center}
    \begin{quantikz}
        \lstick{$\ket{\psi}$} & \targ{} & \meter{} & \setwiretype{c} \;c_0 & \setwiretype{n} \\
        \lstick{$\ket{\phi_T}$} & \ctrl{-1} & & \gate[3]{\Gamma_{c_0}} & & \ctrl{3} & \gate{\Hg} & \gate[3]{\Gamma'_{c_0}} & \setwiretype{n} \\
        \setwiretype{n} & & & & \wireoverride{c} \;c_1 & & & & \wireoverride{c} \;c_2 \\
        \lstick{$\ket{\phi_{\Sg^\dag}}$} & & & & & & \gate{\Hg} & & \setwiretype{n} \\
        \lstick[2]{$\ket{\epr}$}& & & & & \targ{} & & & \meter{} & \setwiretype{c}\; c_3\\
        & & & & &  & & & & & \gate{\Xg^{c_0 \oplus c_3} \Zg^{(c_0 \cdot c_1) \oplus c_2}} & \rstick{$\Tg\ket{\psi}$}
    \end{quantikz}
\end{center}
We remark that the four measurements would completely collapse the state in the top four wires, and the measurement results would decide the Pauli correction on the last wire. We now prove that the computation in the diagram is indeed correct.
\begin{proof}
We write $\ket{\psi} = \alpha \ket{0} + \beta \ket{1}$. After applying the first $\CNOT$, the state on the top two wires becomes
\begin{align*}
    &\frac{1}{\sqrt{2}}\left(\alpha\ket{00} + \beta\ket{10} + \rooti \beta \ket{01} + \rooti \alpha \ket{11}\right)\\
    =&\frac{1}{\sqrt{2}}\ket{0} \ot \left(\alpha\ket{0} + \rooti \beta \ket{1}\right) + \frac{1}{\sqrt{2}}\ket{1} \ot \left( \beta\ket{0} + \rooti \alpha \ket{1}\right)
\end{align*}

When the measurement result of the first wire is $c_0 = 0$, the state on the second wire collapses to $\alpha\ket{0} + \rooti \beta \ket{1} = \Tg \ket{\psi}$. In this case, the next measurement $\Gamma_0$ only measures the third wire, so the state on the second wire stays the same. Then, observe that $\Gamma'_0$ is a standard basis measurement on the second wire, so the operations on the second and fourth wires after $\Gamma_0$ is exactly the teleportation operation. Hence, the fifth wire would hold the state $\Xg^{c_3} \Zg^{c_2} \Tg \ket{\psi}$, and after applying $\Xg^{c_0\oplus c_3} \Zg^{(c_0\cdot c_1) \oplus c_2} = \Xg^{c_3} \Zg^{c_2}$, the state becomes $\Tg \ket{\psi}$ up to a global phase.

When the measurement result of the first wire is $c_0 = 1$, the state on the second wire collapses to $\left( \beta\ket{0} + \rooti \alpha \ket{1}\right)$. The joint state on wires $2$ and $3$ is
\begin{equation}
\begin{aligned}[b]
&\left( \beta\ket{0} + \rooti \alpha \ket{1}\right) \ot \frac{1}{\sqrt{2}}(\ket{0} - i \ket{1})\\
=& \frac{1}{\sqrt{2}} \left( \beta \ket{00} - i \rooti \alpha \ket{11} \right) + \frac{1}{\sqrt{2}} \left( \rooti \alpha \ket{10} - i \beta \ket{01} \right)
\label{eq:T-gate-intermediate-state}
\end{aligned}
\end{equation}
In this case, $\Gamma_1$ measures the parity of wires $2$ and $3$ in the standard basis. When the measurement result is $c_1 = 0$, the state becomes $
    \beta \ket{00} - i \rooti \alpha \ket{11} =  e^{-i\pi/4} \left(\rooti \beta \ket{00} + \alpha \ket{11}\right)$.
After the following $\CNOT$, the state on wires $2,3,4,5$, up to a global phase, becomes
\begin{align*}
    &\frac{1}{\sqrt{2}}\left( \rooti \beta \ket{0000} + \rooti\beta \ket{0011} + \alpha\ket{1110} + \alpha\ket{1101} \right)\\
    =&\frac{1}{2\sqrt{2}}\left(\begin{matrix}
        \;\;\rooti \beta \ket{++00} + \rooti \beta \ket{--00} + \rooti \beta \ket{+-00} + \rooti \beta \ket{-+00}\\
        +\rooti \beta \ket{++11} + \rooti \beta \ket{--11} + \rooti \beta \ket{+-11} + \rooti \beta \ket{-+11}\\
        + \alpha\ket{++10} + \alpha\ket{--10} - \alpha\ket{+-10} - \alpha\ket{-+10} \\
        + \alpha\ket{++01} + \alpha\ket{--01} - \alpha\ket{+-01} - \alpha\ket{-+01} 
    \end{matrix}\right)\\
    =& \;\;\frac{1}{2}
        \left( \frac{\ket{++}+\ket{--}}{\sqrt{2}}\right) \ot \bigg(\ket{0} \ot \left(\rooti \beta \ket{0} + \alpha\ket{1}\right) + \ket{1} \ot \left(\rooti \beta \ket{1} + \alpha\ket{0}\right) \bigg)\\
     & + \frac{1}{2}
        \left( \frac{\ket{+-}+\ket{-+}}{\sqrt{2}}\right) \ot \bigg(\ket{0} \ot \left(\rooti \beta \ket{0} - \alpha\ket{1}\right) + \ket{1} \ot \left(\rooti \beta \ket{1} - \alpha\ket{0}\right) \bigg)
\end{align*}
Next, the Hadamard gates and $\Gamma'_1$ is to measure the parity on wires $2$ and $3$ in the Hadamard basis. One also measures the fourth wire in the standard basis. When these two measurement results are $c_2$ and $c_3$ respectively, from the above expansion we see that the state on the fifth wire becomes 
$ \Xg^{c_3} \Zg^{c_2} (\rooti \beta \ket{0} + \alpha \ket{1}) = \Xg^{c_3} \Zg^{c_2} \Xg \Tg \ket{\psi}$. After applying $\Xg^{c_0 \oplus c_3} \Zg^{(c_0\cdot c_1) \oplus c_2} = \Xg^{c_3 \oplus 1} \Zg^{c_2}$, the state on the fifth wire becomes $\Tg \ket{\psi}$ up to a global phase.

When the measurement results are $c_0 = 1$ and $c_1 = 1$, we see from expression (\ref{eq:T-gate-intermediate-state}) that the state on wires $2$ and $3$ becomes $\rooti \alpha \ket{10} - i \beta \ket{01} = \rooti \left( \alpha \ket{10} - \rooti \beta \ket{01} \right)$. After the following $\CNOT$, the state on wires $2,3,4,5$, up to a global phase, becomes
\begin{align*}
    &\frac{1}{\sqrt{2}}\left( \alpha\ket{1010} + \alpha\ket{1001} -\rooti \beta \ket{0100} - \rooti\beta \ket{0111} \right)\\
    =&\frac{1}{2\sqrt{2}}\left(\begin{matrix}
        \;\; \alpha\ket{++10} - \alpha\ket{--10} + \alpha\ket{+-10} - \alpha\ket{-+10} \\
        + \alpha\ket{++01} - \alpha\ket{--01} + \alpha\ket{+-01} - \alpha\ket{-+01} \\
        -\rooti \beta \ket{++00} + \rooti \beta \ket{--00} + \rooti \beta \ket{+-00} - \rooti \beta \ket{-+00}\\
        -\rooti \beta \ket{++11} + \rooti \beta \ket{--11} + \rooti \beta \ket{+-11} - \rooti \beta \ket{-+11}
    \end{matrix}\right)\\
    =& \;\;\frac{1}{2}
        \left( \frac{\ket{++}-\ket{--}}{\sqrt{2}}\right) \ot \bigg(\ket{0} \ot \left(\alpha\ket{1} - \rooti \beta \ket{0}\right) + \ket{1} \ot \left(\alpha\ket{0} - \rooti \beta \ket{1}\right) \bigg)\\
     & + \frac{1}{2}
        \left( \frac{\ket{+-}-\ket{-+}}{\sqrt{2}}\right) \ot \bigg(\ket{0} \ot \left(\alpha\ket{1} + \rooti \beta \ket{0}\right) + \ket{1} \ot \left(\alpha\ket{0} + \rooti \beta \ket{1}\right) \bigg)
\end{align*}
Next, the Hadamard gates and $\Gamma'_1$ is to measure the parity on wires $2$ and $3$ in the Hadamard basis. One also measures the fourth wire in the standard basis. When these two measurement results are $c_2$ and $c_3$ respectively, we see that the state on the fifth wire becomes
$ \Xg^{c_3 \oplus 1} \Zg^{c_2 \oplus 1} (\alpha\ket{0} + \rooti \beta \ket{1}) = \Xg^{c_3 \oplus 1} \Zg^{c_2 \oplus 1} \Tg \ket{\psi}$. After applying $\Xg^{c_0 \oplus c_3} \Zg^{(c_0\cdot c_1) \oplus c_2} = \Xg^{c_3 \oplus 1} \Zg^{c_2 \oplus 1}$, the state on the fifth wire becomes $\Tg \ket{\psi}$ up to a global phase.
\end{proof}

Our implementation of $\Tg$ gate is designed to admit an orthonormal basis on which the adaptive measurement results are deterministic. We first prove it for a sub-circuit.

\begin{lemma}
\label{lemma:partial-T-gate-basis}
    For every $c_0 \in \zo$, there exists an orthonormal basis on three qubits such that every state in the basis, when applied the following circuit, produces deterministic measurement outcome $(c_1,c_2,c_3)$.
\end{lemma}
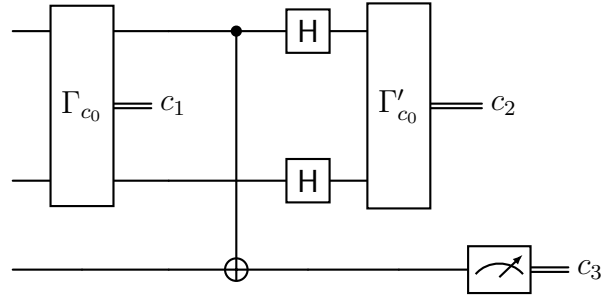
\begin{center}
    \begin{quantikz}
        & \gate[3]{\Gamma_{c_0}} & & \ctrl{3} & \gate{\Hg} & \gate[3]{\Gamma'_{c_0}} & \setwiretype{n} \\
        \setwiretype{n} & & \wireoverride{c} \;c_1 & & & & \wireoverride{c} \;c_2 \\
        & & & & \gate{\Hg} & & \setwiretype{n} \\
        & & & \targ{} & & & \meter{} & \setwiretype{c}\; c_3
    \end{quantikz}
\end{center}
\begin{proof}
    For $c_0 = 0$, the measurement $\Gamma_0$ measures the second wire in the standard basis and $\Gamma'_0$ measures the first wire in the standard basis. Thus, it directly follows that the standard basis on the second wire tensor with the Bell basis on the first and third wires constitute an orthonormal basis with deterministic measurement outcome $(c_1,c_2,c_3)$.

    For $c_0 = 1$, both $\Gamma_1$ and $\Gamma'_1$ are measuring the parity of first and second wires in the standard basis. Consider the following orthonormal basis
    $$\set{
        \frac{\ket{000} \pm \ket{111}}{\sqrt{2}},\;
        \frac{\ket{001} \pm \ket{110}}{\sqrt{2}},\;
        \frac{\ket{101} \pm \ket{010}}{\sqrt{2}},\;
        \frac{\ket{100} \pm \ket{011}}{\sqrt{2}}
    }$$
    Each basis element can be expressed as $\frac{1}{\sqrt{2}}\left(\ket{b_1 b_2 b_3} + (-1)^z \ket{\bar{b}_1 \bar{b}_2 \bar{b}_3}\right)$ for some $b_1,b_2,b_3,z\in\zo$, where $\bar{b} = 1-b$. We now show that this state always produces deterministic measurement results $c_1,c_2,c_3$ in the circuit. The first measurement $\Gamma_1$ measures the parity of the first two wires in the standard basis. Since $b_1 \oplus b_2 = \bar{b}_1 \oplus \bar{b}_2$, the measurement $\Gamma_1$ always produces $c_1 = b_1\oplus b_2$. After applying the $\CNOT$ gate, the state becomes $\frac{1}{\sqrt{2}}\left(\ket{b_1 b_2}\ket{b_1\oplus b_3} + (-1)^z \ket{\bar{b}_1 \bar{b}_2} \ket{\bar{b}_1 \oplus \bar{b}_3}\right) = \frac{1}{\sqrt{2}}\left(\ket{b_1 b_2} + (-1)^z \ket{\bar{b}_1 \bar{b}_2} \right) \ot \ket{b_1\oplus b_3}$. So the measurement on the third wire always produces $c_3 = b_1 \oplus b_3$. Lastly, we apply Hadamard gates on $\frac{1}{\sqrt{2}}\left(\ket{b_1 b_2} + (-1)^z \ket{\bar{b}_1 \bar{b}_2}\right)$. The resulting state will be one of the following four cases.
    \begin{align*}
        &\frac{\ket{++} + \ket{--}}{\sqrt{2}} = \frac{\ket{00} + \ket{11}}{\sqrt{2}},\quad
        &\frac{\ket{+-} + \ket{-+}}{\sqrt{2}} = \frac{\ket{00} - \ket{11}}{\sqrt{2}},\\
        &\frac{\ket{++} - \ket{--}}{\sqrt{2}} = \frac{\ket{10} + \ket{01}}{\sqrt{2}},\quad        
        &\frac{\ket{+-} - \ket{-+}}{\sqrt{2}} = \frac{\ket{10} - \ket{01}}{\sqrt{2}}.
    \end{align*}
    The measurement $\Gamma'_1$ then measures the parity in the standard basis. In the top two cases, it always produces $c_2 = 0$. In the bottom two cases, it always produces $c_2 = 1$.
\end{proof}

\begin{lemma}
\label{lemma:T-gate-basis}
    There exists an orthonormal basis $\beta^{\Tg} := \set{\ket{\wt{\Phi}^{\Tg}_{c_0,c_1,c_2,c_3}} \given c_0,c_1,c_2,c_3 \in \zo}$ on four qubits such that applying \cref{circ:T} with the top four wires being $\ket{\wt{\Phi}^{\Tg}_{c_0,c_1,c_2,c_3}}$ would produce deterministic measurement outcome $(c_0,c_1,c_2,c_3)$.
\end{lemma}
\begin{proof}
    By \cref{lemma:partial-T-gate-basis}, there exist two orthonormal bases $\set{\ket{\Phi^{(0)}_{c_1,c_2,c_3}}}$ and $\set{\ket{\Phi^{(1)}_{c_1,c_2,c_3}}}$ such that applying the sub-circuit of \cref{lemma:partial-T-gate-basis} on $\ket{\Phi^{(c_0)}_{c_1,c_2,c_3}}$ yields deterministic outcome $(c_1,c_2,c_3)$. Define
    $\ket{\wt{\Phi}^{\Tg}_{c_0,c_1,c_2,c_3}} := \onreg{\CNOT}{2,1} \left(\ket{c_0} \ot \ket{\Phi^{(c_0)}_{c_1,c_2,c_3}}\right)$. It is clear that such states form an orthonormal basis on four qubits, and directly applying \cref{circ:T} shows that $\ket{\wt{\Phi}^{\Tg}_{c_0,c_1,c_2,c_3}}$ yields deterministic measurement outcome $(c_0,c_1,c_2,c_3)$.
\end{proof}
It follows from \cref{lemma:T-gate-correctness,lemma:T-gate-basis} that there exist some coefficients $\{\alpha^{\Tg}_{c_0,c_1,c_2,c_3}\}$ with
$$
    \ket{\psi} \ot \ket{\phi'_\Tg} = \sum_{c_0,c_1,c_2,c_3} \alpha^{\Tg}_{c_0,c_1,c_2,c_3} \ket{\wt{\Phi}^{\Tg}_{c_0,c_1,c_2,c_3}} \ot \Xg^{c_0 \oplus c_3} \Zg^{(c_0\cdot c_1) \oplus c_2} \Tg \ket{\psi}
$$

We are now ready to prove the compilation theorem.
\begin{proof}[Proof of \cref{thm:compile}]
Given a quantum circuit $Q$ that also takes a classical input $\inp$, we will introduce ancilla qubits $\ket{\inp}$ and replace the gates of $Q$ with controlled gates whose descriptions do not depend on $\inp$. 
By deferring all measurements and adding extra measurements, we further ensure that the quantum circuit first applies a sequence of gates and then measures all wires in the standard basis.
We collect the sequence of gates as another circuit $\wh{Q}$ and express it using the universal gate set $\set{\CNOT, \Hg, \Tg}$.
    
We now construct our compilation algorithm, which iteratively reads the gates of $\wh{Q}$ and maintains a function $h$, a linear gate $G^*$, and a string $\theta^*$. Here, the function $h$ computes the Pauli corrections $(z_i,x_i)$ of each wire $i$ using the classical input $\inp$ and previous measurement results. At the end of each iteration, the algorithm appends additional states and instructions to the LM quantum program. In addition to the algorithm and throughout the iterations, we will maintain an orthonormal basis $\wt{\gamma}$ on the wires that have encountered measurements, where the basis $\wt{\gamma}$ depends on the classical input $\inp$.

First, the algorithm appends $n_c$ additional wires initialized with $0^{n_c}$. On these wires, we set the value of $h$ to be $(0^{n_c}, \inp)$ to remember that there is a Pauli correction $\Xg^\inp$, and we set the value of $h$ as zero on all other wires. We initialize $G^* = I$ and $\theta^* = \vec{0}$. There is nothing to be included in the orthonormal basis in the beginning. Iteratively, for each gate in the quantum circuit, the algorithm uses the corresponding circuit from \ref{circ:CNOT}, \ref{circ:H}, \ref{circ:T} to instantiate the gate. Note that circuits \ref{circ:CNOT}, \ref{circ:H}, \ref{circ:T} only involve Clifford gates and measurements in the standard or Hadamard basis, so Pauli corrections on the input state can be delayed. The compilation of each gate, including the computation of delayed Pauli corrections through the function $h$, is described as follows.
\begin{itemize}
    \item $\Hg$ on wire $i$: composing \cref{circ:H}.
    \begin{itemize}
        \item Append two new wires $j, k$ to the current set of wires.
        \item Append the state $\ket{\phi_\Hg}$ to the quantum state of the LM quantum program.
        \item Append a $\CNOT$ gate from wire $i$ to wire $j$ to the end of $G^*$, and set $\theta^*_i = 1$.
        \item Append $(d_i: v \mapsto v_i,\; \theta^*, G^*)$ to the instruction of the LM quantum program. Suppose this corresponds to the $e$-th measurement.
        \item Append $(d_j: v \mapsto v_j,\; \theta^*, G^*)$ to the instruction of the LM quantum program. Suppose this corresponds to the $(e+1)$-th measurement.
        \item Update $h$ as follows on top of the original computation result $(z,x)$: Set the values $(z_k, x_k) \gets (z_i \oplus r_e, x_i \oplus r_{e+1})$ and delete the values $(z_i,x_i)$.
        \item Extend $\wt{\gamma}$ by tensoring with the basis $\beta^\Hg$ on wires $i, j$.
        \item Use wire $k$ whenever future operations refer to wire $i$.
    \end{itemize}
    \item $\CNOT$ from wire $i$ to $j$: composing \cref{circ:CNOT}.
    \begin{itemize}
        \item Append four new wires $k,\ell,m,s$ to the current set of wires.
        \item Append the state $\ket{\phi_\CNOT}$ to the quantum state of the LM quantum program.
        \item Append a $\CNOT$ gate from wire $i$ to wire $j$ to the end of $G^*$.
        \item Append a $\CNOT$ gate from wire $i$ to wire $k$ to the end of $G^*$, and set $\theta^*_i = 1$.
        \item Append a $\CNOT$ gate from wire $j$ to wire $\ell$ to the end of $G^*$, and set $\theta^*_j = 1$.
        \item Append $(d_i: v \mapsto v_i,\; \theta^*, G^*)$ to the instruction of the LM quantum program.\\
        Append $(d_j: v \mapsto v_j,\; \theta^*, G^*)$ to the instruction of the LM quantum program.\\
        Append $(d_k: v \mapsto v_k,\; \theta^*, G^*)$ to the instruction of the LM quantum program.\\
        Append $(d_\ell: v \mapsto v_\ell,\; \theta^*, G^*)$ to the instruction of the LM quantum program.\\
        Suppose these correspond to the $e$-th to $(e+3)$-th measurements.
        \item Update $h$ as follows on top of the original computation result $(z,x)$: Set values $(z_m,x_m) \gets (z_i \oplus z_j,x_i)$ and $(z_s,x_s) \gets (z_j,x_i \oplus x_j)$ and then delete $(z_i,x_i), (z_j,x_j)$.
        \item Extend $\wt{\gamma}$ by tensoring with the basis $\beta^\CNOT$ on wires $i,j,k,\ell$.
        \item Use wires $m,s$ whenever future operations refer to wires $i,j$.
    \end{itemize}
    \item $\Tg$ on wire $i$: composing \cref{circ:T}.
    \begin{itemize}
        \item Append four new wires $j,k,\ell,m$ to the current set of wires.
        \item Append the state $\ket{\phi'_\Tg}$ to the quantum state of the LM quantum program.
        \item Append a $\CNOT$ gate from wire $j$ to wire $i$ to the end of $G^*$.
        \item Append $(d_i: v \mapsto v_i,\; \theta^*, G^*)$ to the instruction of the LM quantum program. Suppose this corresponds to the $e$-th measurement.
        \item Compute the output $(z_i,x_i)$ of $h$ on wire $i$.
        \item Define the following functions $d', d''$ that depend on the $e$-th measurement result $r_e$ and the correction bit $x_i$ derived from $\inp, r_1,\dots,r_e$.\\
        $d': v \mapsto \begin{cases}
            v_k & \text{ if } r_e \oplus x_i = 0\\
            v_j \oplus v_k & \text{ if } r_e \oplus x_i = 1
        \end{cases} \qquad d'': v \mapsto \begin{cases}
            v_j & \text{ if } r_e \oplus x_i = 0\\
            v_j \oplus v_k & \text{ if } r_e \oplus x_i = 1
        \end{cases}$
        \item Append $(d', \theta^*, G^*)$ to the instruction of the LM quantum program. Suppose this corresponds to the $(e+1)$-th measurement.
        \item Append a $\CNOT$ gate from wire $j$ to wire $\ell$ to the end of $G^*$, and set $\theta^*_j = \theta^*_k = 1$.
        \item Append $(d'', \theta^*, G^*)$ to the instruction of the LM quantum program. Suppose this corresponds to the $(e+2)$-th measurement.
        \item Append $(d_\ell: v \mapsto v_\ell,\; \theta^*, G^*)$ to the instruction of the LM quantum program. Suppose this corresponds to the $(e+3)$-th measurement.
        \item Update $h$ as follows on top of the original computation result $(z,x)$: Set the values $(z_m, x_m) \gets (z_i \oplus (r_e \cdot r_{e+1}) \oplus r_{e+2},\; x_i \oplus r_e \oplus r_{e+3})$ and delete the values $(z_i,x_i)$.
        \item Extend $\wt{\gamma}$ by tensoring with the basis $\beta^{\Tg}$ on wires $i,j,k,\ell$.
        \item Use wire $m$ whenever future operations refer to wire $i$.
    \end{itemize}
\end{itemize}
Denote the quantum auxiliary state constructed so far as $\ket{\psi_\PLM}$. Since $\wt{\gamma}$ is a tensor of $\beta^{\CNOT},\beta^{\Hg},\beta^{\Tg}$ on locations corresponding to our implementations of $\CNOT,\Hg,\Tg$, we can write $\wt{\gamma} = \{\ket{\wt{\Phi}_{\inp,c}}\}$ for some $\ket{\wt{\Phi}_{\inp,c}}$ that yields deterministic measurement outcome $c$.
By an induction proof, together with the correctness of circuits \ref{circ:H}, \ref{circ:CNOT}, \ref{circ:T}, the basis $\{\ket{\wt{\Phi}_{\inp,c}}\}$, and the correct computation of delayed Pauli corrections, we have the identity
\begin{equation}\label{eq:compilation-middle-structure}
    \ket{\phi} \ot \ket{\psi_\PLM} = \sum_{c} \alpha_{\inp,c} \ket{\wt{\Phi}_{\inp,c}} \ot P_{h(\inp,c)} \wh{Q} \ket{\phi,\inp}
\end{equation}
for some coefficients $\{\alpha_{\inp,c}\}$ that do not depend on $\ket{\phi}$. Recall that $P_{h(\inp,c)}$ is the correction Pauli computed from the classical input $\inp$ and the measurement result $c$.
Note also that $\wh{Q} \ket{\phi,\inp} = U_{Q,\inp} \ket{\phi} \ot \ket{\inp}$ where $U_{Q,\inp}$ is defined in the theorem statement.

The compilation algorithm ends with providing measurement instructions on remaining wires and defining the function $g(\inp,r)$ that computes the classical output.
\begin{itemize}
    \item Standard-basis measurement on wire $i$:
    \begin{itemize}
        \item Append $(d_i: v \mapsto v_i,\; \theta^*, G^*)$ to the instruction of the LM quantum program.
    \end{itemize}
    \item Suppose the original circuit uses the measurement results on wires $w_1,\dots,w_{n'}$ as the classical output. Say they are the $e_1,\dots,e_{n'}$-th measurements in the constructed LM quantum program. Then the algorithm defines the output of $g(\inp,r)$ as $(r_{e_1} \oplus x_{w_1}, \dots, r_{e_i} \oplus x_{w_i})$ where the values $(x_{w_1},\dots,x_{w_i})$ are computed using $h$.
\end{itemize}
Clearly, tensoring the orthonormal basis $\wt{\gamma}$ with the standard basis on the remaining wires yields an orthonormal basis $\set{\ket{\Phi_{\inp,r}}}_r$ on which the adaptive measurements are deterministic. Hence, the compiled LM quantum program is projective. It is clear by construction that $|\psi_\PLM|,t = O(m)$.
Consider any pure state $\sum_j \ket{\psi_j} \ot \ket{\phi_j}$ where $\ket{\psi_j}$ may not be normalized. Since (\ref{eq:compilation-middle-structure}) holds for every $\ket{\phi}$, we have
\begin{align*}
    &\sum_{y,j} \Xg^{y} \ket{\psi_j} \ot \left(U_{Q,\inp}^\dag (\ketbra{y} \ot I) U_{Q,\inp}\right) \ket{\phi_j} \ot \ket{\psi_\PLM} \\
    =& \sum_{y,j} \Xg^{y} \ket{\psi_j} \ot \sum_{c} \alpha_{\inp,c} \ket{\wt{\Phi}_{\inp,c}} \ot P_{h(\inp,c)} \left(U_{Q,\inp} \left(U_{Q,\inp}^\dag (\ketbra{y} \ot I) U_{Q,\inp}\right) \ket{\phi_j} \ot \ket{\inp}\right)\\
    =& \sum_{y,j} \Xg^{y} \ket{\psi_j} \ot \sum_{c} \alpha_{\inp,c} \ket{\wt{\Phi}_{\inp,c}} \ot P_{h(\inp,c)} \left(\left(\ketbra{y} \ot I \right) U_{Q,\inp} \ket{\phi_j} \ot \ket{\inp}\right)\\
    =& \sum_{y,z,j} \Xg^{g(\inp,(c,y,z))} \ket{\psi_j} \ot \sum_{c} \alpha_{\inp,c} \ket{\wt{\Phi}_{\inp,c}} \ot \left(\ketbra{y} \ot \ketbra{z}\right) P_{h(\inp,c)} \left( U_{Q,\inp} \ket{\phi_j} \ot \ket{\inp}\right)\\
    =& \sum_{y',j} \Xg^{y'} \ket{\psi_j} \ot \left(\sum_{r: g(\inp,r) = y'} \ketbra{\Phi_{\inp,r}}\right) \sum_{c} \alpha_{\inp,c} \ket{\wt{\Phi}_{\inp,c}} \ot P_{h(\inp,c)} \left( U_{Q,\inp} \ket{\phi_j} \ot \ket{\inp}\right)\\
    =& \sum_{y,j} \Xg^{y} \ket{\psi_j} \ot \left(\sum_{r: g(\inp,r) = y} \ketbra{\Phi_{\inp,r}}\right) \left(\ket{\phi_j} \ot \ket{\psi_\PLM}\right)
\end{align*}
where the third equality follows because $g(\inp,(c,y,z))$ outputs the XOR of $y$ with the x-correction specified by $h(\inp,c)$ on wires that $\ketbra{y}$ took place, the fourth equality follows from setting $y' = g(\inp,r)$, $r = (c,y,z)$ and that $\ket{\Phi_{\inp,r}} = \ket{\wt{\Phi}_{\inp,c}} \ot \ket{y} \ot \ket{z}$. 
This implies that
$$\sum_y \Xg^{y} \ot \left(U_{Q,\inp}^\dag (\ketbra{y} \ot I) U_{Q,\inp}\right) \ot \ket{\psi_\PLM} = \sum_{y'} \Xg^{y} \ot \left(\sum_{r: g(\inp,r) = y} \ketbra{\Phi_{\inp,r}}\right) (I \ot \ket{\psi_\PLM})$$
Let $(\rho,\tau)$ be any state. Applying the above two operators with the first register being $0^{n'}$ and the second register being $\rho$, we get $\set{(\tau,y) \given  y \from Q(\inp, \rho)} = \set{(\tau,y) \given y \from P(\inp, \rho)}$.
\end{proof}

\section{Approximately Unitary Quantum Program}
\label{sec:approximate-unitary}
This section investigates the power of white-box access to quantum programs $(\psi, Q)$ that approximately implements some unitary transformation $\cU: \rho \mapsto U \rho U^\dagger$.
\begin{dfn} \label{dfn:eps-approximation}
    A quantum program $(\psi,Q)$ is said to be an $\eps$-approximation of a unitary transformation $\cU$ if
    $$\norm{Q(\cdot \ot \psi) - \cU(\cdot)}_\diamond \le \eps.$$
\end{dfn}
The following lemma shows a sufficient condition for $\eps$-approximation of a unitary.
\begin{lemma}\label{lemma:error-transform}
    Let $(\psi,Q)$ be a quantum program and $\cU$ be a unitary transformation such that $\norm{Q(\rho \ot \psi) - \cU(\rho)}_1 \le \eps^2/2$ for every input state $\rho$. Then $(\psi,Q)$ $\eps$-approximates $\cU$.
\end{lemma}
\begin{proof}
    This follows directly from the first item of \cref{lemma:diamond-norm}.
\end{proof}
\begin{dfn}[Approximately Unitary Quantum Programs]
    The class of approximately unitary quantum programs is defined as the set of all sequences of quantum programs $(\psi_\secparam,Q_\secparam)$ that are $\negl(\secparam)$-approximations of some unitary transformations $\cU_\secparam$.
\end{dfn}
We often make the dependency on $\secparam$ implicitly.
By \cref{lemma:error-transform}, $(\psi,Q)$ is an approximately unitary quantum program if for every state $\rho$, it holds that $Q(\rho \ot \psi) \approx_{\negl(\secparam)} \cU(\rho)$.

We have been considering quantum programs $(\psi,Q)$ whose output space has the same dimension as the input space.
Without loss of generality, we will assume that the quantum circuit $Q$ operates as first taking the input and auxiliary registers, applying some gates $G_1,\dots,G_k$ in sequence, and then tracing out the auxiliary register.
Let $U_Q = G_k \cdots G_1$ be the associated unitary operator and $\cU_Q: \rho \mapsto U_Q\; \rho\; U_Q^\dag$ be the associated unitary transformation. The following lemma shows that accessing the code of a quantum program that approximates a unitary enables other computation beyond the implemented unitary itself.
\begin{lemma} \label{lemma:reuse-program-once}
    Let $(\psi,Q)$ be an $\eps$-approximation of a unitary transformation $\cU(\cdot) = U (\cdot) U^\dag$. Let $U_Q$ act on registers $\reg{B,E}$ for placing the input state and the auxiliary state $\psi$ respectively. Let $A$ be a unitary operator on registers $\reg{R,B}$. Let $\reg{C}$ be a register for placing a control qubit. 
    Set $\onreg{W_Q}{C,R,B,E} := \onreg{\CTRL}{C}\!\text{-}\!\left(\onreg{U_Q^\dag}{B,E} \onreg{A}{R,B} \onreg{U_Q}{B,E}\right)$ and $\onreg{W}{C,R,B} := \onreg{\CTRL}{C}\!\text{-}\!\left(\onreg{U^\dag}{B} \onreg{A}{R,B} \onreg{U}{B}\right)$. Then
    $$\norm{W_Q (\cdot \ot \onreg{\psi}{E}) W_Q^\dag - \left(W (\cdot) W^\dag \ot \onreg{\psi}{E} \right)}_\diamond \le O(\sqrt{\eps}).$$
    %When $\psi$ is a pure state $\psi = \ketbra{\psi}$,
    %$$\norm{\onreg{W_Q}{C,R,B,E} \left(\cdot \ot \onreg{\ket{\psi}}{E}\right) - \onreg{W}{C,R,B} \left(\cdot\right) \ot \onreg{\ket{\psi}}{E}}_{\mathsf{op}}\le O(\sqrt{\eps}).$$
\end{lemma}
\begin{proof}
    Suppose $\psi$ is a pure state. The quantum program $(\psi,Q)$ computes $\Tr_{\reg{E}} \left(\cU_Q (\cdot \ot \onreg{\psi}{E})\right)$ and is an $\eps$-approximation of $\cU(\cdot) = \Tr_{\reg{E}} \left(\cU(\cdot) \ot \onreg{\psi}{E}\right)$, which means that $$\norm{\Tr_{\reg{E}} \left(\cU_Q (\cdot \ot \onreg{\psi}{E})\right) - \Tr_{\reg{E}} \left(\cU (\cdot) \ot \onreg{\psi}{E}\right)}_{\diamond} \le \eps$$
    By \cref{thm:continuity-of-dilation}, there exists a unitary operator $V$ on $\reg{E}$ such that
    $$\norm{\onreg{U_Q}{B,E} \left(\cdot \ot \onreg{\ket{\psi}}{E}\right) - \onreg{U}{B} \ot \onreg{V}{E} \left(\cdot \ot \onreg{\ket{\psi}}{E}\right) }_{\mathsf{op}} \le O(\sqrt{\eps})$$
    By setting $\ket{\psi'} = V \ket{\psi}$, we have
    $$\norm{\onreg{U_Q}{B,E} \left(\cdot \ot \onreg{\ket{\psi}}{E}\right) - \left(\onreg{U}{B} \left(\cdot \right) \ot \onreg{\ket{\psi'}}{E}\right) }_{\mathsf{op}} \le O(\sqrt{\eps})$$
    Since composing with unitary operators preserves operator norm, we have
    $$ \norm{\left(\cdot \ot \onreg{\ket{\psi}}{E}\right) - \onreg{U_Q^\dag}{B,E} \left(\onreg{U}{B} \left(\cdot \right) \ot \onreg{\ket{\psi'}}{E}\right) }_{\mathsf{op}} \le O(\sqrt{\eps}) $$
    $$ \norm{\left(\onreg{U^\dag}{B} \onreg{A}{R,B} \onreg{U}{B}(\cdot) \ot \onreg{\ket{\psi}}{E}\right) - \onreg{U_Q^\dag}{B,E} \left( \onreg{A}{R,B} \onreg{U}{B}(\cdot) \ot \onreg{\ket{\psi'}}{E}\right) }_{\mathsf{op}} \le O(\sqrt{\eps}) $$
    By the triangle inequality, we then obtain
    \begin{align*}
        \norm{\left(\onreg{U^\dag}{B} \onreg{A}{R,B} \onreg{U}{B} \left(\cdot\right) \ot \onreg{\ket{\psi}}{E}\right) - \onreg{U_Q^\dag}{B,E} \onreg{A}{R,B} \onreg{U_Q}{B,E} \left(\cdot \ot \onreg{\ket{\psi}}{E}\right) }_{\mathsf{op}} \le O(\sqrt{\eps})
    \end{align*}
    Since both terms within the norm start with tensoring $\ket{\psi}$, we have
    \begin{align*}
        \norm{\left(\onreg{\CTRL}{C}\!\text{-}\!\left(\onreg{U^\dag}{B} \onreg{A}{R,B} \onreg{U}{B}\right) (\cdot) \ot \onreg{\ket{\psi}}{E}\right) \;-\; \left(\onreg{\CTRL}{C}\!\text{-}\!\left(\onreg{U_Q^\dag}{B,E} \onreg{A}{R,B} \onreg{U_Q}{B,E}\right)\right)\left(\cdot \ot \onreg{\ket{\psi}}{E}\right) }_{\mathsf{op}} \le O(\sqrt{\eps})
    \end{align*}
    By definition, this is equivalent to $\norm{\onreg{W}{C,R,B}(\cdot) \ot \onreg{\ket{\psi}}{E} - \onreg{W_Q}{C,R,B,E}\left(\cdot \ot \onreg{\ket{\psi}}{E}\right)}_{\mathsf{op}} \le O(\sqrt{\eps})$.
    Applying \cref{lemma:easy-norm}, we have
    $$\norm{W_Q (\cdot \ot \onreg{\psi}{E}) W_Q^\dag - \left(W (\cdot) W^\dag \ot \onreg{\psi}{E} \right)}_\diamond \le O(\sqrt{\eps})$$
    The last inequality holds even if $\psi$ is not a pure state because we can first obtain the inequality for the purification of $\psi$ and then use the property that taking partial trace does not increase the diamond norm to derive the inequality that refers only to $\psi$.
\end{proof}

The following lemma shows that a quantum program approximating a unitary allows repeated evaluation of the unitary and its inverse.

\begin{lemma} \label{lemma:reuse-program-many-times}
    There is an efficient classical algorithm $\cT$ that, given quantum circuits $Q',Q$ of the quantum programs $(\psi',Q'), (\psi,Q)$ where
    \begin{itemize}
        \item $Q'$ uses the gate set\footnote{We assume that every gate set under consideration is closed under inversion.} $\cG' \cup \set{U, U^\dag}$ and applies the gates $U, U^\dag$ at most $q$ times,
        \item $Q$ uses the gate set $\cG$ and $(\psi,Q)$ is an $\eps$-approximation of $\cU: \rho \mapsto U \rho U^\dag$,
    \end{itemize}
    $\cT$ outputs a quantum circuit $Q''$ using the gate set $\cG' \cup \cG \cup \set{\SWAP}$ such that with the auxiliary state $\psi'' = \psi' \ot \psi \ot \ketbra{0}^{\ot n}$ where $n$ is the input length of $U$, we have
    $$\norm{Q''(\cdot \ot \psi'') - Q'(\cdot \ot \psi')}_{\diamond} \le O\left(q \sqrt{\eps}\right)$$
\end{lemma}

\begin{proof}
    Let $G_1,\dots,G_k \in \cG$ be the gates that $Q$ applies in sequence and set $U_Q = G_k \cdots G_1$. 
    The algorithm $\cT$ is described as follows:
    \begin{enumerate}
        \item Take the circuit $Q'$, which uses the gate set $\cG' \cup \set{U,U^\dag}$ and specifies a register for $\psi'$. 
        \item Specify an additional register $\reg{E}$ for storing $\psi$ and a register $\reg{B}$ for storing $\ketbra{0}^{\ot n}$.
        \item Replace each $U$ gate (say on $\reg{R}$) with
        $\onreg{(G_1^\dag \cdots G_k^\dag)}{B,E} \onreg{\SWAP}{B,R} \onreg{(G_k \cdots G_1)}{B,E} \onreg{\SWAP}{B,R} $.
        \item Replace each $U^\dag$ gate (say on $\reg{R}$) with
        $\onreg{\SWAP}{B,R} \onreg{(G_1^\dag \cdots G_k^\dag)}{B,E} \onreg{\SWAP}{B,R} \onreg{(G_k \cdots G_1)}{B,E}$.
        \item Output the resulting circuit $Q''$.
    \end{enumerate}
    The resulting circuit $Q''$ uses the gate set $\cG' \cup \cG \cup \set{\SWAP}$ by construction. Now we analyze the behavior of the quantum program $(\psi'',Q'')$. 
    Consider the circuit $\wt{Q}$ generated similarly as $Q''$ except that
    \begin{itemize}
        \item In step $3$, replace each $U$ gate (say on $\reg{R}$) with $\onreg{U^\dag}{B} \onreg{\SWAP}{B,R} \onreg{U}{B}  \onreg{\SWAP}{B,R}$ instead.
        \item In step $4$, replace each $U^\dag$ gate (say on $\reg{R}$) with $\onreg{\SWAP}{B,R} \onreg{U^\dag}{B} \onreg{\SWAP}{B,R} \onreg{U}{B} $ instead.
    \end{itemize}
    Define $W_Q := \onreg{(G_1^\dag \cdots G_k^\dag)}{B,E} \onreg{\SWAP}{B,R} \onreg{(G_k \cdots G_1)}{B,E}$ and 
    $W := \onreg{U^\dag}{B} \onreg{\SWAP}{R,B} \onreg{U}{B} $.
    Applying \cref{lemma:reuse-program-once} with the control qubit being $\ket{1}$ and $A=\SWAP$, we have 
    $$\norm{W_Q (\cdot \ot \onreg{\psi}{E}) W_Q^\dag - \left(W (\cdot) W^\dag \ot \onreg{\psi}{E} \right)}_\diamond \le O(\sqrt{\eps}).$$
    By the triangle inequality and noting that $\psi''$ initializes $\psi$ on register $\reg{E}$, we have
    $$\norm{Q''(\cdot \ot \psi'') - \wt{Q}(\cdot \ot \psi'')}_{\diamond} \le O(q \sqrt{\eps}).$$ 
    Since $\onreg{U^\dag}{B} \onreg{\SWAP}{B,R} \onreg{U}{B}  \onreg{\SWAP}{B,R} = \onreg{U^\dag}{B} \onreg{U}{R}$ and $\onreg{\SWAP}{B,R} \onreg{U^\dag}{B} \onreg{\SWAP}{B,R} \onreg{U}{B} = \onreg{U^\dag}{R} \onreg{U}{B}$,
    the program $\wt{Q}(\cdot \ot \psi'')$ only differs from the original program $Q'(\cdot \ot \psi')$ by having additional tensored auxiliary registers $\reg{E,B}$ and extra computation on $\reg{B}$. These additional registers are not used for the output, so the two programs implement the same quantum mapping. Thus, we conclude that
    $$\norm{Q''(\cdot \ot \psi'') - Q'(\cdot \ot \psi')}_{\diamond} \le O(q \sqrt{\eps})$$
\end{proof}
We summarize the capability of approximately unitary quantum programs as follows.
\begin{cor} \label{cor:unitary-functionality}
    Let quantum program $(\psi,Q)$ be a $\negl(\secparam)$-approximation of $\cU: \rho \mapsto U \rho U^\dag$. Then the quantum program can be used to apply $U$ and $U^\dag$ for a polynomial number of times with negligible error in diamond distance. Moreover, suppose the gate set $\cG$ used by $Q$ satisfy that for every $G \in \cG$, the operator $\CTRL\text{-}G$ can be efficiently implemented from $\cG$. Then for every operator $A$ that can be efficiently implemented from $\cG$, one can efficiently construct a quantum program that $\negl(\secparam)$-approximates $\CTRL\text{-}(U^\dag A U)$ using the quantum program $(\psi,Q)$.
\end{cor}

The following lemmas show how $\CTRL\text{-}(U^\dag A U)$ is connected to $U,U^\dag$ through black-box reductions. Note that black-box access to a unitary operator $U$ is equivalent to black-box access to its associated unitary transformation $\cU: \rho \mapsto U \rho U^\dag$.

\begin{lemma}\label{lemma:the-single-swap-oracle}
    Let $U$ be a unitary operator on $n$ qubits and $A$ be a unitary operator. Then
    $\CTRL\text{-}(U^\dag A U)$ can be computed from $\CTRL\text{-}(U^\dag \SWAP^{\ot n} U)$ and $\CTRL\text{-}A$ in a black-box manner.
\end{lemma}
\begin{proof}
    Observe that $\left(\onreg{U^\dag}{B} \onreg{\SWAP^{\ot n}}{B,C} \onreg{U}{B}\right) \onreg{A}{C,D} \left(\onreg{U^\dag}{B} \onreg{\SWAP^{\ot n}}{B,C} \onreg{U}{B}\right) = \onreg{U^\dag}{B} \onreg{A}{B,D} \onreg{U}{B}$ holds. The lemma follows from taking the controlled version of the identity.
\end{proof}

\begin{lemma}\label{lemma:U-and-U-inverse-suffice}
    $\CTRL\text{-}(U^\dag \SWAP^{\ot n} U)$ can be computed from $U$ and $U^\dag$ in a black-box manner.
\end{lemma}

\begin{proof}$\CTRL\text{-}(U^\dag \SWAP^{\ot n} U)$ can be computed by the following circuit.
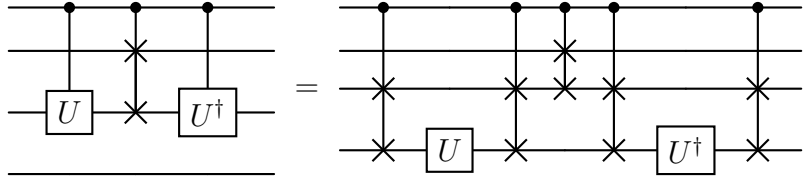
\begin{center}
    \begin{quantikz}
        & \ctrl{2} & \ctrl{2} & \ctrl{2} &\\
        & & \swap{1} & & \\
        & \gate{U} & \targX{} & \gate{U^\dag} &\\
        & &  & & 
    \end{quantikz}
    $=$
    \begin{quantikz}
        & \ctrl{2} & & \ctrl{2} & \ctrl{2} & \ctrl{2} & & \ctrl{2} &\\
        & & & & \swap{1} & & & & \\
        & \swap{1} & & \swap{1} &\targX{} & \swap{1} & & \swap{1} &\\
        & \targX{} & \gate{U} & \targX{} & & \targX{} & \gate{U^\dag} & \targX{} &
    \end{quantikz}
\end{center}

\end{proof}

\section{Quantum State Obfuscation}
\label{sec:obfuscation}

We now present an obfuscation scheme for quantum programs that approximately implement unitary transformations. This setting captures a broad and expressive class of quantum functionalities while supporting program reusability. We begin by formalizing the notion of quantum state ideal obfuscation in the classical oracle model, followed by the description of our construction. The security of our obfuscation scheme critically relies on the functional authentication scheme and PLM quantum programs developed in previous sections.
%%%%%%%%%%%%%%%%%%%%%%%%%%%%%%
% Definition
%%%%%%%%%%%%%%%%%%%%%%%%%%%%%%
\begin{dfn}[Quantum State Obfuscation]
\label{dfn:obf}
 A \emph{quantum state ideal obfuscation} for the class of approximately-unitary quantum programs in the classical oracle model is a pair of QPT algorithms $(\QObf,\QEval)$ with the following syntax:
\begin{itemize}
    \item $\QObf(1^\secparam, (\psi, Q)) \to (\wt{\psi},\sF)$: The obfuscator takes as input the security parameter $1^\secparam$ and a quantum program $(\psi, Q)$ in the plain model\footnote{$Q$ is described without reference to oracles, so as to circumvent the impossibility result of \cite{barak2001possibility}.}, and outputs an obfuscated program specified by a state $\wt{\psi}$ and a classical function $\sF$. 
    \item $\QEval^{\sF}(\rho_{\intext},\wt{\psi}) \to \rho_{\out}$: The evaluation algorithm executes the obfuscated program on quantum input $\rho_{\intext}$ by making use of the state $\wt{\psi}$ and making superposition queries to the classical oracle $\sF$, and produces the quantum output $\rho_{\out}$.
\end{itemize}
These algorithms have to satisfy the following properties.%\footnote{We require the quantum program to be separable from the quantum input and quantum advice of $\adv$.}
\begin{itemize}
    \item \textbf{Functionality-Preserving}: 
    For every quantum program $(\psi,Q)$ which is an $\negl(\secparam)$-approximation of some unitary transformation $\cU$, the quantum program
    $$\E_{(\widetilde{\psi}, \sF) \leftarrow \QObf(1^\lambda, (\psi,Q))} \left(\wt{\psi},\; \QEval^\sF\right)$$
    is also a $\negl(\secparam)$-approximation of $\cU$. Note that our convention for classical oracles (\cref{Sec:preliminary}) ensures that  $\CTRL\text{-}\sF$ can be efficiently computed by querying $\sF$. Therefore, the resulting quantum program would satisfy all properties described in \cref{cor:unitary-functionality}.
    \item \textbf{Ideal Obfuscation}:
    There exists a stateful\footnote{The simulator can produce an oracle that carries a quantum state across queries as auxiliary input.} QPT simulator $\Sim$ such that for every QPT adversary $\adv$ and quantum program $(\psi,Q)$ that $\negl(\secparam)$-approximates some unitary transformation $\cU$ associated with a unitary operator $U$,
    {\small
    \[
    \abs{\Pr_{(\widetilde{\psi}, \sF) \leftarrow \QObf(1^\secparam, (\psi,Q))} \left[1 \gets \adv^\sF\left(\wt{\psi} \right) \right] - 
    \Pr_{(\widetilde{\psi}, \sF_\Sim) \leftarrow \Sim(1^\secparam, 1^{n}, 1^{m})} \left[1 \gets \adv^{\sF_{\Sim}^{\CTRL\text{-}(U^\dag \cdot U)}}\left(\wt{\psi}\right) \right]} = \negl(\secparam)
    \]
    }Here, $n$ is the input length of the quantum program, $m$ is the size of the quantum program, and $\sF_{\Sim}$ makes black-box oracle access to $\CTRL\text{-}(U^\dag A U)$ for some unitary $A$ that can be efficiently implemented.
\end{itemize}
\end{dfn}

\begin{rmk}
    While our definition allows $A$ to be chosen freely, one can also set $A = \SWAP^{\ot n}$ explicitly without loss of generality by \cref{lemma:the-single-swap-oracle}, 
    where $n$ is the program's input length. In this case, the definition of ideal obfuscation admits an intuitive interpretation: the obfuscated program only enables the controlled computation $\CTRL\text{-}(U^\dag \SWAP^{\ot n} U)$ that corresponds to the combined operation of applying $U$, swapping the result with any state, and applying $U^\dag$.
\end{rmk}

\begin{rmk}\label{rmk:2}
    An alternative way of defining ideal obfuscation is to require that $\sF_{\Sim}$ makes black-box oracle access to $\cU$ and $\cU^\dag$ instead. This definition is automatically satisfied by our \cref{dfn:obf}, according to \cref{lemma:the-single-swap-oracle,lemma:U-and-U-inverse-suffice}. This definition also provides an intuitive interpretation: the obfuscated program only enables black-box computation of $\cU$ and $\cU^\dag$.
\end{rmk}

\begin{theorem}[Main theorem]
\label{thm:main}
There exists a quantum state ideal obfuscation for the class of approximately-unitary quantum programs with quantum inputs and outputs in the classical oracle model, assuming post-quantum one-way functions.
\end{theorem}

\begin{rmk} 
    Similar to \cite{bartusek2024quantum}, the assumption of a post-quantum one-way function can be replaced with a truly random function (or a random oracle) if the classical oracle is not required to be efficiently computable.
\end{rmk}
\begin{rmk}
    We also improve the tolerated approximation error when obfuscating $\varepsilon$-pseudo-deterministic quantum programs. \cite{bartusek2024quantum} only handles the case $\varepsilon = 2^{-2n} \negl(\secparam)$, whereas our result applies whenever $\varepsilon = \negl(\secparam)$. This follows from our main theorem and \cref{lemma:error-transform}, which shows that any $\negl(\secparam)$-pseudo-deterministic quantum program is also an $\negl(\secparam)$-approximation of the unitary associated with the classical function it implements.
\end{rmk}

\subsection{Construction}
\label{sec:construction}
Our obfuscation scheme is constructed using the compilation algorithm of \cref{thm:compile}, the quantum authentication scheme introduced in \cref{Sec:auth}, a tokenized signature scheme (\cref{sec:token-sig}), and a post-quantum PRF with $\kappa = O(\secparam n)$ bits of output (\cref{sec:prf}). 

Without loss of generality, the input of our obfuscation scheme will be a quantum program $(\psi, Q^{\mathsf{univ}})$ where $Q^{\mathsf{univ}}$ is a universal circuit from the universal gate set $\CliffordGroup_2 \cup \set{\Tg}$ and the output qubits will be placed at the same position as the input qubits. The state $\psi$ encodes the actual content of the program. From now on, we will omit writing $Q^{\mathsf{univ}}$ as part of the input arguments to $\QObf$. Using $Q^{\mathsf{univ}}$, we construct the following quantum circuit $\wt{Q}^\mathsf{univ}$ with classical outputs:
    \begin{enumerate}
        \item Take a quantum state on register $\reg{V} = \reg{(V_\intext, V_\outtext, V_\aux})$, of length $n,n,m$ respectively.
        \item Take a classical input $\inp \in \zo^{2n}$.
        \item Apply the teleportation receiving operation $\TPRecv(P_\inp, \reg{V_\intext})$.
        \item Apply $Q^{\mathsf{univ}}$ on $\reg{(V_\intext, V_\aux)}$.
        \item Perform quantum teleportation $\TPSend(\reg{V_\intext},\reg{V_\outtext})$ and output its outcome.
    \end{enumerate}
Applying \cref{thm:compile}, we compile $\wt{Q}^{\mathsf{univ}}$ into a PLM quantum program $P$ with auxiliary state $\psi_{\PLM}$ (on some register $\reg{V_{\PLM}}$) and instructions $\set{f_j,\theta_j,G_j}_{j\in [\nstep]}$ with $|\psi_\PLM|,\nstep = O(m)$. Our obfuscation scheme is described as follows.

\begin{figure}[H]
\label{fig:protocol}
\begin{framed}
\begin{center}
    \textbf{Construction of Quantum State Obfuscation}
\end{center}
\vspace{-3mm}
{
%%%%%%%%%%%%%%%%%%%%%%%%%%%%%%
% QObf
%%%%%%%%%%%%%%%%%%%%%%%%%%%%%%
{\small
$\QObf\left(1^\secparam,\psi \right)$:
\begin{itemize}
    \item Generate input EPR pairs $(\epr_{(\intext,\pub)},\epr_{(\intext,\priv)})$ of length $n$.\\
    Generate output EPR pairs $(\epr_{(\outtext,\priv)},\epr_{(\outtext,\pub)})$ of length $n$.
    \item Prepare the auxiliary state $\psi_{\PLM}$ for the PLM quantum program.
    \item Sample an authentication key $k \from \Auth.\KeyGen(1^{O(\secparam+m)}, 1^{O(m)})$.\\
    Compute the authenticated state $\onreg{\psi_\auth}{\wt{V}} \from \Auth.\Enc_k((\epr_{(\intext,\priv)}, \epr_{(\outtext,\priv)}, \psi, \psi_{\PLM}))$.
    \item Sample a signature token $(\vk,\tau_{\token}) \from \Token.\Gen(1^\secparam, 1^{2n})$.
    \item Sample a PRF key $k_0 \from \zo^\secparam$ and set $H \equiv \PRF_{k_0} : \zo^* \to \zo^\kappa$.
    
    \item Define the classical function $\sF\left(j,\widetilde{v}, \inp,s,\ell_1,\dots,\ell_\nstep \right)$ where $j \in [\nstep]$:
    \begin{enumerate}
        \item Compute $v \from \Auth.\Dec_{k,\theta_j,G_j}(\widetilde{v})$. If it rejects, output $\bot$.
        \item Compute $\Token.\Verify_\vk(\inp,s)$. If it rejects, output $\bot$.
        \item For $i = 1,\dots, j-1$, reconstruct $r_i$ as follows:
        \begin{itemize}
            \item Set $r_i \in \zo$ so that $\ell_i = H(i,r_i,\inp,s)$.
            \item If there are two or no solutions for $r_i$, output $\bot$. 
        \end{itemize}
        
        \item Compute $r_j = f_j^{\inp,r_1,\dots,r_{j-1}}(v)$.
        \item Output $\begin{cases}
            \text{the label } H(j,r_j,\inp, s) & \text{ if } j < \nstep.\\
            \text{the value } g(\inp, r_1,\dots,r_\nstep) & \text{ if } j = \nstep.
        \end{cases}$
    \end{enumerate}

    %\item Define the classical function $\sG\left(\inp,s,\ell_1,\dots,\ell_{\nstep} \right)$:
    %\begin{enumerate}
        %\item Compute $\Token.\Verify_\vk(\inp,s)$. If it rejects, output $\bot$.
        %\item For $i = 1,\dots, \nstep$, reconstruct $r_i$ as follows:
        %\begin{itemize}
            %\item Set $r_i \in \zo$ so that $\ell_i = H_0(i,r_\inp,s)$.
            %\item If there are two or no solutions for $r_i$, output $\bot$. 
        %\end{itemize}
        %\item Output $g(\inp,r_1,\dots,r_\nstep)$.
    %\end{enumerate}
    %\item Output $\ket{\widetilde{\psi}} =  \ket{\psi_\auth} \ket{\epr_{(\intext,\pub)}} \ket{\epr_{(\outtext,\pub)}} \ket{\token}$ and classical oracle $O$.
    \item Output $\wt{\psi} =  (\onreg{\psi_\auth}{\wt{V}}, \epr_{(\intext,\pub)}, \epr_{(\outtext,\pub)}, \tau_{\token})$ and the classical function $\sF$.
\end{itemize}

%%%%%%%%%%%%%%%%%%%%%%%%%%%%%%
% QEval
%%%%%%%%%%%%%%%%%%%%%%%%%%%%%%
\noindent
$\QEval^{\sF}\left(\rho_\intext,\wt{\psi}\right)$:
\begin{itemize}
    \item Parse $\wt{\psi} =  (\psi_\auth, \epr_{(\intext,\pub)}, \epr_{(\outtext,\pub)}, \tau_{\token})$.
    \item Perform quantum teleportation, obtaining the result $\inp \from \TPSend(\rho_\intext, \epr_{(\intext,\pub)})$.
    \item Sign the teleportation result, obtaining its signature $s \from \Token.\Sign \left(\inp,\tau_{\token} \right)$.
    \item Place $\wt{\psi}$ on register $\reg{\wt{V}}$ and do the following iteratively for $j = 1,\dots, \nstep$:
    \begin{itemize}
        \item Apply ${\Hg^{\wt{\theta}_j} \wt{G_j}}$ on register $\reg{\wt{V}}$.
        \item Measure the outcome $\ell_j \from \sF\left(j,\reg{\wt{V}},\inp,s,\ell_1,\dots,\ell_{j-1},0^\kappa,\dots,0^\kappa\right)$.
        \item Apply $\left({\Hg^{\wt{\theta}_j} \wt{G_j}}\right)^\dag$ on register $\reg{\wt{V}}$.
    \end{itemize}
    \item Output $\rho_\outtext \from \TPRecv(\ell_\nstep, \epr_{(\outtext,\pub)})$.
    %\item Compute $\ell_{\out} \from \sG\left(\inp, s, \ell_1,\dots, \ell_\nstep\right)$.
    %\item Output $\rho_\outtext \from \TPRecv(\ell_\out, \epr_{(\outtext,\pub)})$.
    %\item Output $\rho_{\out} \from \TPRecv(\ell_m, \epr_{\out, R})$
\end{itemize}
}}
\end{framed}
\end{figure}

\subsection{Security}\label{sec:security}
We prove \cref{thm:main} by showing that Construction \ref{fig:protocol} is a quantum state ideal obfuscation for approximately-unitary quantum programs. We start by constructing the following hybrids, where we may change the way we describe quantumly accessible oracles, from a purely classical interface to a partially quantum interface. For example, outputting the classical value $x$ to the output register $\reg{Y}$ will be written as $\onreg{\Write(x)}{Y}$. Furthermore, all ``If–Then'' expressions are to be interpreted as describing quantum controlled operations, whereas all measurements are to be applied coherently.

We extensively use the notations $\onreg{\WriteH(x)}{Y,D}, \onreg{\EqualH(x)}{Y,D}, \ket{\cL_\emptyset}, \Has_{S}, \Only_S$
introduced in \cref{sec:purified-ro} to handle purified random oracles. We also make the following abbreviations:
$\Has_{(i,b,\inp,s)} := \Has_{\set{(i,b,\inp,s)}},\;
\Has_{\neq \inp} := \Has_{\set{(i,b,\inp',s) | i,b,s, \inp'\neq \inp}},\;
\Only_{\inp} := \Only_{\set{(i,b,\inp,s) | i,b,s}}.$

{\small
\begin{framed}
\label{fig:main-hybrids}
\noindent
$\QObf(1^\secparam, \psi)$: Output $\wt{\psi}$ and the oracle $\sF$.
\vspace{1mm}\hrule\vspace{1mm}

\noindent
$\hybrid_1(1^\secparam, \psi)$: 
\begin{itemize}
    \item Generate input EPR pairs $(\epr_{(\intext,\pub)},\epr_{(\intext,\priv)})$ of length $n$.\\
    Generate output EPR pairs $(\epr_{(\outtext,\priv)},\epr_{(\outtext,\pub)})$ of length $n$.
    \item Sample keys $k, (\vk,\tau_{\token}), k_0 \from \zo^\secparam$ and set $H(\cdot) \equiv \PRF_{k_0}(\cdot)$ as in $\QObf$.
    \item \mark{Set $\onreg{\psi_\auth}{\wt{V}} \from \Auth.\Enc_k(0^{O(m)})$ and $\reg{V} \gets (\epr_{(\intext,\priv)}, \epr_{(\outtext,\priv)}, \psi, \psi_\PLM)$.}
    \item Define the oracle $\onreg{\sF_1\left(j,\wt{v}, \inp,s,\ell_1,\dots,\ell_\nstep, \reg{Y} \right)}{V}$ where $\reg{Y}$ denotes the output register:
    \begin{enumerate}
        \item \mark{Compute $\Auth.\Ver_{k,\theta_j,G_j}(\widetilde{v})$.} If it rejects, apply $\onreg{\Write(\bot)}{Y}$ and return.
        \item Compute $\Token.\Verify_\vk(\inp,s)$. If it rejects, apply $\onreg{\Write(\bot)}{Y}$ and return.
        \item For $i = 1,\dots, j-1$, reconstruct $r_i$ as follows:
        \begin{itemize}
            \item Set $r_i \in \zo$ so that $\ell_i = H(i,r_i,\inp,s)$.
            \item If there are two or no solutions for $r_i$, apply $\onreg{\Write(\bot)}{Y}$ and return.
        \end{itemize}
        
        \item \mark{Compute $r_j \gets \meas{f_j^{\inp,r_1,\dots,r_{j-1}}, \theta_j, G_j}(\reg{V})$.}
        \item Apply $\begin{cases}
            \onreg{\Write(H(j,r_j,\inp, s))}{Y} & \text{ if } j < \nstep.\\
            \onreg{\Write(g(\inp, r_1,\dots,r_\nstep))}{Y} & \text{ if } j = \nstep.
        \end{cases}$
    \end{enumerate}
    \item Output $\wt{\psi} =  (\onreg{\psi_\auth}{\wt{V}}, \epr_{(\intext,\pub)}, \epr_{(\outtext,\pub)}, \tau_{\token})$ and the oracle $\sF_1$.
\end{itemize}
\vspace{1mm}\hrule\vspace{1mm}

\noindent
$\hybrid_2(1^\secparam, \psi)$: 
\begin{itemize}
    \item Generate $\wt{\psi},k,\vk$ and initialize $\reg{V} \gets (\epr_{(\intext,\priv)}, \epr_{(\outtext,\priv)}, \psi, \psi_{\PLM})$ as in $\hybrid_1$.
    \item \mark{Initialize a database register $\reg{D} \gets \ket{\cL_{\emptyset}}$.}
    \item Define the oracle $\sF_2\left(j,\widetilde{v}, \inp,s,\reg{L_1},\dots,\reg{L_{\nstep}},\reg{Y} \right)^{\reg{V,D}}$ where $\reg{Y}$ denotes the output register:
    \begin{enumerate}
        \item Compute $\Auth.\Ver_{k,\theta_j,G_j}(\widetilde{v})$. If it rejects, apply $\onreg{\Write(\bot)}{Y}$ and return.
        \item Compute $\Token.\Verify_\vk(\inp,s)$. If it rejects, apply $\onreg{\Write(\bot)}{Y}$ and return.
        \item For $i = 1,\dots, j-1$, reconstruct $r_i$ as follows:\footnotemark
        \begin{itemize}
            \item \mark{Set $r_i = 0$ if $\onreg{\EqualH(i,0,\inp,s)}{L_i, D}$ holds}.
            \item \mark{Set $r_i = 1$ if $\onreg{\EqualH(i,1,\inp,s)}{L_i, D}$ holds}.
            \item If none or both of the above hold, apply $\onreg{\Write(\bot)}{Y}$ and return.
        \end{itemize}
        \item Compute $r_j \gets \meas{f_j^{\inp,r_1,\dots,r_{j-1}}, \theta_j, G_j}(\reg{V})$.
        \item Apply $\begin{cases}
            \mark{\onreg{\WriteH(j,r_j,\inp, s)}{Y,D}} & \text{ if } j < \nstep.\\
            \onreg{\Write(g(\inp, r_1,\dots,r_\nstep))}{Y} & \text{ if } j = \nstep.
        \end{cases}$
    \end{enumerate}
    \item Output $\wt{\psi}$ and the oracle $\sF_2$.
\end{itemize}
\hrule\vspace{1mm}

\noindent
$\hybrid_3(1^\secparam, \psi)$: 
\begin{itemize}
    \item Generate $\wt{\psi},k,\vk$ and initialize $\reg{V,D}$ as in $\hybrid_2$.
    \item Define the oracle $\sF_3\left(j,\widetilde{v}, \inp,s,\reg{L_1},\dots,\reg{L_{\nstep}},\reg{Y} \right)^{\reg{V,D}}$ where $\reg{Y}$ denotes the output register:
    \begin{enumerate}
        \item If $\Auth.\Ver_{k,\theta_j,G_j}(\widetilde{v})$ rejects, apply $\onreg{\Write(\bot)}{Y}$ and return.
        \item If \mark{$\onreg{\Has_{\neq \inp}}{D}$ holds or} $\Token.\Verify_\vk(\inp,s)$ rejects, apply $\onreg{\Write(\bot)}{Y}$ and return.
        \item For $i = 1,\dots, j-1$, reconstruct $r_i$ as follows:\footnotemark
        \begin{itemize}
            \item Set $r_i = 0$ if $\onreg{\EqualH(i,0,\inp,s)}{L_i, D}$ holds.
            \item Set $r_i = 1$ if $\onreg{\EqualH(i,1,\inp,s)}{L_i, D}$ holds.
            \item If none or both of the above hold, apply $\onreg{\Write(\bot)}{Y}$ and return.
        \end{itemize}
        \item Compute $r_j \gets \meas{f_j^{\inp,r_1,\dots,r_{j-1}}, \theta_j, G_j}(\reg{V})$.
        \item Apply $\begin{cases}
            \onreg{\WriteH(j,r_j,\inp, s)}{Y,D} & \text{ if } j < \nstep.\\
            \onreg{\Write(g(\inp, r_1,\dots,r_\nstep))}{Y} & \text{ if } j = \nstep.
        \end{cases}$
    \end{enumerate}
    \item Output $\wt{\psi}$ and the oracle $\sF_3$.
\end{itemize}
\hrule\vspace{1mm}

\noindent
$\hybrid_4(1^\secparam, \psi)$: 
\begin{itemize}
    \item Generate $\wt{\psi},k,\vk$ and initialize $\reg{V,D}$ as in $\hybrid_2$.
    \item Define the oracle $\sF_4\left(j,\widetilde{v}, \inp,s,\reg{L_1},\dots,\reg{L_{\nstep}},\reg{Y} \right)^{\reg{V,D}}$ where $\reg{Y}$ denotes the output register:
    \begin{enumerate}
        \item If $\Auth.\Ver_{k,\theta_j,G_j}(\widetilde{v})$ rejects, apply $\onreg{\Write(\bot)}{Y}$ and return.
        \item If $\onreg{\Has_{\neq \inp}}{D}$ holds or $\Token.\Verify_\vk(\inp,s)$ rejects, apply $\onreg{\Write(\bot)}{Y}$ and return.
        \item For $i = 1,\dots, j-1$, 
        \begin{itemize}
            \item If none or both of the conditions $\onreg{\EqualH(i,0,\inp,s)}{L_i, D}$ and $ \onreg{\EqualH(i,1,\inp,s)}{L_i, D}$ hold, apply $\onreg{\Write(\bot)}{Y}$ and return.
            \item \mark{Compute $r_i \gets \meas{f_i^{\inp,r_1,\dots,r_{i-1}}, \theta_i, G_i}(\reg{V})$}.
        \end{itemize}
        \item Compute $r_j \gets \meas{f_j^{\inp,r_1,\dots,r_{j-1}}, \theta_j, G_j}(\reg{V})$.
        \item Apply $\begin{cases}
            \onreg{\WriteH(j,r_j,\inp, s)}{Y,D} & \text{ if } j < \nstep.\\
            \onreg{\Write(g(\inp, r_1,\dots,r_\nstep))}{Y} & \text{ if } j = \nstep.
        \end{cases}$
    \end{enumerate}
    \item Output $\wt{\psi}$ and the oracle $\sF_4$.
\end{itemize}
\hrule\vspace{1mm}

\noindent
$\hybrid_5(1^\secparam, \psi)$: 
\begin{itemize}
    \item Generate $\wt{\psi},k,\vk$ and initialize $\reg{V,D}$ as in $\hybrid_2$.
    \item Define the oracle $\sF_5\left(j,\widetilde{v}, \inp,s,\reg{L_1},\dots,\reg{L_{\nstep}},\reg{Y} \right)^{\reg{V,D}}$ where $\reg{Y}$ denotes the output register:
    \begin{enumerate}
        \item If $\Auth.\Ver_{k,\theta_j,G_j}(\widetilde{v})$ rejects, apply $\onreg{\Write(\bot)}{Y}$ and return.
        \item If $\onreg{\Has_{\neq \inp}}{D}$ holds or $\Token.\Verify_\vk(\inp,s)$ rejects, apply $\onreg{\Write(\bot)}{Y}$ and return.
        \item For $i = 1,\dots, j-1$, 
        \begin{itemize}
            \item If none or both of the conditions $\onreg{\EqualH(i,0,\inp,s)}{L_i, D}$ and $ \onreg{\EqualH(i,1,\inp,s)}{L_i, D}$ hold, apply $\onreg{\Write(\bot)}{Y}$ and return.
        \end{itemize}
        \item For $i = 1,\dots, j$, compute $r_i \gets \meas{f_i^{\inp,r_1,\dots,r_{i-1}}, \theta_i, G_i}(\reg{V})$.
        \item Apply $\begin{cases}
            \mark{\onreg{\WriteH(j,0,\inp, s)}{Y,D}} & \text{ if } j < \nstep.\\
            \onreg{\Write(g(\inp, r_1,\dots,r_\nstep))}{Y} & \text{ if } j = \nstep.
        \end{cases}$
    \end{enumerate}
    \item Output $\wt{\psi}$ and the oracle $\sF_5$.
\end{itemize}
\hrule\vspace{1mm}

\noindent
$\hybrid_6(1^\secparam, \psi)$: 
\begin{itemize}
    \item Generate $\wt{\psi},k,\vk$ and initialize $\reg{V,D}$ as in $\hybrid_2$.
    \item Define the oracle $\sF_6\left(j,\widetilde{v}, \inp,s,\reg{L_1},\dots,\reg{L_{\nstep}},\reg{Y} \right)^{\reg{V,D}}$ where $\reg{Y}$ denotes the output register:
    \begin{enumerate}
        \item If $\Auth.\Ver_{k,\theta_j,G_j}(\widetilde{v})$ rejects, apply $\onreg{\Write(\bot)}{Y}$ and return.
        \item If \mark{\sout{$\onreg{\Has_{\neq \inp}}{D}$ holds or}} $\Token.\Verify_\vk(\inp,s)$ rejects, apply $\onreg{\Write(\bot)}{Y}$ and return.
        \item \mark{Whenever $\onreg{\EqualH(i,0,\inp,s)}{L_i,D}$ fails for some $i < j$}, apply $\onreg{\Write(\bot)}{Y}$ and return.
        \item For $i = 1,\dots, j$, compute $r_i \gets \meas{f_i^{\inp,r_1,\dots,r_{i-1}}, \theta_i, G_i}(\reg{V})$.
        \item Apply $\begin{cases}
            \onreg{\WriteH(j,0,\inp, s)}{Y,D} & \text{ if } j < \nstep.\\
            \onreg{\Write(g(\inp, r_1,\dots,r_\nstep))}{Y} & \text{ if } j = \nstep.
        \end{cases}$
    \end{enumerate}
    \item Output $\wt{\psi}$ and the oracle $\sF_6$.
\end{itemize}
\hrule\vspace{1mm}

\noindent
$\hybrid_7(1^\secparam, \psi)$: 
\begin{itemize}
    \item Generate $\wt{\psi},k,\vk$ and initialize $\reg{V,D}$ as in $\hybrid_2$.
    \item \mark{Parse $\reg{V} = (\reg{V_{\intext}}, \reg{V_{\outtext}}, \reg{V_\aux}, \reg{V_{\PLM}})$ as four parts.}
    \item Define the oracle $\sF_7\left(j,\widetilde{v}, \inp,s,\reg{L_1},\dots,\reg{L_{\nstep}},\reg{Y} \right)^{\reg{V,D}}$ where $\reg{Y}$ denotes the output register:
    \begin{enumerate}
        \item If $\Auth.\Ver_{k,\theta_j,G_j}(\widetilde{v})$ rejects, apply $\onreg{\Write(\bot)}{Y}$ and return.
        \item If $\Token.\Verify_\vk(\inp,s)$ rejects, apply $\onreg{\Write(\bot)}{Y}$ and return.
        \item Whenever $\onreg{\EqualH(i,0,\inp,s)}{L_i,D}$ fails for some $i < j$, apply $\onreg{\Write(\bot)}{Y}$ and return.
        \item \mark{\sout{For $i = 1,\dots, j$, compute $r_i \gets \meas{f_i^{\inp,r_1,\dots,r_{i-1}}, \theta_i, G_i}(\reg{V})$.}}
        \item Apply $\begin{cases}
            \onreg{\WriteH(j,0,\inp, s)}{Y,D} & \text{if } j < \nstep.\\
            \mark{\left(\onreg{P_{\inp}}{V_{\intext}} \onreg{U^\dag}{V_{\intext}} \onreg{\TP^\dag}{V_{\intext},V_{\out}}\right) \onreg{\Write(\reg{V_{\intext},V_{\outtext}})}{Y} \left(\onreg{\TP}{V_{\intext},V_{\out}} \onreg{U}{V_{\intext}} \onreg{P_{\inp}^{\dag}}{V_{\intext}}\right)} & \text{if } j = \nstep.
        \end{cases}$
    \end{enumerate}
    \item Output $\wt{\psi}$ and the oracle $\sF_7$.
\end{itemize}
\hrule\vspace{1mm}

\noindent
$\Sim(1^\secparam,1^n,1^m)$:
\begin{itemize}
    \item Generate input EPR pairs $(\epr_{(\intext,\pub)},\epr_{(\intext,\priv)})$ of length $n$.\\
    Generate output EPR pairs $(\epr_{(\outtext,\priv)},\epr_{(\outtext,\pub)})$ of length $n$.
    \item Sample an authentication key $k \gets \Auth.\KeyGen(1^{O(\secparam + m)}, 1^{O(m)})$.\\
    Compute the authenticated state $\onreg{\psi_\auth}{\wt{V}} \from \Auth.\Enc_k(0^{O(m)})$.
    \item Sample a signature token $(\vk,\tau_{\token}) \gets \Token.\Gen(1^\secparam, 1^{2n})$.
    \item \mark{Sample a PRF key $k_0 \from \zo^\secparam$ and set $H \equiv \PRF_{k_0} : \zo^* \to \zo^\kappa$.}
    \item \mark{Initialize registers $\reg{S_{\intext}} \gets \epr_{(\intext,\priv)}$ and $\reg{S_{\outtext}} \gets \epr_{(\outtext,\priv)}$.}
    \item Define the oracle $\sF_{\Sim}\left(j,\widetilde{v}, \inp,s,\ell_1,\dots,\ell_{\nstep}, \reg{Y} \right)^{\reg{S_{\intext},S_{\outtext}}}$ which makes black-box queries to $\CTRL\text{-}\!\left(U^\dag \left(\TP^\dag\; \Write\; \TP\right) U\right)$, where $\reg{Y}$ denotes the output register:
    \begin{enumerate}
        \item If $\Auth.\Ver_{k,\theta_j,G_j}(\widetilde{v})$ rejects, apply $\onreg{\Write(\bot)}{Y}$ and return.
        \item If $\Token.\Verify_\vk(\inp,s)$ rejects, apply $\onreg{\Write(\bot)}{Y}$ and return.
        \item Whenever $\mark{\ell_i \neq H(i,0,\inp,s)}$ for some $i < j$, apply $\onreg{\Write(\bot)}{Y}$ and return.
        \item Apply $\begin{cases}
            \mark{\onreg{\Write(H(j,0,\inp, s))}{Y}} & \text{if } j < \nstep.\\
            \mark{\left(\onreg{P_{\inp}}{S_{\intext}} \onreg{U^\dag}{S_{\intext}} \onreg{\TP^\dag}{S_{\intext},S_{\out}}\right) \onreg{\Write(\reg{S_{\intext},S_{\outtext}})}{Y} \left(\onreg{\TP}{S_{\intext},S_{\out}} \onreg{U}{S_{\intext}} \onreg{P_{\inp}^{\dag}}{S_{\intext}}\right)} & \text{if } j = \nstep.
        \end{cases}$
    \end{enumerate}
\end{itemize}
\end{framed}
}
\footnotetext{The operator is well-defined because $\EqualH(x)$ and $\EqualH(x')$ commute for every $x,x'$ (\cref{sec:purified-ro}).}

Let $\adv$ be a quantum polynomial-time adversary throughout the following lemmas. 
The output of the adversary will be denoted as $\adv^{\cO}(\wt{\psi})$. For simplicity, we sometimes write $\ket{\wt{\psi}}$ to denote the purified state that includes $\wt{\psi}$ and the state on registers $\reg{D,V}$.
\begin{lemma} \label{lemma:hybrid:use-auth-security}
$\left\{ \adv^{\sF}(\wt{\psi}) \middle| (\wt{\psi},\sF) \gets \QObf(1^\secparam,\psi) \right\}
\underset{2^{-\Omega(\secparam)}}{\approx}
\left\{ \adv^{\sF_1}(\wt{\psi}) \middle| (\wt{\psi},\sF_1) \gets \hybrid_1(1^\secparam,\psi) \right\}$
\end{lemma}
\begin{proof}
    The lemma follows directly from the functional security (\cref{dfn:FuncAuth}) of the authentication scheme (\cref{thm:FuncAuth}) since we have set its security parameter to be $O(\secparam + m)$.
\end{proof}

\begin{lemma}
$\abs{\Pr_{(\wt{\psi},\sF_1) \gets \hybrid_1(1^\secparam,\psi)} \left[1 \gets \adv^{\sF_1}(\wt{\psi}) \right]
-
\Pr_{(\wt{\psi},\sF_2) \gets \hybrid_2(1^\secparam,\psi)} \left[1 \gets \adv^{\sF_2}(\wt{\psi}) \right]} = \negl(\secparam)$
\end{lemma}
\begin{proof}
    The lemma follows from the computational indistinguishability between post-quantum PRFs and random oracles under polynomially-many quantum queries (\cref{dfn:PRF}) and that random oracles can be perfectly simulated in a purified form (\cref{sec:purified-ro}). 
\end{proof}

\begin{lemma} \label{lemma:hybrid:sig-token}
$\left\{ \adv^{\sF_2}(\wt{\psi}) \middle| (\wt{\psi},\sF_2) \gets \hybrid_2(1^\secparam,\psi) \right\}
\underset{2^{-\Omega(\secparam)}}{\approx}
\left\{ \adv^{\sF_3}(\wt{\psi}) \middle| (\wt{\psi},\sF_3) \gets \hybrid_3(1^\secparam,\psi) \right\}$
\end{lemma}
\begin{proof}
    From the definitions of the hybrids, the oracles $\sF_2, \sF_3$ only differ on inputs that have overlapping with the $1$-eigenspace of the projector
    $$\Pi_{\token} := \sum_{\inp,s: \Token.\Ver_{\vk}(\inp,s) = \top} \onreg{\ketbra{\inp}}{I} \ot \onreg{\ketbra{s}}{S} \ot \onreg{\Has_{\neq \inp}}{D}.$$
    By the oracle hybrid lemma (\cref{lemma:oracleswitch}), it suffices to show that
    \begin{equation*}
    \label{eq:no-double-inp}
        \E_{(\wt{\psi},\sF_3) \gets \hybrid_3} \norm{\Pi_{\token} \adv^{\sF_3} \ket{\wt{\psi}}} \le 2^{-\Omega(\secparam)}.
    \end{equation*}
    Given any $\poly(\secparam)$-query adversary $\adv$, let us consider the following algorithm $\cB$ that attempts to break the unforgeability of the tokenized signature scheme (\cref{dfn:token}).
    \begin{enumerate}
        \item Receive a challenge $(\vk, \tau_{\token})$ from the unforgeability game.
        \item Run $(\wt{\psi}, \sF_3) \gets \hybrid_{3}(1^{\secparam}, \psi)$ but using the challenge $(\vk, \tau_{\token})$ instead.
        \item Run $\adv^{\sF_3}$ and measure register $\reg{(I, S)}$ in the standard basis to get $(\inp, s)$.
        \item Measure registers $\reg{D \setminus D_{(\cdot,\cdot,\inp,\cdot)}}$ in the Hadamard basis. 
        \item Output $(\inp,s,\inp',s')$ if there is a measured entry $\reg{D_{(i',b',\inp',s')}}$ that is not $\ket{+^\kappa}$.
    \end{enumerate}
    From the definition of $\sF_3$, applying $\adv^{\sF_3}$ will keep entries $\reg{D_{(i',b',\inp',s')}}$ in the state $\ket{+^\kappa}$ unless $\Token.\Ver_\vk(\inp',s') = \top$. Also, measuring if there is an entry in $\reg{D \setminus D_{(\cdot,\cdot,\inp,\cdot)}}$ that is orthogonal to $\ket{+^\kappa}$ is equivalent to applying the projector $\onreg{\Has_{\neq \inp}}{D}$. Therefore, $\cB$ wins the unforgeability game with probability at least $\E_{(\wt{\psi},\sF_3) \gets \hybrid_3} \norm{\Pi_{\token} \adv^{\sF_3}\ket{\wt{\psi}}}$. Note that $\cB$ only makes $\poly(\secparam)$ queries to $\Token.\Ver_{\vk}$. By the unforgeability of the tokenized signature scheme, the winning probability of $\cB$ will be at most $2^{-\Omega(\secparam)}$, which implies the desired inequality.

\end{proof}

\begin{lemma} \label{lemma:hybrid:guess-label}
$\left\{ \adv^{\sF_3}(\wt{\psi}) \middle| (\wt{\psi},\sF_3) \gets \hybrid_3(1^\secparam,\psi) \right\}
\underset{2^{-\Omega(\secparam)}}{\approx}
\left\{ \adv^{\sF_4}(\wt{\psi}) \middle| (\wt{\psi},\sF_4) \gets \hybrid_4(1^\secparam,\psi) \right\}$
\end{lemma} 
\begin{proof}
    By \cref{thm:compile}, making adaptive measurements $r_i \gets \meas{f_i^{\inp,r_1,\dots,r_{i-1}}, \theta_i, G_i}(\reg{V})$ is equivalent to making the projective measurement $\{\onreg{\ketbra{\Phi_{\inp,r}}}{V}\}_{r}$.
    From the definitions of the hybrids, the oracles $\sF_3, \sF_4$ only differ on inputs that have overlapping with the $1$-eigenspace of one of the following projectors for some $i,\inp,s$.
    $$\onreg{\Pi'_{i,\inp,s}}{L_i,D,V} := \sum_{r} \onreg{\EqualH(i,r_i\oplus 1, \inp, s)}{L_i,D} \ot \onreg{\ketbra{\Phi_{\inp,r}}}{V}$$
    Consider another projector
    $\onreg{\Pi_{i,\inp,s}}{D,V} := \sum_{r} \onreg{\Has_{(i,r_i\oplus 1, \inp, s)}}{D} \ot  \onreg{\ketbra{\Phi_{\inp,r}}}{V}$.
    \begin{claim}\label{claim:database-has-no-wrong-label}
        $\norm{ \onreg{\Pi_{i,\inp,s}}{D,V} \adv^{\sF_4}\ket{\wt{\psi}} } \le 2^{-\Omega(\kappa)}$ for every $i, \inp, s$, every instantiation of $(\wt{\psi},\sF_4) \gets \hybrid_4$, and every $q$-query adversary $\adv$ where $q = \poly(\secparam)$.
    \end{claim}
    The claim will be proved shortly after. Following the claim, we have
    {\small
    \begin{align*}
        \norm{\onreg{\Pi'_{i,\inp,s}}{L_i,D,V} \adv^{\sF_4}\ket{\wt{\psi}} }
        \le& \norm{\onreg{\Pi'_{i,\inp,s}}{L_i,D,V} (I - \onreg{\Pi_{i,\inp,s}}{D,V}) \adv^{\sF_4}\ket{\wt{\psi}} } + \norm{\onreg{\Pi'_{i,\inp,s}}{L_i,D,V}  \onreg{\Pi_{i,\inp,s}}{D,V} \adv^{\sF_4}\ket{\wt{\psi}} }\\
        \le& \norm{\onreg{\Pi'_{i,\inp,s}}{L_i,D,V} (I - \onreg{\Pi_{i,\inp,s}}{D,V})}_{\mathsf{op}} + 2^{-\Omega(\kappa)}\\
        =& \norm{\sum_{r} \onreg{\EqualH(i,r_i \oplus 1,\inp,s)}{L_i,D} \onreg{(I - \Has_{i,r_i\oplus 1,\inp,s})}{D} \ot \onreg{\ketbra{\Phi_{\inp,r}}}{V}}_{\mathsf{op}} + 2^{-\Omega(\kappa)}\\
        \le& \sum_{a \in \zo} \norm{\onreg{\EqualH(i,a,\inp,s)}{L_i,D} \onreg{(I - \Has_{i,a,\inp,s})}{D}}_{\mathsf{op}} + 2^{-\Omega(\kappa)} \le 2^{-\Omega(\kappa)}
    \end{align*}
    }where the first inequality uses the triangle inequality, the second inequality follows from the claim, the third inequality follows from the triangle inequality and that the norm is multiplicative under tensor, and the last inequality follows from a random oracle analysis (\cref{lemma:ro-unpredictable}). Now a union bound over $(i, \inp, s) \in [t] \times \zo^{2n} \times \zo^{O(2n\cdot \secparam)}$ and the oracle hybrid argument (\cref{lemma:oracleswitch}) shows that the output states of $\hybrid_3$ and $\hybrid_4$ are within trace distance $$O\left( q\cdot t \cdot 2^{2n} \cdot 2^{O(2n\cdot\secparam)} \cdot 2^{-\Omega(\kappa)}\right) = 2^{-\Omega(\secparam)}$$

\end{proof}
\begin{proof}[Proof of \cref{claim:database-has-no-wrong-label}]
    We will write $\Pi_{i',\inp',s'}$ instead of $\Pi_{i,\inp,s}$.
    By \cref{thm:compile}, making adaptive measurements $r_i \gets \meas{f_i^{\inp,r_1,\dots,r_{i-1}}, \theta_i, G_i}(\reg{V})$ is equivalent to making the projective measurement $\{\onreg{\ketbra{\Phi_{\inp,r}}}{V}\}_{r}$.
    Looking at $\hybrid_4$, we can express $\sF_4 \equiv U_{P} P + \onreg{\Write(\bot)}{Y} (I-P)$ with the following projector $P$ and unitary $U_{P}$.
    {\footnotesize
    $$P := \sum_{j} \onreg{\ketbra{j}}{J} \ot \sum_{\begin{smallmatrix} \wt{v}: \Auth.\Ver_{k,\theta_j,G_j}(\wt{v}) = \top\\ \inp, s: \Token.\Ver_{\vk}(\inp,s) = \top \end{smallmatrix}}
    \onreg{\ketbra{\wt{v}}}{\wt{V}} \ot \onreg{\ketbra{\inp}}{I} \ot \onreg{\ketbra{s}}{S} \ot \prod_{i<j} \onreg{\left(\begin{smallmatrix}
        \EqualH(i,0,\inp,s)(I - \EqualH(i,1,\inp,s))\\
        +\EqualH(i,1,\inp,s)(I - \EqualH(i,0,\inp,s))
    \end{smallmatrix}\right)}{L_i, D}$$
    $$\begin{multlined}U_{P} :=  \left(
        \sum_{\inp,s,r} \onreg{\ketbra{\inp}}{I} \ot \onreg{\ketbra{s}}{S} \ot \onreg{\ketbra{\Phi_{\inp,r}}}{V} \ot \left( \begin{smallmatrix}
        \sum_{j<t} \onreg{\ketbra{j}}{J} \ot \onreg{\WriteH(j,r_j,\inp,s)}{Y,D}\\
        + \onreg{\ketbra{t}}{J} \ot \onreg{\Write(g(\inp,r))}{Y}
    \end{smallmatrix} \right)  \onreg{(I - \Has_{\neq \inp})}{D} \right)\\
    + \onreg{\Write(\bot)}{Y} \ot \sum_{\inp} \onreg{\ketbra{\inp}}{I} \ot \onreg{\Has_{\neq \inp}}{D} 
    \end{multlined}$$
    }We will show that the operator norm of $\Pi_{i',\inp',s'} \sF_4 (I - \Pi_{i',\inp',s'})$ is exponentially small, which means that $\sF_4$ does not help much with landing into the $1$-eigenspace of $\Pi_{i',\inp',s'}$. Intuitively, this is because on input $\inp = \inp'$, $\sF_4$ does not query the random oracle at the positions favored by $\Pi_{i',\inp',s'}$, whereas on input $\inp \neq \inp'$, $\sF_4$ will refuse to make changes to register $\reg{V}$ if the adversary has learned random oracle values on positions $(\cdot,\cdot,\inp',\cdot)$.
    To make the argument precise, we first show that $\Pi_{i',\inp',s'} U_{P} (I - \Pi_{i',\inp',s'}) = 0$. We start with some observations.
    \begin{itemize}
        \item $\Pi_{i',\inp',s'}$ only acts on registers $\reg{(D_{(i',\cdot,\inp',s')},V)}$.
        \item $\Has_{\neq \inp}$ only acts on registers $\reg{D \setminus D_{(\cdot,\cdot,\inp,\cdot)}}$.
        \item For $\inp' \neq \inp$ and any $i,b,s$, we have $\Has_{(i,b,\inp',s)} \Has_{\neq \inp} = \Has_{(i,b,\inp',s)} = \Has_{\neq \inp} \Has_{(i,b,\inp',s)}$.
    \end{itemize}
    Hence, $\Pi_{i',\inp',s'} = \sum_{r} \onreg{\Has_{(i',r_{i'}\oplus 1, \inp', s')}}{D} \ot  \onreg{\ketbra{\Phi_{\inp',r}}}{V}$ commutes with every $\Has_{\neq \inp}$, and thus it commutes also with $\left(\sum_{\inp} \onreg{\ketbra{\inp}}{I} \ot \onreg{\Has_{\neq \inp}}{D} \right)$. This shows that
    {\scriptsize
    \begin{align*}
        \Pi_{i',\inp',s'} \left(\onreg{\Write(\bot)}{Y} \left(\sum_{\inp} \onreg{\ketbra{\inp}}{I} \ot \onreg{\Has_{\neq \inp}}{D} \right) \right) (I - \Pi_{i',\inp',s'})
        = \onreg{\Write(\bot)}{Y} \left(\sum_{\inp} \onreg{\ketbra{\inp}}{I} \ot \onreg{\Has_{\neq \inp}}{D} \right) \Pi_{i',\inp',s'} (I - \Pi_{i',\inp',s'}) = 0
    \end{align*}
    }For $\inp' \neq \inp$, we also derive that $\Has_{(i,b,\inp',s)} (I - \Has_{\neq \inp}) = 0$, and thus $\Pi_{i',\inp',s'} (I - \Has_{\neq \inp}) = 0$. Furthermore, $\WriteH(j,b,\inp,s)$ only acts on register $\reg{(Y,D_{(j,b,\inp,s)})}$, so we have
    {\scriptsize
    \begin{align*}
    & \Pi_{i',\inp',s'} \left(\sum_{\inp,s,r} \onreg{\ketbra{\inp}}{I} \ot \onreg{\ketbra{s}}{S} \ot \onreg{\ketbra{\Phi_{\inp,r}}}{V} \ot \left( \begin{smallmatrix}
        \sum_{j<t} \onreg{\ketbra{j}}{J} \ot \onreg{\WriteH(j,r_j,\inp,s)}{Y,D}\\
        + \onreg{\ketbra{t}}{J} \ot \onreg{\Write(g(\inp,r))}{Y}
        \end{smallmatrix} \right)  \onreg{(I - \Has_{\neq \inp})}{D} \right) (I - \Pi_{i',\inp',s'})\\
    =& \Pi_{i',\inp',s'} \left(\sum_{\inp,s,r} \onreg{\ketbra{\inp}}{I} \ot \onreg{\ketbra{s}}{S} \ot \onreg{\ketbra{\Phi_{\inp,r}}}{V} \ot \onreg{(I - \Has_{\neq \inp})}{D} \left( \begin{smallmatrix}
        \sum_{j<t} \onreg{\ketbra{j}}{J} \ot \onreg{\WriteH(j,r_j,\inp,s)}{Y,D}\\
        + \onreg{\ketbra{t}}{J} \ot \onreg{\Write(g(\inp,r))}{Y}
        \end{smallmatrix} \right) \right) (I - \Pi_{i',\inp',s'})\\
    =& \Pi_{i',\inp',s'} \left(\sum_{s,r} \onreg{\ketbra{\inp',s,\Phi_{\inp',r}}}{I,S,V} \ot \onreg{(I - \Has_{\neq \inp'})}{D} \left( \begin{smallmatrix}
        \sum_{j<t} \onreg{\ketbra{j}}{J} \ot \onreg{\WriteH(j,r_j,\inp',s)}{Y,D}\\
        + \onreg{\ketbra{t}}{J} \ot \onreg{\Write(g(\inp',r))}{Y}
        \end{smallmatrix} \right) \right) (I - \Pi_{i',\inp',s'})\\
    =& \sum_{s,r} \onreg{\ketbra{\inp',s,\Phi_{\inp',r}}}{I,S,V} \ot \onreg{(I - \Has_{\neq \inp'})}{D} \left(\onreg{\Has_{(i', r_{i'} \oplus 1, \inp', s')}}{D} \left( \begin{smallmatrix}
        \sum_{j<t} \onreg{\ketbra{j}}{J} \ot \onreg{\WriteH(j,r_j,\inp',s)}{Y,D}\\
        + \onreg{\ketbra{t}}{J} \ot \onreg{\Write(g(\inp',r))}{Y}
        \end{smallmatrix} \right) \onreg{(I - \Has_{(i', r_{i'} \oplus 1, \inp', s')})}{D} \right)\\
    =& \sum_{s,r} \onreg{\ketbra{\inp',s,\Phi_{\inp',r}}}{I,S,V} \ot \onreg{(I - \Has_{\neq \inp'})}{D} \left( \left( \begin{smallmatrix}
        \sum_{j<t} \onreg{\ketbra{j}}{J} \ot \onreg{\WriteH(j,r_j,\inp',s)}{Y,D}\\
        + \onreg{\ketbra{t}}{J} \ot \onreg{\Write(g(\inp',r))}{Y}
        \end{smallmatrix} \right) \onreg{\Has_{(i', r_{i'} \oplus 1, \inp', s')}}{D} \onreg{(I - \Has_{(i', r_{i'} \oplus 1, \inp', s')})}{D} \right)\\
        =& 0
    \end{align*}
    }Combining the equalities gives $\Pi_{i',\inp',s'} U_{P} (I - \Pi_{i',\inp',s'}) = 0$. From commutativity, we also have $\Pi_{i',\inp',s'} \onreg{\Write(\bot)}{Y} (I - \Pi_{i',\inp',s'}) = 0$. Next, we derive an upper bound for $\norm{\Pi_{i',\inp',s'} P (I - \Pi_{i',\inp',s'})}_{\mathsf{op}}$. Since $\EqualH(i,\cdot,\inp,s)$ only acts on register $(\reg{L_i, D_{(i,\cdot,\inp,s)}})$, several terms in $P$ commutes with $\Pi_{i',\inp',s'}$ and allows us to simplify $\Pi_{i',\inp',s'} P (I - \Pi_{i',\inp',s'})$ as
    
    {\scriptsize
    \begin{align*}
    & \Pi_{i',\inp',s'} \left(\sum_{j>i'} \onreg{\ketbra{j}}{J} \ot \sum_{\wt{v}: \Ver(\wt{v}) = \top}
    \onreg{\ketbra{\wt{v}}}{\wt{V}} \ot \onreg{\ketbra{\inp'}}{I} \ot \onreg{\ketbra{s'}}{S} \ot  \onreg{\left(\begin{smallmatrix}
        \EqualH(i',0,\inp',s')(I - \EqualH(i',1,\inp',s'))\\
        +\EqualH(i',1,\inp',s')(I - \EqualH(i',0,\inp',s'))
    \end{smallmatrix}\right)}{L_{i'}, D}\right) (I - \Pi_{i',\inp',s'})\\
    =& \sum_{j>i'}\; \sum_{\wt{v}: \Ver(\wt{v}) = \top}\; \sum_r\; \begin{multlined}[t]
        \onreg{\ketbra{j}}{J} \ot \onreg{\ketbra{\wt{v}}}{\wt{V}} \ot \onreg{\ketbra{\inp'}}{I} \ot \onreg{\ketbra{s'}}{S} \ot \onreg{\ketbra{\Phi_{\inp',r}}}{V} \ot \\
        \onreg{\Has_{(i', r_{i'} \oplus 1, \inp', s')}}{D} \onreg{\left(\begin{smallmatrix}
        \EqualH(i',0,\inp',s')(I - \EqualH(i',1,\inp',s'))\\
        +\EqualH(i',1,\inp',s')(I - \EqualH(i',0,\inp',s'))
    \end{smallmatrix}\right)}{L_{i'}, D} \onreg{(I - \Has_{(i', r_{i'} \oplus 1, \inp', s')})}{D} 
    \end{multlined}
    \end{align*}
    }By the fact that the operator norm is multiplicative under tensor and the triangle inequality, the expression above has operator norm at most
    {\small
    \begin{align*}
    & \sum_{r_{i'} \in \zo} \norm{\onreg{\Has_{(i', r_{i'} \oplus 1, \inp', s')}}{D} \onreg{\left(\begin{smallmatrix}
        \EqualH(i',0,\inp',s')(I - \EqualH(i',1,\inp',s'))\\
        +\EqualH(i',1,\inp',s')(I - \EqualH(i',0,\inp',s'))
    \end{smallmatrix}\right)}{L_{i'}, D} \onreg{(I - \Has_{(i', r_{i'} \oplus 1, \inp', s')})}{D} }_{\mathsf{op}}\\
    =& \sum_{a \in \zo} \norm{\onreg{\Has_{(i', a, \inp', s')}}{D} \onreg{\left(\begin{smallmatrix}
        \EqualH(i',a,\inp',s')(I - \EqualH(i',a \oplus 1,\inp',s'))\\
        +\EqualH(i',a \oplus 1,\inp',s')(I - \EqualH(i',a,\inp',s'))
    \end{smallmatrix}\right)}{L_{i'}, D} \onreg{(I - \Has_{(i', a, \inp', s')})}{D} }_{\mathsf{op}}\\
    \le& \sum_{a \in \zo} 2 \norm{ \onreg{\EqualH(i',a,\inp',s')}{L_{i'}, D} \onreg{\left(I - \Has_{(i', a, \inp', s')}\right)}{D} }_{\mathsf{op}} \le 2^{-\Omega(\kappa)}
    \end{align*}
    }where we use the commutativity of $\EqualH(i',a,\inp',s'), \EqualH(i',a \oplus 1,\inp',s')$, the triangle inequality, that norm is sub-multiplicative, followed by a random oracle analysis (\cref{lemma:ro-unpredictable}). Hence, we have $\norm{\Pi_{i',\inp',s'} P (I - \Pi_{i',\inp',s'})}_{\mathsf{op}} \le 2^{-\Omega(\kappa)}$. Together with $\Pi_{i',\inp',s'} U_{P} (I - \Pi_{i',\inp',s'}) = 0$ and $\Pi_{i',\inp',s'} \onreg{\Write(\bot)}{Y} (I - \Pi_{i',\inp',s'}) = 0$, we derive that
    {\small
    \begin{align*}
        \norm{\Pi_{i',\inp',s'} \sF_4 (I - \Pi_{i',\inp',s'})}_{\mathsf{op}} 
        =& \norm{\Pi_{i',\inp',s'} \left(U_{P} P + \onreg{\Write(\bot)}{Y} (I-P)\right) (I - \Pi_{i',\inp',s'})}_{\mathsf{op}}\\
        =& \norm{\Pi_{i',\inp',s'} \left(U_{P} \Pi_{i',\inp',s'} P + \onreg{\Write(\bot)}{Y} \Pi_{i',\inp',s'} (I-P)\right) (I - \Pi_{i',\inp',s'})}_{\mathsf{op}}\\
        \le& \norm{\Pi_{i',\inp',s'} P (I - \Pi_{i',\inp',s'})}_{\mathsf{op}} + \norm{\Pi_{i',\inp',s'} (I-P) (I - \Pi_{i',\inp',s'})}_{\mathsf{op}} \le 2^{-\Omega(\kappa)}
    \end{align*}
    }If $\adv$ makes $q>0$ queries, we can write $\adv^{\sF_4} = U_q \sF_4 \adv_{<q}^{\sF_4}$ where $\adv_{<q}$ makes $q-1$ queries. Since $U_q$ commutes with $\Pi_{i,\inp,s}$, we have
    \begin{align*}
        \norm{\Pi_{i,\inp,s} \adv^{\sF_4}\ket{\wt{\psi}}} =& \norm{\Pi_{i,\inp,s} U_q \sF_4 \adv_{<q}^{\sF_4}\ket{\wt{\psi}}}\\
        \le& \norm{ U_q \Pi_{i,\inp,s} \sF_4 (I - \Pi_{i,\inp,s}) \adv_{<q}^{\sF_4}\ket{\wt{\psi}}} +  \norm{U_q \Pi_{i,\inp,s} \sF_4 \Pi_{i,\inp,s} \adv_{<q}^{\sF_4}\ket{\wt{\psi}}}\\
        \le& \norm{\Pi_{i,\inp,s} \sF_4 (I - \Pi_{i,\inp,s})}_{\mathsf{op}} + \norm{\Pi_{i,\inp,s} \adv_{<q}^{\sF_4}\ket{\wt{\psi}}}
    \end{align*}
    Before any queries have been made, the state on $\reg{D}$ is $\ket{\cL_\emptyset}$ so we have 
    $$\onreg{\Pi_{i,\inp,s}}{D,V} \adv_{<1}^{\sF_4}\ket{\wt{\psi}} = \onreg{\Pi_{i,\inp,s}}{D,V} \onreg{\ketbra{\cL_\emptyset}}{D} \adv_{<1}^{\sF_4}\ket{\wt{\psi}} = 0$$
    By induction on $q = \poly(\secparam)$, we get
    $$\norm{\Pi_{i,\inp,s} \adv^{\sF_4}\ket{\wt{\psi}}} \le q \norm{\Pi_{i,\inp,s} \sF_4 (I - \Pi_{i,\inp,s})}_{\mathsf{op}} \le 2^{-\Omega(\kappa)}$$
\end{proof}

\begin{lemma}
$\left\{ \adv^{\sF_4}(\wt{\psi}) \middle| (\wt{\psi},\sF_4) \gets \hybrid_4(1^\secparam,\psi) \right\}
=
\left\{ \adv^{\sF_5}(\wt{\psi}) \middle| (\wt{\psi},\sF_5) \gets \hybrid_5(1^\secparam,\psi) \right\}$
\end{lemma}
\begin{proof} Define the projector
{\small
    $$P := \sum_{j} \onreg{\ketbra{j}}{J} \ot \sum_{\begin{smallmatrix} \wt{v}: \Auth.\Ver_{k,\theta_j,G_j}(\wt{v}) = \top\\ \inp, s: \Token.\Ver_{\vk}(\inp,s) = \top \end{smallmatrix}}
    \onreg{\ketbra{\wt{v}}}{\wt{V}} \ot \onreg{\inp}{I} \ot \onreg{s}{S} \ot \prod_{i<j} \onreg{\left(\begin{smallmatrix}
        \EqualH(i,0,\inp,s)(I - \EqualH(i,1,\inp,s))\\
        +\EqualH(i,1,\inp,s)(I - \EqualH(i,0,\inp,s))
    \end{smallmatrix}\right)}{L_i, D}$$
}
By definition of $\hybrid_4, \hybrid_5$, we can express $\sF_4, \sF_5$ as
{\small
    $$\sF_4 = \begin{multlined}[t]
        \left(\sum_{\inp,s,r} \onreg{\ketbra{\inp}}{I} \ot \onreg{\ketbra{s}}{S} \ot \onreg{\ketbra{\Phi_{\inp,r}}}{V} \ot \left( \begin{smallmatrix}
            \sum_{j<t} \onreg{\ketbra{j}}{J} \ot \onreg{\WriteH(j,r_j,\inp,s)}{L_i,D}\\
            + \onreg{\ketbra{t}}{J} \ot \onreg{\Write(g(\inp,r))}{L_i}
        \end{smallmatrix} \right) \onreg{(I - \Has_{\neq \inp})}{D} \right) P\\
    + \onreg{\Write(\bot)}{L_i} \left(\sum_{\inp} \onreg{\ketbra{\inp}}{I} \ot \onreg{\Has_{\neq \inp}}{D}\right) P + \onreg{\Write(\bot)}{L_i} (I-P)
    \end{multlined}
    $$
    $$\sF_5 = \begin{multlined}[t]
        \left(\sum_{\inp,s,r} \onreg{\ketbra{\inp}}{I} \ot \onreg{\ketbra{s}}{S} \ot \onreg{\ketbra{\Phi_{\inp,r}}}{V} \ot \left( \begin{smallmatrix}
            \sum_{j<t} \onreg{\ketbra{j}}{J} \ot \onreg{\WriteH(j,0,\inp,s)}{L_i,D}\\
            + \onreg{\ketbra{t}}{J} \ot \onreg{\Write(g(\inp,r))}{L_i}
        \end{smallmatrix} \right) \onreg{(I - \Has_{\neq \inp})}{D} \right) P\\
    + \onreg{\Write(\bot)}{L_i} \left(\sum_{\inp} \onreg{\ketbra{\inp}}{I} \ot \onreg{\Has_{\neq \inp}}{D}\right) P + \onreg{\Write(\bot)}{L_i} (I-P)
    \end{multlined}
    $$
}Consider the following unitary $W$ that conditionally swaps the database entries of $(i,r_i,\inp,s)$ and $(i,0,\inp,s)$. Here, we make the convention that $\onreg{\SWAP}{D_x, D_{x'}} := \onreg{I}{D_x}$ when $x=x'$.
    $$\onreg{W}{V,D} := \onreg{I}{V} \ot \onreg{\big(I - \sum_{\inp} \Only_{\inp}\big)}{D} + \sum_{\inp, r} \onreg{\ketbra{\Phi_{\inp,r}}}{V} \ot \left(\prod_{i,s} \onreg{\SWAP}{D_{(i,r_i,\inp,s)},D_{(i,0,\inp,s)}}\right) \onreg{\Only_{\inp}}{D} $$
We will show that $W \sF_4 W = \sF_5$. First, we have $W^2 = I$ because $\SWAP^2 = I$ and the projectors $\Only_{\inp}$ are pairwise orthogonal.
Also, since $I - \Has_{\neq \inp} = \ketbra{\cL_\emptyset} + \Only_{\inp}$, we have
{\small
\begin{align*}
    W \left(\sum_{\inp} \onreg{\ketbra{\inp}}{I} \ot \onreg{\Has_{\neq \inp}}{D}\right) W &= W \left(\sum_{\inp} \onreg{\ketbra{\inp}}{I} \ot \onreg{(I - \ketbra{\cL_\emptyset} - \Only_\inp)}{D}\right) W\\
    &= \sum_{\inp} \onreg{\ketbra{\inp}}{I} \ot \onreg{(I - \ketbra{\cL_\emptyset} - \Only_\inp)}{D}
    = \sum_{\inp} \onreg{\ketbra{\inp}}{I} \ot \onreg{\Has_{\neq \inp}}{D}
\end{align*}
}Next, observe that for any $r_i \in \zo$, the term $\onreg{\left(\begin{smallmatrix}
        \EqualH(i,0,\inp,s)(I - \EqualH(i,1,\inp,s))\\
        +\EqualH(i,1,\inp,s)(I - \EqualH(i,0,\inp,s))
    \end{smallmatrix}\right)}{L_i, D}$ is equal to
$$\onreg{\SWAP}{D_{(i,r_i,\inp,s)},D_{(i,0,\inp,s)}} \onreg{\left(\begin{smallmatrix}
        \EqualH(i,0,\inp,s)(I - \EqualH(i,1,\inp,s))\\
        +\EqualH(i,1,\inp,s)(I - \EqualH(i,0,\inp,s))
    \end{smallmatrix}\right)}{L_i, D} \onreg{\SWAP}{D_{(i,r_i,\inp,s)},D_{(i,0,\inp,s)}}$$
since it is symmetric in the second argument of $(i,\cdot,\inp,s)$. This implies that $W P W = P$. 
Next, observe that $W \onreg{(I-\Has_{\neq \inp})}{D} = W \onreg{(\ketbra{\cL_\emptyset} + \Only_\inp)}{D}$ is equal to
\begin{align*}
    & \onreg{I}{V} \ot \onreg{\ketbra{\cL_\emptyset}}{D} + \sum_{r} \onreg{\ketbra{\Phi_{\inp,r}}}{V} \ot  \left(\prod_{i,s} \onreg{\SWAP}{D_{(i,r_i,\inp,s)},D_{(i,0,\inp,s)}}\right) \onreg{\Only_{\inp}}{D}\\
    =& \sum_{r} \onreg{\ketbra{\Phi_{\inp,r}}}{V} \ot \left(\prod_{i,s} \onreg{\SWAP}{D_{(i,r_i,\inp,s)},D_{(i,0,\inp,s)}}\right) \onreg{(\ketbra{\cL_\emptyset} + \Only_{\inp})}{D}\\
    =& \sum_{r} \onreg{\ketbra{\Phi_{\inp,r}}}{V} \ot \left(\prod_{i,s} \onreg{\SWAP}{D_{(i,r_i,\inp,s)},D_{(i,0,\inp,s)}}\right) \onreg{(I - \Has_{\neq \inp})}{D}
\end{align*}
where the first equality follows because $\onreg{\SWAP}{D_{(i,r_i,\inp,s)},D_{(i,0,\inp,s)}} \ket{\cL_\emptyset} = \ket{\cL_\emptyset}$. Similarly, the expression is also equal to $\onreg{(I-\Has_{\neq \inp})}{D} W$. Also, simple calculations show that
{\footnotesize
$$\left( \begin{smallmatrix}
            \sum_{j<t} \onreg{\ketbra{j}}{J} \ot \onreg{\WriteH(j,r_j,\inp,s)}{L_i,D}\\
            + \onreg{\ketbra{t}}{J} \ot \onreg{\Write(g(\inp,r))}{L_i}
        \end{smallmatrix} \right) \onreg{(I - \Has_{\neq \inp})}{D} = \onreg{(I - \Has_{\neq \inp})}{D} \left( \begin{smallmatrix}
            \sum_{j<t} \onreg{\ketbra{j}}{J} \ot \onreg{\WriteH(j,r_j,\inp,s)}{L_i,D}\\
            + \onreg{\ketbra{t}}{J} \ot \onreg{\Write(g(\inp,r))}{L_i}
        \end{smallmatrix} \right) \onreg{(I - \Has_{\neq \inp})}{D} $$
}and that $\left( 
            \sum_{j<t} \onreg{\ketbra{j}}{J} \ot \onreg{\WriteH(j,0,\inp,s)}{L_i,D}
            + \onreg{\ketbra{t}}{J} \ot \onreg{\Write(g(\inp,r))}{L_i}
        \right)$ is equal to
{\small
    $$\left(\prod_{i,s} \onreg{\SWAP}{D_{(i,r_i,\inp,s)},D_{(i,0,\inp,s)}}\right) \left( \begin{smallmatrix}
            \sum_{j<t} \onreg{\ketbra{j}}{J} \ot \onreg{\WriteH(j,r_j,\inp,s)}{L_i,D}\\
            + \onreg{\ketbra{t}}{J} \ot \onreg{\Write(g(\inp,r))}{L_i}
        \end{smallmatrix} \right) \left(\prod_{i,s} \onreg{\SWAP}{D_{(i,r_i,\inp,s)},D_{(i,0,\inp,s)}}\right)$$
}Combining these observations, we derive that
{\small
\begin{align*}
    &W\left(\sum_{\inp,s,r} \onreg{\ketbra{\inp}}{I} \ot \onreg{\ketbra{s}}{S} \ot \onreg{\ketbra{\Phi_{\inp,r}}}{V} \ot \left( \begin{smallmatrix}
            \sum_{j<t} \onreg{\ketbra{j}}{J} \ot \onreg{\WriteH(j,r_j,\inp,s)}{L_i,D}\\
            + \onreg{\ketbra{t}}{J} \ot \onreg{\Write(g(\inp,r))}{L_i}
        \end{smallmatrix} \right) \onreg{(I - \Has_{\neq \inp})}{D} \right)W\\
    =& \left(\sum_{\inp,s,r} \onreg{\ketbra{\inp}}{I} \ot \onreg{\ketbra{s}}{S} \ot \onreg{\ketbra{\Phi_{\inp,r}}}{V} \ot \left( \begin{smallmatrix}
            \sum_{j<t} \onreg{\ketbra{j}}{J} \ot \onreg{\WriteH(j,0,\inp,s)}{L_i,D}\\
            + \onreg{\ketbra{t}}{J} \ot \onreg{\Write(g(\inp,r))}{L_i}
        \end{smallmatrix} \right) \onreg{(I - \Has_{\neq \inp})}{D} \right)
\end{align*}
}From all of the above, we obtain that $W \sF_4 W = \sF_5$. Hence, we have 
$$\adv^{\sF_5}(\wt{\psi}) = \onreg{W}{V,D} \adv^{\sF_4} \onreg{W}{V,D} (\onreg{\ketbra{\cL_{\emptyset}}}{D} \wt{\psi}) = \onreg{W}{V,D} \adv^{\sF_4} (\wt{\psi})$$
where we recall that register $\reg{D}$ was initialized as $\ket{\cL_\emptyset}$ during the preparation of $\wt{\psi}$ and we use the relation $\onreg{W}{V,D} \onreg{\ketbra{\cL_{\emptyset}}}{D} = \onreg{\ketbra{\cL_{\emptyset}}}{D}$. This proves the lemma because the adversary's output state does not contain registers $\reg{V,D}$. \end{proof}

\begin{lemma}
$\left\{ \adv^{\sF_5}(\wt{\psi}) \middle| (\wt{\psi},\sF_5) \gets \hybrid_5(1^\secparam,\psi) \right\}
\underset{2^{-\Omega(\secparam)}}{\approx}
\left\{ \adv^{\sF_6}(\wt{\psi}) \middle| (\wt{\psi},\sF_6) \gets \hybrid_6(1^\secparam,\psi) \right\}$
\end{lemma}
\begin{proof}
    First, by a similar argument as the proof of \cref{lemma:hybrid:sig-token}, removing the condition ``If $\Has_{\neq \inp}$ holds'' from $\sF_5$ would only affect the output state by at most $2^{-\Omega(\secparam)}$ in trace distance. The modified oracle and $\sF_6$ only differ on inputs that have overlapping with the $1$-eigenspace of $\onreg{\EqualH(i,1,\inp,s)}{L_i,D}$ for some $i,\inp,s$. Similar to the proof of \cref{lemma:hybrid:guess-label}, by a union bound and the oracle hybrid lemma (\cref{lemma:oracleswitch}), it suffices to show that
    $$ \norm{\onreg{\EqualH(i,1,\inp,s)}{L_i,D} \adv^{\sF_6}\ket{\wt{\psi}}} \le 2^{-\Omega(\kappa)}$$
    for every $i,\inp,s$, every instantiation of $(\wt{\psi},\sF_6) \gets \hybrid_6$, and polynomial-query adversary $\adv$. By the definition of $\hybrid_6$, we have $\onreg{\Has_{(i,1,\inp,s)}}{D} \adv^{\sF_6}(\wt{\psi}) = 0$. Hence, we have
    $$
        \norm{\onreg{\EqualH(i,1,\inp,s)}{L_i,D} \adv^{\sF_6}\ket{\wt{\psi}}}
        = \norm{\onreg{\EqualH(i,1,\inp,s)}{L_i,D} \onreg{(I-\Has_{(i,1,\inp,s)})}{D} \adv^{\sF_6}\ket{\wt{\psi}}} \le 2^{-\Omega(\kappa)}
    $$
    as desired by a random oracle analysis (\cref{lemma:ro-unpredictable}).
\end{proof}

\begin{lemma}
$\left\{ \adv^{\sF_6}(\wt{\psi}) \middle| (\wt{\psi},\sF_6) \gets \hybrid_6(1^\secparam,\psi) \right\}
\underset{\negl(\secparam)}{\approx}
\left\{ \adv^{\sF_7}(\wt{\psi}) \middle| (\wt{\psi},\sF_7) \gets \hybrid_7(1^\secparam,\psi) \right\}$
\end{lemma}
\begin{proof}
    Since $\adv$ only makes $\poly(\secparam)$ queries, by a standard hybrid argument, it suffices to prove that replacing the last query to $\sF_7$ with $\sF_6$ one at a time only affects the output state by $\negl(\secparam)$ trace distance. Therefore, we just have to prove that for every instantiation of $(\wt{\psi},\sF_7) \gets \hybrid_7$ together with $\sF_6$ defined using the same key as $\sF_7$, and every adversary $\adv$,
    $$\sF_7 \adv^{\sF_7}(\wt{\psi}) \underset{\negl(\secparam)}{\approx} \sF_6 \adv^{\sF_7}(\wt{\psi})$$
    To compare the difference between $\sF_6$ and $\sF_7$, observe that when the query input $j<\nstep$, the internal results $r_i$ computed in line $4$ of $\sF_6$ are never used, so removing this step does not affect oracle behavior in this case. When the query input $j = \nstep$ and controlled on all verifications pass, $\sF_6$ computes $g(\inp,r)$ according to the PLM quantum program compiled from $\wt{Q}^{\mathsf{univ}}$ and applies $\onreg{\Write(g(\inp,r))}{Y}$. It is equivalent to applying 
    $$\sum_{y} \onreg{(\Xg^y)}{Y} \ot \sum_{r: g(\inp,r) = y} \onreg{\ketbra{\Phi_{\inp,r}}}{V}$$
    Note that in $\adv^{\sF_7}(\wt{\psi})$, registers $\reg{V_{\aux}, V_{\PLM}}$ still hold the states $\psi, \psi_{\PLM}$ because $\sF_7$ does not use these registers at all. By \cref{thm:compile},
    {\small
    $$\sum_y \Xg^y \ot \sum_{r: g(\inp,r) = y}\ketbra{\Phi_{\inp,r}} (I \ot \ket{\psi_{\PLM}}) = \sum_y \Xg^y \ot \left(U_{\wt{Q}^{\mathsf{univ}},\inp}^\dag \left(\ketbra{y} \ot I\right) U_{\wt{Q}^{\mathsf{univ}},\inp}\right) \ot \ket{\psi_{\PLM}}$$
    }where $\onreg{U_{\wt{Q}^{\mathsf{univ}},\inp}}{V} = \onreg{\TP}{V_\intext, V_\outtext} \onreg{Q^{\mathsf{univ}}}{V_\intext, V_\aux} \onreg{P_\inp^\dag}{V_\intext}$. 
    Hence, it is equivalent if $\sF_6$ instead applies
    \begin{align*}
        & \onreg{U_{\wt{Q}^{\mathsf{univ}},\inp}^\dag}{V}\; \onreg{\Write(\reg{V_\intext, V_\outtext})}{Y} \onreg{U_{\wt{Q}^{\mathsf{univ}},\inp}}{V} \\
        =&\left(\onreg{P_\inp}{V_\intext} \onreg{Q^{\mathsf{univ}\;\dag}}{V_\intext, V_\aux} \onreg{\TP^\dag}{V_\intext, V_\outtext}\right) \onreg{\Write(\reg{V_\intext, V_\outtext})}{Y} \left(\onreg{\TP}{V_\intext, V_\outtext} \onreg{Q^{\mathsf{univ}}}{V_\intext, V_\aux} \onreg{P_\inp^\dag}{V_\intext} \right) 
    \end{align*}
    In comparison, when the query input $j = \nstep$ and controlled on all verifications pass, $\sF_7$ applies
    $$\left(\onreg{P_\inp}{V_\intext} \onreg{U^{\dag}}{V_\intext} \onreg{\TP^\dag}{V_\intext, V_\outtext}\right) \onreg{\Write(\reg{V_\intext, V_\outtext})}{Y} \left(\onreg{\TP}{V_\intext, V_\outtext} \onreg{U}{V_\intext} \onreg{P_\inp^\dag}{V_\intext}\right) $$
    As noted above, the state $\adv^{\sF_7}(\wt{\psi})$ still holds $\psi$ on register $\reg{V_{\aux}}$ . By assumption, $Q^{\mathsf{univ}}(\cdot, \psi)$ is an $\negl(\secparam)$-approximation of $U$, so it follows by \cref{lemma:reuse-program-once} that $\sF_6 \adv^{\sF_7}(\wt{\psi})$ is negligibly close to $\sF_7 \adv^{\sF_7}(\wt{\psi})$ in trace distance.
\end{proof}

\begin{lemma}
{\footnotesize
$\abs{ \underset{(\wt{\psi},\sF_7) \gets \hybrid_{7}(1^\secparam,\psi)}{\Pr} \left[1\gets \adv^{\sF_{7}}(\wt{\psi}) \right] - 
\underset{(\wt{\psi},\sF_\Sim) \gets \Sim(1^\secparam,1^n,1^m)}{\Pr} \left[1\gets \adv^{\sF_\Sim^{\CTRL\text{-}(U^\dag\cdot U)}}(\wt{\psi}) \right] } = \negl(\secparam)$.
}\end{lemma}
\begin{proof}
    Observe that the registers $\reg{V_{\aux}, V_{\PLM}}$ storing $\psi, \psi_{\PLM}$ are never used in $\hybrid_7$, so they can be completely removed. The remaining difference is that the purified random oracle in $\hybrid_7$ is changed to a post-quantum PRF in $\Sim$, and they are computationally indistinguishable under polynomially-many quantum queries.
\end{proof}

\begin{proof}[Proof of \cref{thm:main}]
    It is clear by definition that $\Sim$ runs in polynomial time. From the above lemmas which prove indistinguishability between consecutive hybrids, it follows that for every polynomial-time adversary $\adv$,
    $$\abs{\Pr_{(\wt{\psi},\sF) \gets \QObf(1^\secparam,\psi)} \left[1 \gets  \adv^{\sF}(\wt{\psi}) \right] - \Pr_{(\wt{\psi},\sF_\Sim) \gets \Sim(1^\secparam,1^n,1^m)} \left[ 1 \gets \adv^{\sF_\Sim^{\CTRL\text{-}(U^\dag \cdot U)}}(\wt{\psi}) \right]} = \negl(\secparam).$$
    This proves the ideal obfuscation property. To prove the functionality-preserving property, we apply \cref{lemma:error-transform,lemma:hybrid:use-auth-security} with $\adv = \QEval$, so it suffices to prove that for every state $\rho_{\intext}$,
\[
    \E_{(\widetilde{\psi}, \sF_1) \leftarrow \hybrid_1(1^\lambda, (\psi,Q^{\mathsf{univ}}))}\left[ \QEval^{\sF_1}\left(\rho_{\intext}, \wt{\psi} \right) \right] \underset{\negl(\secparam)}{\approx} 
    Q^{\mathsf{univ}}(\rho_{\intext}, \psi).
\]
First, $\QEval$ performs teleportation $\inp \gets \TPSend(\rho_{\intext}, \epr_{(\intext,\pub)})$, and the state $\epr_{(\intext,\priv)}$ on register $\reg{V_\intext}$ collapses to $P_\inp \rho_{\intext} P_\inp^\dag$. Then $\QEval$ obtains a signature $s \gets \Token.\Sign(\inp, \tau_\token)$. When $\QEval$ calls $\sF_1$, the first two verification steps of $\sF_1$ always succeed by the correctness of the authentication code and the tokenized signature. Furthermore, for every $\inp$ and $s$, the third verification step of $\sF_1$ encounters a collision only with negligible probability over the choice of PRF key. When no collision occurs, the value $\ell_\nstep$ is computed as
{\small
$$\ell_{\nstep} \gets g(\inp,r_1,\dots,r_\nstep) \text{ where } \reg{V} \gets \left(P_\inp \rho_{\intext} P_\inp^\dag, \epr_{(\outtext,\priv)}, \psi, \psi_{\PLM}\right) \text{ and } r_j \gets \meas{f_j^{\inp,r_j,\dots,r_{j-1}}, \theta_j, G_j} (\reg{V})$$
}In this case, by \cref{thm:compile}, the value $\ell_\nstep$ can also be computed as
$$\ell_{\nstep} \gets \wt{Q}^{\mathsf{univ}}\left(P_\inp \rho_{\intext} P_\inp^\dag, \epr_{(\outtext,\priv)}, \psi, \psi_{\PLM}\;;\; \inp \right) 
\equiv \TPSend \left(Q^{\mathsf{univ}}\left(\rho_\intext, \psi\right),\epr_{(\outtext,\priv)}\right)$$
Hence, the output $\TPRecv(\ell_\nstep, \epr_{(\outtext,\pub)})$ of $\QEval^{\sF_1}$ is negligibly close in trace distance to
\begin{align*}
\TPRecv\left(\TPSend \left(Q^{\mathsf{univ}}\left(\rho_\intext, \psi\right),\epr_{(\outtext,\priv)}\right), \epr_{(\outtext,\pub)} \right) 
= Q^{\mathsf{univ}}\left(\rho_\intext, \psi\right)
\end{align*}
by the correctness of quantum teleportation since $(\epr_{(\outtext,\priv)}, \epr_{(\outtext,\pub)})$ is an EPR .
\end{proof}

\ifdefined\AckSection
\section*{Acknowledgement}
The authors deeply appreciate Andrea Coladangelo for his selfless devotion in his line-by-line feedback in polishing the paper and discussing it with them. The authors are grateful to Rachel Lin for deepening their understanding of classical obfuscation. The authors also thank Jiapeng Zhang, Shanghua Teng, Kai-Min Chung, Eli Goldin, Aparna Gupte, and anonymous FOCS reviewers for useful discussions.

Miryam Huang is supported by NSF CAREER award 2141536. Part of the work was done when Miryam Huang was visiting to UW Quantum group and Cryptography group. Er-Cheng Tang is supported by NSF NRT-QL award 2021540. This work was initiated during Miryam Huang and Er-Cheng Tang's visit to the Simons Institute for the Theory of
Computing.
\fi

\ifdefined\LLNCS
\bibliographystyle{splncs04}
\else
\bibliographystyle{alpha}
\fi
\bibliography{references}

\ifdefined\IncludeAppendix
\appendix
\fi

\end{document}